\documentclass[aps,prd,twocolumn,superscriptaddress,preprintnumbers,nofootinbib,longbibliography,floatfix]{revtex4-1}

\usepackage[utf8]{inputenc}
\usepackage{amssymb}
\usepackage{graphicx}
\usepackage{amsmath}
\usepackage{hyperref}
\usepackage{xspace}
\usepackage{hepunits}
\usepackage[caption=false]{subfig}

\usepackage{csquotes}
\MakeOuterQuote{"}

\usepackage{multirow}

\newcommand{\zcut}{z_\text{cut}}

\DeclareRobustCommand{\Sec}[1]{Sec.~\ref{#1}}
\DeclareRobustCommand{\Secs}[2]{Secs.~\ref{#1} and \ref{#2}}
\DeclareRobustCommand{\App}[1]{App.~\ref{#1}}
\DeclareRobustCommand{\Tab}[1]{Table~\ref{#1}}

\DeclareRobustCommand{\Fig}[1]{Fig.~\ref{#1}}

\DeclareRobustCommand{\Figss}[3]{Figs.~\ref{#1}, \ref{#2} and \ref{#3}}

\DeclareRobustCommand{\Eq}[1]{Eq.~(\ref{#1})}
\DeclareRobustCommand{\Eqs}[2]{Eqs.~(\ref{#1}) and (\ref{#2})}
\DeclareRobustCommand{\Ref}[1]{Ref.~\cite{#1}}
\DeclareRobustCommand{\Refs}[1]{Refs.~\cite{#1}}

\newcommand{\df}{\mathrm{d}}

\renewcommand{\as}{\alpha_s}

\newcommand{\be}{\begin{equation}}
\newcommand{\ee}{\end{equation}}

\newcommand{\pythia}{\textsc{Pythia 8.219}}
\newcommand{\herwig}{\textsc{Herwig 7.0.3}}
\newcommand{\sherpa}{\textsc{Sherpa 2.2.1}}

\begin{document}

\preprint{MIT-CTP 4890}

\title{Jet Substructure Studies with CMS Open Data}

\author{Aashish Tripathee}
\email{aashisht@mit.edu}

\affiliation{Center for Theoretical Physics, Massachusetts Institute of Technology, Cambridge, MA 02139, USA}

\author{Wei Xue}
\email{weixue@mit.edu}

\affiliation{Center for Theoretical Physics, Massachusetts Institute of Technology, Cambridge, MA 02139, USA}

\author{Andrew Larkoski}
\email{larkoski@reed.edu}

\affiliation{Physics Department, Reed College, Portland, OR 97202, USA}

\author{Simone Marzani}
\email{smarzani@buffalo.edu}
\affiliation{University at Buffalo, The State University of New York, Buffalo, NY 14260-1500, USA}


\author{Jesse Thaler}
\email{jthaler@mit.edu}

\affiliation{Center for Theoretical Physics, Massachusetts Institute of Technology, Cambridge, MA 02139, USA}


\begin{abstract}
We use public data from the CMS experiment to study the 2-prong substructure of jets.  The CMS Open Data is based on 31.8~pb$^{-1}$ of 7 TeV proton-proton collisions recorded at the Large Hadron Collider in 2010, yielding a sample of 768,687 events containing a high-quality central jet with transverse momentum larger than 85 GeV.  Using CMS's particle flow reconstruction algorithm to obtain jet constituents, we extract the 2-prong substructure of the leading jet using soft drop declustering.  We find good agreement between results obtained from the CMS Open Data and those obtained from parton shower generators, and we also compare to analytic jet substructure calculations performed to modified leading-logarithmic accuracy.  Although the 2010 CMS Open Data does not include simulated data to help estimate systematic uncertainties, we use track-only observables to validate these substructure studies.
\end{abstract}

\maketitle

\tableofcontents

\section{Introduction}
\label{sec:intro}

In November 2014, the CMS experiment at the Large Hadron Collider (LHC) announced the CMS Open Data project \cite{CERNOpenDataPortal}.  To our knowledge, this is the first time in the history of particle physics that research-grade collision data has been made publicly available for use outside of an official experimental collaboration.  The CMS Open Data was reconstructed from 7 TeV proton-proton collisions in 2010, corresponding to a unique low-luminosity running environment where pileup contamination was minimal and trigger thresholds were relatively low.  The CMS Open Data presents an enormous opportunity to the particle physics community, both for performing physics studies that would be more difficult at higher luminosities as well as for demonstrating the scientific value of open data releases.

In this paper, we use the CMS Open Data to analyze the substructure of jets.  Jets are collimated sprays of particles that are copiously produced in LHC collisions, and by studying the substructure of jets, one can gain valuable information about their parentage \cite{Seymour:1991cb,Seymour:1993mx,Butterworth:2002tt,Butterworth:2007ke,Butterworth:2008iy, Abdesselam:2010pt,Altheimer:2012mn,Altheimer:2013yza,Adams:2015hiv}.  A key application of jet substructure is tagging boosted heavy objects like top quarks \cite{Brooijmans:2008zza,Kaplan:2008ie,Thaler:2008ju,Almeida:2008yp,Almeida:2008tp,CMS:2009lxa,CMS:2009fxa,ATLAS:2009hdz,Plehn:2009rk,Plehn:2010st,Almeida:2010pa,Thaler:2010tr,Thaler:2011gf,Jankowiak:2011qa,Soper:2012pb,Larkoski:2013eya,Anders:2013oga,Freytsis:2014hpa,Kasieczka:2015jma,Lapsien:2016zor,Moult:2016cvt} and electroweak bosons \cite{Seymour:1993mx,Butterworth:2002tt,Butterworth:2008iy,Almeida:2008yp,Thaler:2010tr,Kribs:2009yh,Kribs:2010hp,Chen:2010wk,Hackstein:2010wk,Falkowski:2010hi,Katz:2010mr,Cui:2010km,Kim:2010uj,Gallicchio:2010dq,Gallicchio:2010sw,Hook:2011cq,Soper:2011cr,Almeida:2011aa,Ellis:2012sn,Larkoski:2014wba,Cogan:2014oua,Izaguirre:2014ira,Larkoski:2014gra,deOliveira:2015xxd,Dasgupta:2015yua,Larkoski:2015kga,Stewart:2015waa,Dasgupta:2015lxh,Baldi:2016fql,Conway:2016caq,Lapsien:2016zor,Barnard:2016qma,Roy:2016qfv,Dasgupta:2016ktv,Moult:2016cvt}.  To successfully tag such objects, though, one first has to understand the radiation patterns of ordinary quark and gluon jets \cite{Nilles:1980ys,Jones:1988ay,Fodor:1989ir,Jones:1990rz,Lonnblad:1990qp,Pumplin:1991kc,Gallicchio:2011xq,Gallicchio:2012ez,Larkoski:2013eya,Larkoski:2014pca,Bhattacherjee:2015psa,Badger:2016bpw,FerreiradeLima:2016gcz,Bhattacherjee:2016bpy,FerreiradeLima:2016gcz,Komiske:2016rsd,Davighi:2017hok,Gras:2017jty}, which are the main backgrounds to boosted objects.  The CMS Open Data is a fantastic resource for performing these baseline quark/gluon studies.  Using the Jet Primary Dataset \cite{CMS:JetPrimary}, we perform initial investigations of the 2-prong substructure of jets as well as present a general analysis framework to facilitate future studies.  This effort is complementary to the growing catalog of jet substructure measurements performed within the ATLAS and CMS collaborations 
\cite{CMS:2011bqa,CMS:2011xsa,Miller:2011qg,Chatrchyan:2012mec,ATLAS:2012jla,Aad:2012meb,ATLAS:2012kla,ATLAS:2012am,Chatrchyan:2012ku,Chatrchyan:2012sn,Chatrchyan:2013vbb,Aad:2013gja,Aad:2013fba,TheATLAScollaboration:2013tia,TheATLAScollaboration:2013sia,TheATLAScollaboration:2013ria,TheATLAScollaboration:2013pia,TheATLAScollaboration:2013qia,CMS:2013cda,CMS:2013uea,CMS:2013kfa,CMS:2013wea,Fleischmann:2013woa,Pilot:2013bla,CMS-PAS-QCD-10-041,Aad:2014gea,LOCH:2014lla,CMS:2014fya,CMS:2014joa,Aad:2014haa,CMS:2014afa,CMS:2014aka,Khachatryan:2014vla,Khachatryan:2015axa,CMS:1900uua,Khachatryan:2015bma,Aad:2015owa,Aad:2015cua,Aad:2015lxa,Aad:2015rpa,Aad:2015ina,Aad:2015typ,atlas2015035,atlas2015037,ATLAS-CONF-2015-071,ATLAS-CONF-2015-073,Khachatryan:2015scf,Aad:2015eax,ATLAS-CONF-2015-036,Khachatryan:2015gza,CMS:2015nmz,CMS:2015naa,CMS:2015jha,Aad:2016pux,ATLAS-CONF-2016-008,ATLAS:2016wzt,ATLAS:2016vmy,ATLAS:2016wlr,ATLAS:2016cuv,CMS:2016ldu,CMS:2016bnj,ATLAS:2016jct,CMS:2016djf,CMS:2016pfl,CMS:2016pod,CMS:2016flr,CMS:2016rfr,Khachatryan:2016mdm,CMS:2016jdj,CMS:2016yuu,ATLAS-CONF-2016-016,CMS:2016rtp,CMS:2016pnc,Khachatryan:2016zcu,Khachatryan:2016cfa,ATLAS-CONF-2016-014,Khachatryan:2016oia,CMS:2016bja,CMS:2016ehh,CMS:2016wev,CMS:2016tns,CMS:2016ydv,CMS:2016qwm,CMS:2016kkf,CMS:2016zte,CMS:2016ude,CMS:2016hxa,CMS:2016pwo,CMS:2016mjh,CMS:2016jog,CMS:2016jys,CMS:2016usi,CMS:2016ccy,ATLAS:2016kxc,ATLAS:2016npe,ATLAS:2016yqq,ATLAS:2016btu,ATLAS:2016dpc,CMS:2016dmr,CMS:2016knm,CMS:2016ete,CMS:2016tvk,CMS:2014ata,Khachatryan:2016cfx,Khachatryan:2016vph,CMS:2016mwi,Khachatryan:2016whc,Sirunyan:2016ipo,Sirunyan:2016cao,Sirunyan:2016wqt,CMS:2016pul,Sirunyan:2017ezy,CMS:2017mrw,Sirunyan:2017hnk,Sirunyan:2017bfa,Sirunyan:2017yar,Sirunyan:2017bey,Sirunyan:2017hci,CMS:2017orf,CMS:2017vbf,CMS:2017skt,CMS:2017eme,CMS:2017oef}.\footnote{To highlight the vibrancy of the field, we have attempted to list all published jet substructure measurements from ATLAS and CMS.  Please contact us if we missed a reference.}

The core of our analysis is based on soft drop declustering \cite{Larkoski:2014wba}, which is a jet grooming technique \cite{Butterworth:2008iy,Ellis:2009su,Ellis:2009me,Krohn:2009th} that mitigates jet contamination from initial state radiation (ISR), underlying event (UE), and pileup.  For the studies in this paper, we set the soft drop parameter $\beta$ equal to zero, such that soft drop behaves like the modified mass drop tagger (mMDT) \cite{Dasgupta:2013ihk,Dasgupta:2013via}.\footnote{The original mass drop tagger \cite{Butterworth:2008iy} was a pioneering technique in jet substructure; see also precursor work in \Refs{Seymour:1991cb,Seymour:1993mx,Butterworth:2002tt,Butterworth:2007ke}.}  After soft drop, a jet is composed of two well-defined subjets, which can then be used to derive various 2-prong substructure observables.  In addition to comparing the CMS Open Data to parton shower generators, we perform first-principles calculations of soft-dropped observables using recently-developed analytic techniques \cite{Larkoski:2014wba,Larkoski:2013paa,Larkoski:2015lea}.  In a companion paper, we use soft drop to expose the QCD splitting function using the CMS Open Data \cite{Larkoski:2017bvj}; a similar strategy was used in preliminary CMS \cite{CMS:2016jys}, STAR \cite{StarTalk}, and ALICE \cite{AlicePoster} heavy ion studies to test for possible modifications to the splitting function from the dense QCD medium \cite{Chien:2016led,Mehtar-Tani:2016aco}. 

For studying jet substructure, the key feature of the CMS Open Data is that it contains full information about particle flow candidates (PFCs).  The particle flow algorithm \cite{CMS-PAS-PFT-09-001,CMS-PAS-PFT-10-001} synthesizes information from multiple detector elements to create a unique particle-like interpretation of each collision event.  Within CMS, these PFCs are used directly in jet reconstruction \cite{Khachatryan:2016kdb}.  Here, we can exploit the PFC information to perform detailed jet substructure studies, using standard particle-based jet analysis tools.

The main limitation of the 2010 CMS Open Data release is that it only provides minimal calibration information, and therefore we cannot properly estimate systematic uncertainties from detector effects.  Ideally, we would like a detector simulation or a smearing parametrization to account for finite resolution and granularity.   Absent that, we cannot make a direct comparison of CMS Open Data to properly folded particle-level distributions.  With that caveat in mind, our plots will overlay detector-level CMS Open Data (without further calibration) and particle-level theory distributions (without detector simulation).  The overall agreement turns out to be rather good, highlighting the excellent performance of the CMS detector and CMS's particle flow reconstruction.  One must always keep in mind, though, that our plots cannot be interpreted like standard LHC experimental plots, both because of the absence of detector (un)folding and the absence of systematic uncertainties in the error bars.

To gain confidence in the robustness of our substructure analysis, we perform cross checks using track-based variants.  Distributions using only charged particles are expected to exhibit better resolution than those using all particles, and we indeed find better qualitative agreement with parton showers using these track-based observables.  We also attempted to estimate detector effects using the \textsc{Delphes} fast simulation tool \cite{deFavereau:2013fsa}, but we found that the default CMS-like detector settings led to over-smearing of the distributions, so no \textsc{Delphes} results will be shown in this paper.  For the future, we plan to repeat these studies using the 2011 CMS Open Data \cite{CMS:JetPrimary2011}, which does come accompanied by detector-simulated Monte Carlo files.

The remainder of this paper is organized as follows.  In \Sec{sec:CMSOpen}, we give an overview of the CMS Open Data and corresponding analysis tools.  In \Sec{sec:hardest_jet}, we present basic kinematic and substructure properties of the hardest jet in the event, comparing the CMS Open Data to parton shower generators.  In \Sec{sec:2prong}, we review the soft drop algorithm and compare analytic calculations of 2-prong substructure to open data and parton shower distributions.  Based on our experience with the CMS Open Data, we provide recommendations to CMS and to the broader particle physics community in \Sec{sec:recommend}.  We conclude in \Sec{sec:conclude}, leaving additional details and plots to the appendices.

\section{The CMS Open Data}
\label{sec:CMSOpen}

The CMS Open Data is available from the CERN Open Data Portal \cite{CERNOpenDataPortal}, with the initial release corresponding to Run 2010B of the LHC.  The primary datasets are in the form of Analysis Object Data (AOD) files, which is a file format used internally within CMS based on the ROOT framework \cite{Brun:1997pa}.  To process the CMS data, one first has to install a virtual machine (VM) with \textsc{CernVM} running \textsc{Scientific Linux CERN 5}.  Within the VM, one can then run the official CMS software framework (CMSSW), which provides access to the complete analysis tools needed to parse the AOD files.

Our jet substructure study is based on the Jet Primary Dataset \cite{CMS:JetPrimary}, which is a subset of the full open data release with events that pass a predefined set of single-jet and multi-jet triggers.  There are 1664 AOD files in the Jet Primary Dataset, corresponding to 20,022,826 events and 2.0 Terabytes of disk space.  Within CMSSW, it is possible to access the AOD files remotely through the \textsc{XRootD} interface \cite{XRootD}.  We found it more convenient to first download the AOD files and then process them locally, being careful to maintain the same directory structure as on the Open Data servers in order to ensure consistency of the workflow.  We then converted AOD files into a text-based MIT Open Data (MOD) format to facilitate the use of external analysis tools.

\subsection{The CMS Software Framework}

CMSSW is a hybrid \textsc{Python/C++} analysis framework where event processing takes place through user-defined modules.   The version provided with the CMS Open Data is \textsc{4.2.8}, which was also used internally by CMS in 2010 (as of this writing, the current CMSSW version is 9.0.0).  In principle, we could have used CMSSW directly to perform our jet substructure studies, but we found it more convenient to simply use CMSSW for data extraction and then use external tools for analysis, described in \Sec{sec:MODAnalyzer}.

Within CMS, there are multiple tiers of data, but only AOD files are provided by the CMS Open Data.  Starting from RAW detector-level data, CMS derives RECO (reconstructed) data which includes both low-level objects (like reconstructed tracks) and high-level objects (like clustered jets).  For most CMS analyses, only a subset of the RECO data is required, and this is the basis for the AOD files.  For our open data analysis, the AOD files contain far more information than needed, so we use CMSSW to isolate only the required physics objects and event information.

To use CMSSW for data extraction, we rely on a chain of user-defined modules.  We use a \texttt{Source} module to read in events from the AOD files and an \texttt{EDProducer} called \texttt{MODProducer} to convert the AOD format into our own text-based MOD format (see \Sec{subsection:MOD}).\footnote{Here, \texttt{ED} refers to ``event data''.  Strictly speaking, since we are not modifying the AOD files directly, we could have used an \texttt{EDAnalyzer} instead of an \texttt{EDProducer}.  We decided to use \texttt{EDProducer} because the name aligns better with what the module is actually doing, namely ``creating data'', albeit in the MOD format.  Also, CMS recommends using an \texttt{OutputModule} when writing to an external file, but we instead used the standard \textsc{C++} libraries for output.}   The \texttt{MODProducer} software is available through a \textsc{GitHub} repository \cite{MODProducer}.

In order to maintain reasonable file sizes and enable easier data validation, we wanted \texttt{MODProducer} to generate a separate MOD file for each of the 1664 AOD files, rather than one monolithic MOD file.  While we could have run \texttt{MODProducer} separately for each AOD file, it turns out that \texttt{MODProducer} has to load \texttt{FrontierConditions\_GlobalTag\_cff} and the appropriate global tag (\texttt{GR\_R\_42\_V25::All}) in order to properly extract trigger information from the AOD file.  Loading this information takes around 10 minutes at the beginning of a CMSSW run, so to save computing time, we wanted to process multiple AOD files in series in the same run.  To the best of our knowledge, though, CMSSW does not allow an \texttt{EDProducer} to know which AOD file is being processed.  To circumvent this limitation, we created a lightweight \texttt{FilenameMapProducer} that only runs on one file at a time and creates a map relating event and run numbers to the corresponding AOD filename.  This filename map is then read in by \texttt{MODProducer}, along with a list provided by CMS of validated runs suitable for physics analyses.

From the AOD files, \texttt{MODProducer} extracts PFCs, jets clustered from these PFCs, associated jet calibration information, trigger information, luminosity information, and basic event identification information like event and run numbers.  The PFCs provide a unique reference event interpretation in terms of reconstructed photons, electrons, muons, charged hadrons, and neutral hadrons \cite{CMS-PAS-PFT-09-001,CMS-PAS-PFT-10-001}.  Each PFC has a particle identification flag and a full Lorentz four-vector, with non-zero invariant mass when available.  We use AK5 jets provided by CMS \cite{Khachatryan:2016kdb}, corresponding to the anti-$k_t$ jet clustering algorithm \cite{Cacciari:2008gp} with $R = 0.5$ and a minimum jet threshold of $p_T > 3.0~\GeV$, and we later validate the anti-$k_t$ clustering by running \textsc{FastJet 3.1.3} \cite{Cacciari:2011ma} ourselves on the PFCs.  The jet calibration information includes both jet quality criteria as well as jet energy corrections (JEC) factors, discussed further in \App{app:opendatainfo}.  We discuss trigger and luminosity information in more detail next.

\subsection{The Jet Primary Dataset}
\label{subsection:jet_primary_dataset}

\begin{table}[t]
\begin{tabular}{r @{$\quad$} l @{$\quad$} r @{$\quad$} r}
\hline
\hline
&Trigger & Present? & Fired? \\
\hline
Single-jet & \texttt{HLT\_Jet15U} & 16,341,190 & 1,342,155 \\
&\texttt{* HLT\_Jet15U\_HNF} & 16,341,190 & 1,341,930 \\
&\texttt{* HLT\_Jet30U} & 16,341,190 & 604,287 \\
&\texttt{* HLT\_Jet50U} & 16,341,190 & 870,649 \\
&\texttt{* HLT\_Jet70U} & 16,341,190 & 5,257,339 \\
&\texttt{* HLT\_Jet100U} & 16,341,190 & 3,689,951 \\
&\texttt{* HLT\_Jet140U} & 5,989,945 & 1,898,874 \\
&\texttt{HLT\_Jet180U} & 2,595,038 & 553,331 \\
\hline
Di-jet & \texttt{HLT\_DiJetAve15U} & 16,341,191 & 1,067,561 \\
&\texttt{HLT\_DiJetAve30U} & 16,341,191 & 648,000 \\
&\texttt{HLT\_DiJetAve50U} & 16,341,191 & 859,292 \\
&\texttt{HLT\_DiJetAve70U} & 16,341,191 & 2,310,033 \\
&\texttt{HLT\_DiJetAve100U} & 5,989,945 & 1,252,661 \\
&\texttt{HLT\_DiJetAve140U} & 2,595,038 & 452,222 \\
\hline
Quad-jet & \texttt{HLT\_QuadJet20U} & 10,351,245 & 677,451 \\
& \texttt{HLT\_QuadJet25U} & 10,351,244 & 219,256 \\
\hline
$H_T$ & \texttt{HLT\_HT100U} & 10,351,245 & 7,369,985 \\
&\texttt{HLT\_HT120U} & 10,351,245 & 4,090,218 \\
&\texttt{HLT\_HT140U} & 10,351,245 & 2,430,208 \\
&\texttt{HLT\_EcalOnly\_SumEt160} & 10,351,246 & 208,718 \\
\hline
\hline
\end{tabular}
\caption{Jet triggers provided in the Jet Primary Dataset \cite{CMS:JetPrimary}, including the number of events for which the trigger was present and/or fired.  Entries marked by \texttt{*} are used in this analysis (see \Tab{tab:trigger_table}).  \texttt{HNF} stands for \texttt{HcalNoiseFiltered}.  We do not separate out the different versions of the same trigger in our analysis.}
\label{tab:trigger_names}
\end{table}

\begin{figure*}[t]
\subfloat[]{
\label{fig:trigger_turn_on_curve}
\includegraphics[width=\columnwidth]{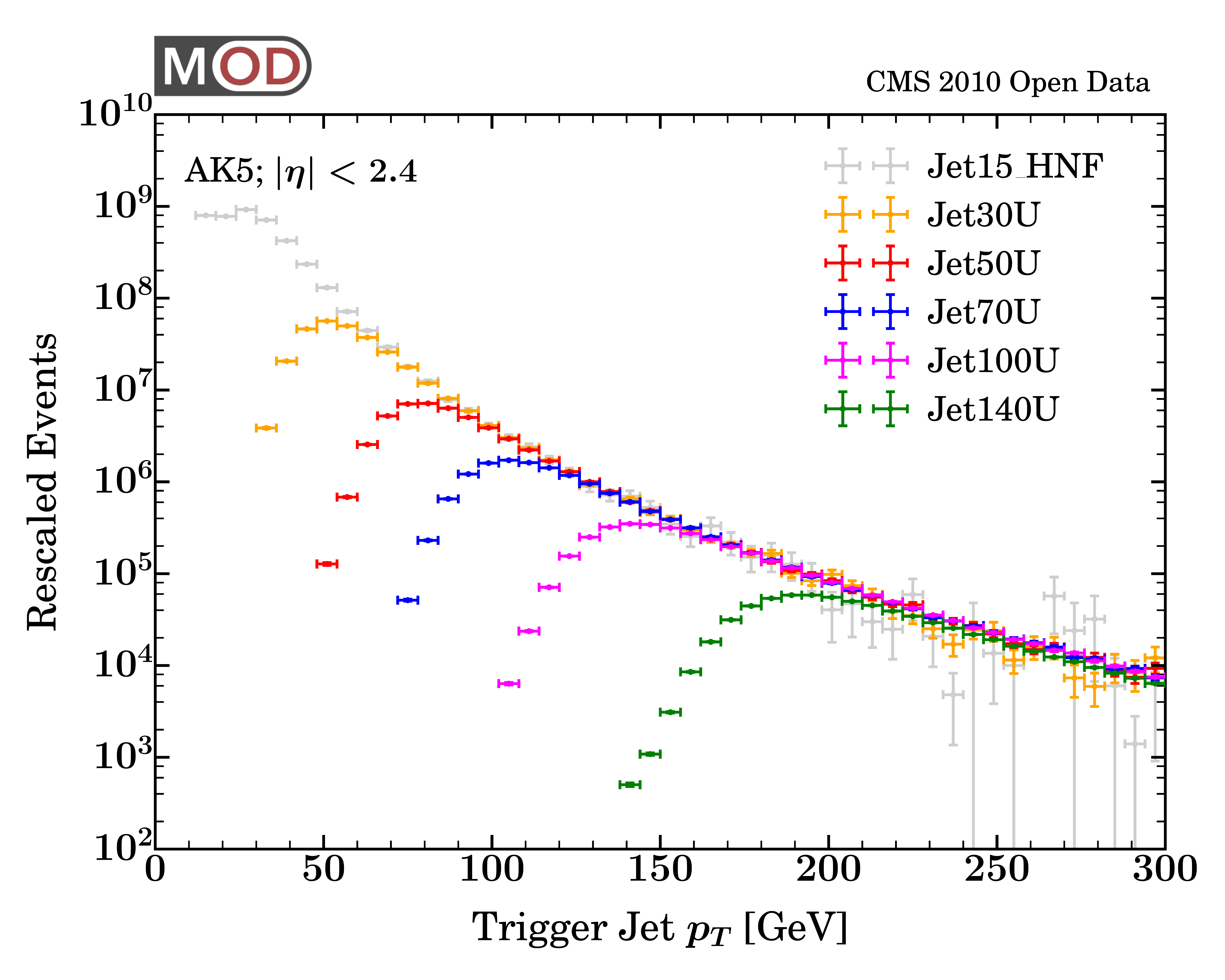}
}
\subfloat[]{
\label{fig:trigger_efficiency_curve}
\includegraphics[width=\columnwidth]{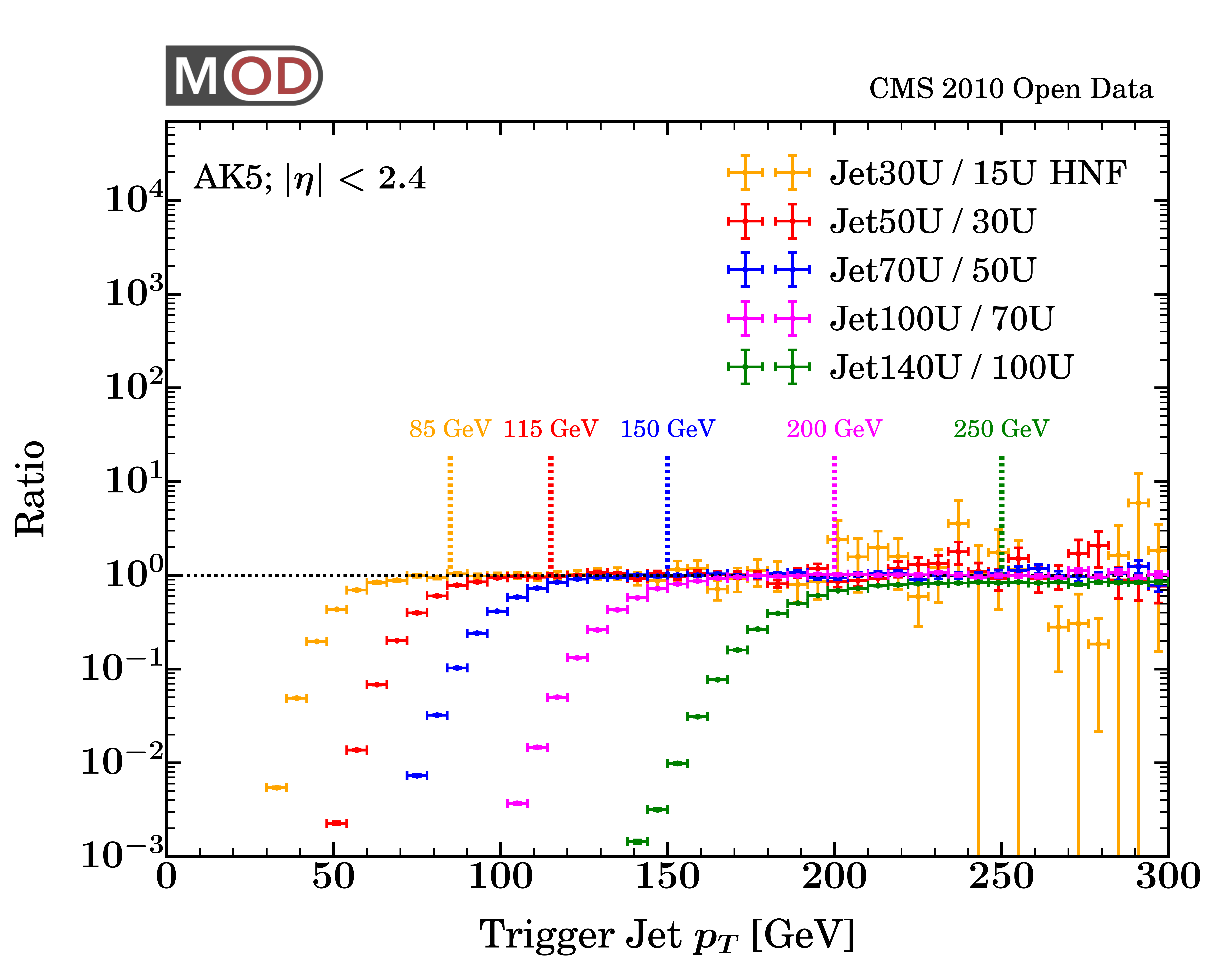}
}
\caption{(a) Hardest jet $p_T$ spectrum in the CMS Open Data from the six triggers used in this analysis (see \Tab{tab:trigger_names}).  (b) Ratios of the jet $p_T$ spectra from adjacent triggers used to determine when the triggers are nearly 100\% efficient, which determine the jet trigger boundaries in \Tab{tab:trigger_table}.  Because the \texttt{Jet\_140U} trigger was not present for the entirety of the run, it artificially appears systematically low in these plots.}
\label{fig:trigger_all}
\end{figure*}

\begin{table}[t]
\begin{tabular}{r@{$\quad$}r@{$\quad$}r@{$\quad$}r}
\hline
\hline
Hardest Jet $p_T$ & Trigger Name & Events  & $\langle$Prescale$\rangle$ \\
\hline
$[85, 115]~\GeV$ & \texttt{HLT\_Jet30U} & 33,375 & 851.514 \\
$[115, 150]~\GeV$ & \texttt{HLT\_Jet50U} & 66,412 & 100.320 \\
$[150, 200]~\GeV$ & \texttt{HLT\_Jet70U}  & 365,821 & 5.362 \\
$[200, 250]~\GeV$ & \texttt{HLT\_Jet100U} & 216,131 & 1.934\\
\hline
\multirow{2}{*}{$> 250~\GeV$} & \texttt{HLT\_Jet100U}   & 34,736 & 1.000\\
& \texttt{HLT\_Jet140U} & 177,891 & 1.000\\
\hline
\hline
\end{tabular}
\caption{ Assigned triggers for the hardest jet in a given $p_T$ range, along with the average prescale value that determines subsequent histogram weights.  Since the \texttt{Jet140U} trigger was not present for all of Run 2010B, we use \texttt{Jet100U} when needed for the highest $p_T$ bin.}
\label{tab:trigger_table}
\end{table}

\begin{table}[t]
\begin{tabular}{r@{$\quad$}r@{$\quad$}l}
\hline
\hline
& Events & Fraction\\
\hline
Jet Primary Dataset & 20,022,826 & 1.000 \\
Validated Run & 16,341,187 & 0.816 \\
Assigned Trigger Fired (\Tab{tab:trigger_table}) & 894,366 & 0.045 \\
\hline
Loose Jet Quality (\Tab{tab:jet_quality}) & 843,129 & 0.042 \\
AK5 Match & 843,128 & 0.042\\
$\left| \eta \right| < 2.4$ & 768,687 & 0.038 \\
\hline
Passes Soft Drop ($z_g > z_{\rm cut}$) & 760,055 & 0.038 \\
\hline
\hline
\end{tabular}
\caption{Overall workflow to go from the events in the Jet Primary Dataset to the events used in our jet substructure analysis.  The three steps above the first horizontal line indicate the steps included as part of event skimming.  The next three steps are used for the \texttt{Hardest\_Jet\_Selection}.  The final line is for events that pass the soft drop requirement in \Sec{sec:2prong}.}
\label{tab:event_numbers}
\end{table}

The CMS Open Data is grouped into primary datasets, corresponding to the types of triggers that were used for event selection.  Our analysis is based exclusively on the Jet Primary Dataset \cite{CMS:JetPrimary}.\footnote{In order to study lower $p_T$ jets, we would have to incorporate the MinimumBias Primary Dataset \cite{CMS:MinimumBias}.  Because primary datasets are overlapping, one has to be careful not to double count events when using multiple primary datasets.}  
As listed in \Tab{tab:trigger_names}, this dataset has single-jet, di-jet, quad-jet, and $H_T$ triggers, though we only use single-jet triggers for our study.  
Each trigger has an associated prescale factor, which is the ratio of how often the triggering criteria are met compared to how many events the trigger actually records.  A prescale factor of 1 indicates that all triggered events are kept, whereas larger prescale factors are assigned to frequently-encountered event categories that would otherwise overwhelm data acquisition.  The prescale factor used in the analysis is the product of the prescale factors from the underlying Level 1 Trigger (based on low-level objects) and the final High Level Trigger (HLT).
There are various versions of the triggers, indicated by suffixes like \texttt{\_v2} and \texttt{\_v3}, but we do not distinguish between the versions in our analysis.

The CMS single-jet triggers are designed to fire whenever \emph{any} jet in the event is above a given $p_T$ threshold.  Since our substructure study is based only on the hardest jet in an event, we have to make sure that the correct ``assigned'' trigger fired for the hardest AK5 jet in an event.  We also have to check that this trigger is nearly 100\% efficient for jets of the given $p_T$.

\begin{figure*}
\subfloat[]{
\label{fig:intg_rec_lumi}
\includegraphics[width=\columnwidth]{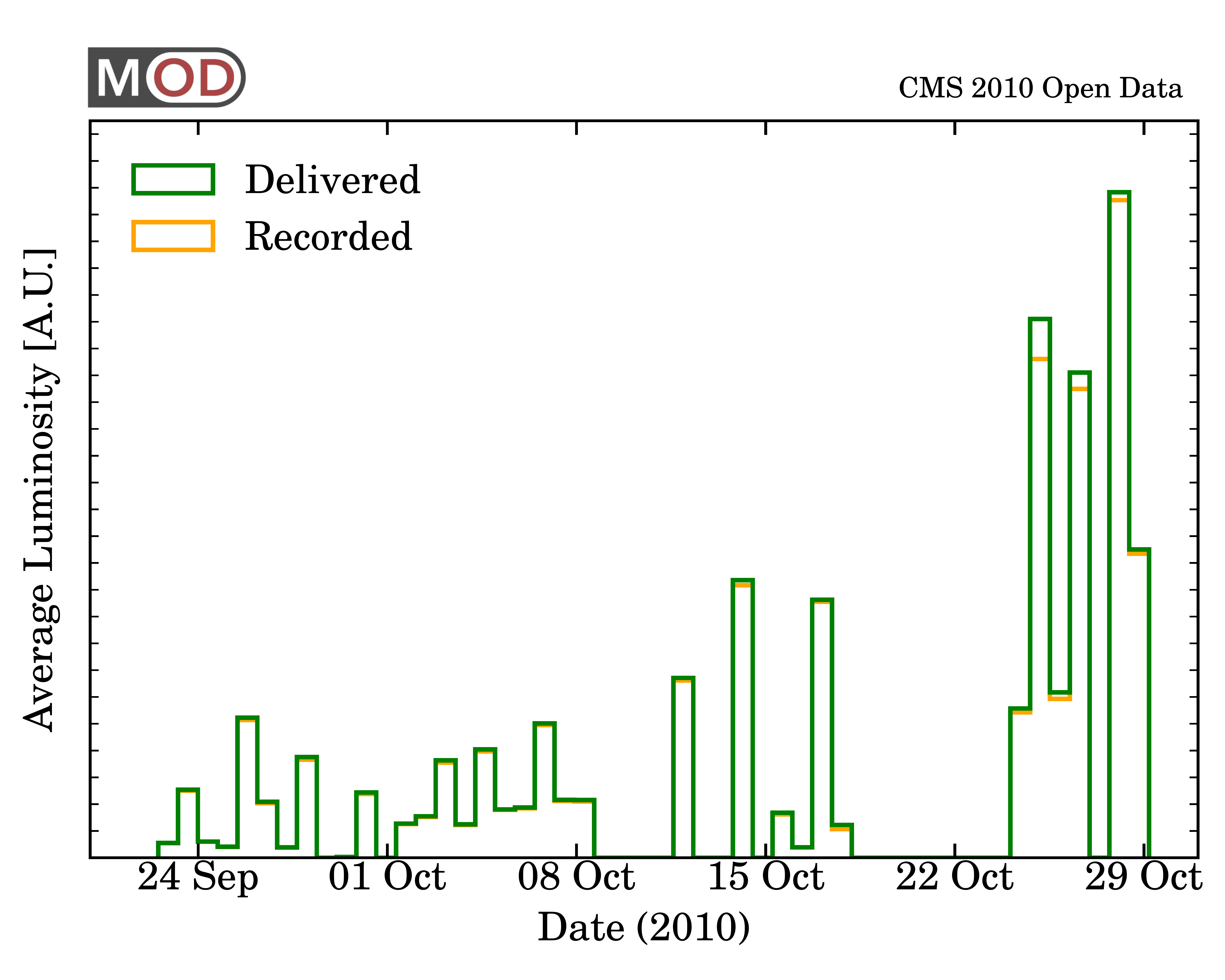}
}
\subfloat[]{
\label{fig:integrated_lumi_cumulative}
\includegraphics[width=\columnwidth]{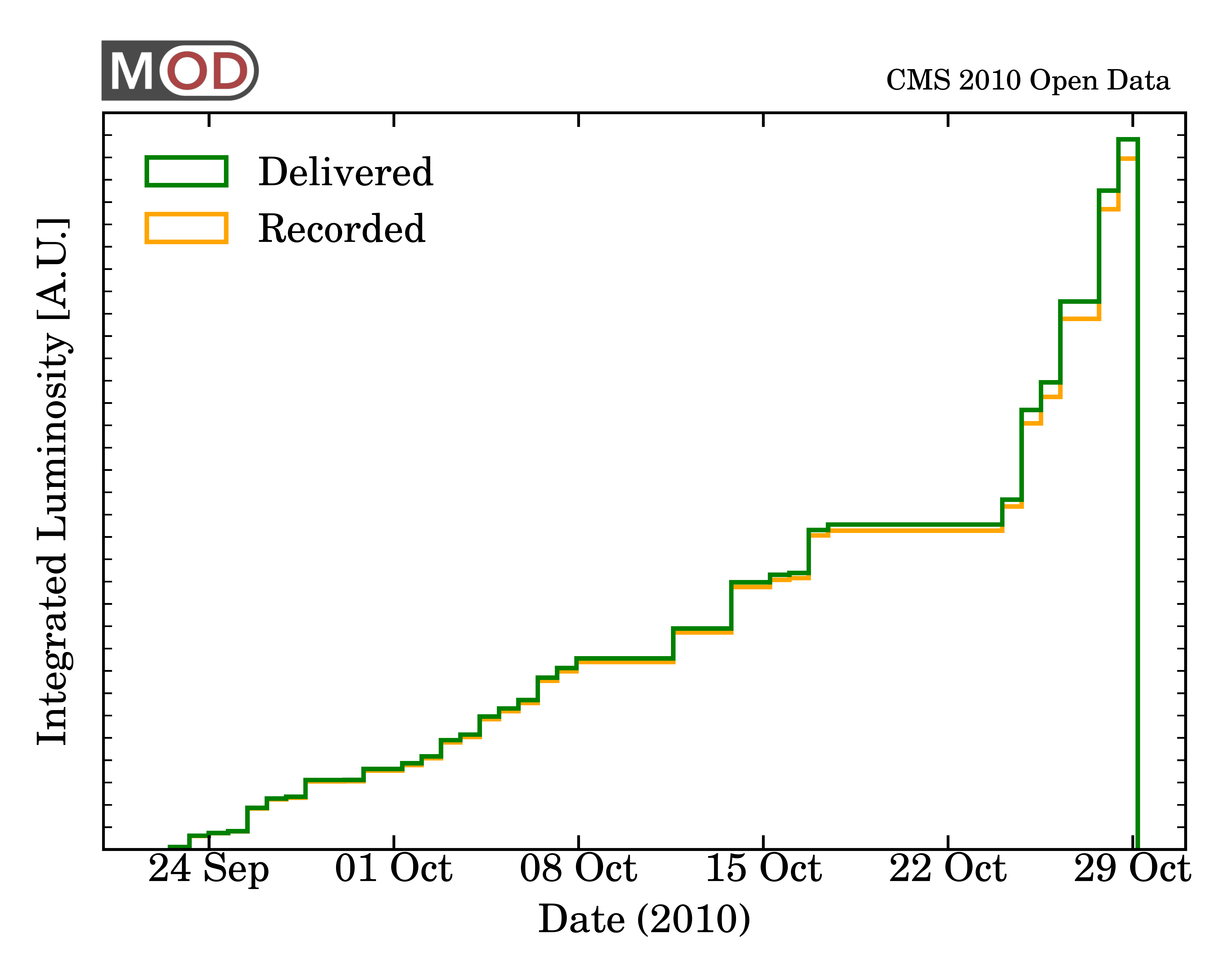}
}
\caption{Integrated luminosity collected by the CMS experiment during Run 2010B, plotted (a) per day and (b) cumulative.  Because the luminosity information provided in the AOD files does not match the official recorded integrated luminosity of $31.8~\mathrm{pb}^{-1}$, we suppress the vertical normalization in these plots.  The qualitative features shown here do agree with the official Run 2010B luminosity profile.}
\label{fig:integrated_lumi}
\end{figure*}

In \Fig{fig:trigger_turn_on_curve}, we show the $p_T$ spectrum of the hardest jet for the six triggers used in our analysis.   All jets have passed a ``loose'' jet quality cut with appropriate JEC factors applied; see \Tab{tab:jet_quality} and \Fig{fig:JEC} in \App{app:opendatainfo}.  We further impose a pseudorapidity cut of $|\eta| < 2.4$ to ensure that jets are reconstructed in the central part of the CMS detector where tracking information is available.  With prescale factors included, we see good overlap of the $p_T$ spectra as desired, except for the \texttt{Jet140U} trigger which is systematically low.  The reason is that the \texttt{Jet140U} trigger was not present for the entirety of Run 2010B, so we revert to the \texttt{Jet100U} trigger when needed.

Using \texttt{HLT\_Jet15U\_HcalNoiseFiltered} as the baseline, the trigger efficiencies of the five remaining triggers are shown in \Fig{fig:trigger_efficiency_curve}. For our analysis, we want to work with single triggers that are nearly 100\% efficient when the hardest jet is in a given $p_T$ range.  Crosschecking \Fig{fig:trigger_efficiency_curve} with \Ref{CMS-PAS-QCD-10-011}, we define the trigger boundaries in \Tab{tab:trigger_table}, where the $p_T > 250~\GeV$ bin uses either \texttt{Jet100U} or \texttt{Jet140U} depending on whether the latter is present.  We see that lower $p_T$ triggers have higher average prescale values as expected.  Because each trigger selects a homogenous event sample, we can use the average prescale value for the assigned trigger when filling histograms, which is statistically preferable to using the individual event prescale values.  For completeness, in \Fig{fig:prescale_values} in \App{app:opendatainfo}, we show the distribution of prescale values encountered for each trigger within their assigned $p_T$ range.

Our event selection workflow is summarized in \Tab{tab:event_numbers}.  Starting from the 20 million events in the Jet Primary Dataset, we reduce the dataset to about 82\% by only including events that are in the official list of validated runs.  Restricting to events that pass their assigned trigger in \Tab{tab:trigger_table} drops the event sample to around 900 thousand events, and this is used to define a skimmed dataset.  Requiring the loose jet quality criteria removes a small number of events, as does verifying that the AK5 jet provided by CMS matches those clustered by \textsc{FastJet} on the PFCs directly (see \Secs{subsection:MOD}{sec:MODAnalyzer}).  If the hardest jet passes $|\eta| < 2.4$, then it is used for substructure analyses (see \Sec{sec:MODAnalyzer}).  For later reference, \Tab{tab:event_numbers} shows the number of events where the hardest jet has valid 2-prong substructure as determined by soft drop declustering (see \Sec{sec:2prong}).

In the plots below, we always present normalized histograms in order to suppress fixed-order QCD corrections to the overall jet production rate.  While knowledge of the total luminosity is therefore not needed for our study, it is still instructive to try to extract luminosity information from the CMS Open Data.  The AOD files provide the integrated luminosities achieved during each luminosity block, such that the sum over blocks should give the total luminosity.  Unfortunately, the AOD-extracted value of $309.5~\mathrm{pb}^{-1}$ does not match the official recorded luminosity value of $31.79~\mathrm{pb}^{-1}$ during Run 2010B \cite{DeGruttola:2010kha,CMS-PAS-EWK-10-004}.\footnote{It is suspicious that the difference is very close to a factor of 10, but as far as we can tell, this is a coincidence.}  This turns out to be a known limitation of the provided AOD files, though the AOD-extracted values do have the expected qualitative structures.  Removing the overall vertical normalization to avoid confusion, the delivered and recorded integrated luminosities are shown in \Fig{fig:intg_rec_lumi} and the cumulative distributions in \Fig{fig:integrated_lumi_cumulative}.   As expected, we see that Run 2010B occurred from September 22 to October 29 in 2010, with a substantial ramp up of collected data over that two month period.


\subsection{The MIT Open Data Format}
\label{subsection:MOD}

The output of \texttt{MODProducer} is a text-based MOD file, which contains a subset of the AOD data, similar in spirit to the Mini-AOD format being developed internally within CMS \cite{Petrucciani:2015gjw}.  The MOD format is intended to be lightweight, easy to parse, and human readable, so it uses space-separated entries with keyword labels.  While there are other text-based file formats used within high energy physics, such as \textsc{HepMC} \cite{Dobbs:2001ck} and \textsc{LHEF} \cite{Boos:2001cv,Alwall:2006yp}, they are primarily intended for use with Monte Carlo generators and therefore do not have a standard way to incorporate CMS-specific information like triggers and JEC factors.  Instead of trying to augment these existing file formats and risk breaking backward compatibility, we decided to develop our own MOD format.  Ultimately, one could envision a standard file format for open collider data, since the MOD file already contains much of the information common to all collider analyses.  In our analysis, we use the MOD format not only for experimental data but also for data generated from parton showers (see \Sec{sec:parton_shower}).  As a cross check of the results in this paper, we also performed an independent analysis using an internal ROOT-based framework.  

A typical MOD event consists of the following six keywords:
\begin{itemize}
	\item \texttt{BeginEvent}:  A header that indicates the source of the event:  CMS Open Data or parton shower generator.\footnote{We also generated samples using fast detector simulation.  As already mentioned, because of apparent oversmearing by the default CMS-like \textsc{Delphes} configuration \cite{deFavereau:2013fsa}, we do not show any fast simulation results in this paper.}  It also includes the version number of the MOD format (currently version 5).
	\item \texttt{Cond}:  Basic information about the run and event conditions, including run and event numbers, a timestamp, the number of reconstructed primary vertices, and information about the luminosity block.
	\item \texttt{Trig}:  List of all triggers used in the Jet Primary Dataset, their associated prescale factors, and flags indicating whether a given trigger fired for that event.
	\item \texttt{AK5}: List of anti-$k_t$ $R = 0.5$ jets provided by CMS.  In addition to the jet four-momentum, CMS provides a JEC factor, a jet area value \cite{Cacciari:2008gn}, and information about jet quality.  
	\item \texttt{PFC}: List of PFCs, with their four-momenta and particle identification codes.
	\item \texttt{EndEvent}: A footer indicating the end of an event.
\end{itemize}
An example MOD event is included in the \texttt{arXiv} source files of this paper.  For MOD files coming from parton shower generators, we replace \texttt{Cond} and \texttt{Trig} with event weight information and rename \texttt{PFC} to \texttt{Part} to indicate truth-level particles.  The MOD format can be easily extended to accommodate additional information in the future.

\begin{table}[t]
\begin{tabular}{r @{$\quad$} l @{$\quad$} r @{$\quad$} r}
\hline
\hline
Code & Candidate & Total Count & $p_T > 1~\GeV$ \\
\hline
$11$ & electron ($e^-$) & 32,917 & 32,900\\
$-11$ & positron ($e^+$) & 32,984 & 32,968\\
$13$ & muon ($\mu^-$) & 12,941 & 12,653 \\
$-13$ & antimuon ($\mu^+$) & 13,437 & 13,110 \\
$211$ & positive hadron ($\pi^+$) & 6,908,914 & 5,183,048 \\
$-211$ & negative hadron ($\pi^-$) & 6,729,328 & 5,027,146 \\
\hline
22 & photon ($\gamma$) & 9,436,530 & 4,805,173 \\
130 & neutral hadron ($K_L^0$) & 2,214,385 & 1,658,892\\
\hline
\hline
\end{tabular}
\caption{Valid particle identification codes for PFCs, with their most likely hadron interpretation.  The total counts are taken from the sample of hard central jet with $p_T > 85~\GeV$ and $|\eta| < 2.4$.  In the forward region with $|\eta| > 2.4$, one also finds code $1$ (for forward hadron candidate) and code $2$ (for forward electron/photon candidate).  The last column lists the counts after the $p_T^{\rm min} = 1.0~\GeV$ cut derived in \Fig{fig:PFC}.}
\label{tab:pfc_names}
\end{table}

The list of valid particle identification codes for the PFCs is given in \Tab{tab:pfc_names}, along with their prevalence in the hardest jet sample ($p_T > 85~\GeV$, $|\eta| < 2.4$).  These codes, determined by the CMS particle flow algorithm, are inspired by the Monte Carlo particle number scheme in \Ref{Olive:2016xmw}.  For example, all charged hadron candidates are assigned a code of $\pm 211$ corresponding to charged pions, which are more prevalent than charged kaons.  Neutral pions, which decay as $\pi^0 \to \gamma \gamma$, are typically reconstructed as one or two photon candidates with code 22.  Neutral hadron candidates are assigned code 130 corresponding to $K$-long.  Electrons ($\pm 11$) and muons ($\pm 13$) are relatively rare in our jet sample.

Although the AK5 jets are derived from clustering the PFCs, we need to separately extract the AK5 jets provided by CMS in order to obtain JEC factors and impose jet quality cuts.  Throughout our analysis, we impose the recommended ``loose'' jet quality cut; see \Tab{tab:jet_quality} in \App{app:opendatainfo}.  Due to numerical rounding issues when outputting the MOD text file, the AK5 jets from CMS and ones we cluster ourselves from the PFCs can be subtly different, though if we restrict our attention to the hardest jet, this is a rare effect that has almost no impact in our analysis (see further discussion in \Sec{sec:MODAnalyzer}).\footnote{Alternatively, we could have decided to directly identify the PFC constituents of the AK5 jet using CMSSW.  This leads to a different numerical rounding issue where the jet is not the four-vector sum of its constituents.  These issues could have been avoided by not relying on text-based output, at the expense of requiring ROOT dependencies in \textsc{MODAnalyzer}.  Our internal ROOT-based analysis framework encounters no numerical rounding issues.} 


After running \texttt{gzip} for compression, the final MOD files are roughly 10 times smaller than the corresponding AOD files (which are already in a compressed ROOT format).  Furthermore, if we restrict to a skimmed dataset where the hardest jet has $p_T > 85~\GeV$ and the assigned trigger fired, we reduce the 198.8 gigabytes of compressed MOD files down to 11.6 gigabytes.  This is small enough to easily fit on a flash drive.

\subsection{Analysis Tools}
\label{sec:MODAnalyzer}

With the MOD files in hand, we are no longer tied to CMSSW.  In order to leverage existing jet substructure tools, we built an external analysis framework in \textsc{C++} based on the \textsc{FastJet} package \cite{Cacciari:2011ma}.  This framework, called \textsc{MODAnalyzer}, is available from a \textsc{GitHub} repository \cite{MODAnalyzer}, which also includes the \textsc{Python} histogramming and plotting tools used for this paper.  For the soft drop studies in \Sec{sec:2prong}, we use the \textsc{RecursiveTools} 1.0.0 package from \textsc{FastJet contrib} 1.019 \cite{fjcontrib}.

The structure of \textsc{MODAnalyzer} mirrors the structure of the MOD files.  The core class is \texttt{Event}, which is not only a container for all of the event information but also handles parsing of the MOD files and selecting the assigned trigger for the hardest jet.  The \texttt{Cond} and \texttt{Trig} MOD entries are stored in \texttt{Condition} and \texttt{Trigger} classes.  The \texttt{AK5} and  \texttt{PFC} MOD entries are stored as \textsc{FastJet} \texttt{PseudoJet} objects.   To amend these \texttt{PseudoJet}s with additional MOD-specific information, we define \texttt{InfoCalibratedJet} and \texttt{InfoPFC} classes that inherit from \textsc{FastJet}'s \texttt{UserInfoBase}.  Apart from the \texttt{Event} class, the elements of \textsc{MODAnalyzer} are relatively lightweight, since much of the required functionality is already provided by \textsc{FastJet}.

The main complication in processing the MOD files is handling the duplicate jet information.  Within \textsc{MODAnalyzer}, we have two types of jets:  AK5 jets clustered by CMS and anti-$k_t$ $R = 0.5$ jets clustered internally from the PFCs. Note that the AK5 jets are associated with JEC factors and jet quality criteria, whereas the internally clustered jets are not, so we cannot discard the AK5 jets completely.  To define the hardest jet in the event (i.e.~the ``trigger jet''), we use the AK5 jet sample from CMS, rescaling the jet $p_T$ values by the appropriate JEC factors and checking whether the assigned trigger fired.  At this point, we remove events where the trigger jet fails the loose jet quality cut.  We then find the internal PFC jet that is closest in rapidity-azimuth to the trigger jet.  If this internal jet has the same number of constituents as the trigger jet and if the four-momenta agree up to $1~\MeV$ precision (after rescaling the internal jet by the same JEC factor), then we declare a match and perform all subsequent analyses on the internal jet.  In rare cases where there is no match, we discard the event, though this only affects 1 event out of 843,129 in our analysis (see \Tab{tab:event_numbers}).

\begin{figure*}[t]
\subfloat[]{
\label{fig:PFC_neutral}
\includegraphics[width=\columnwidth, page=1]{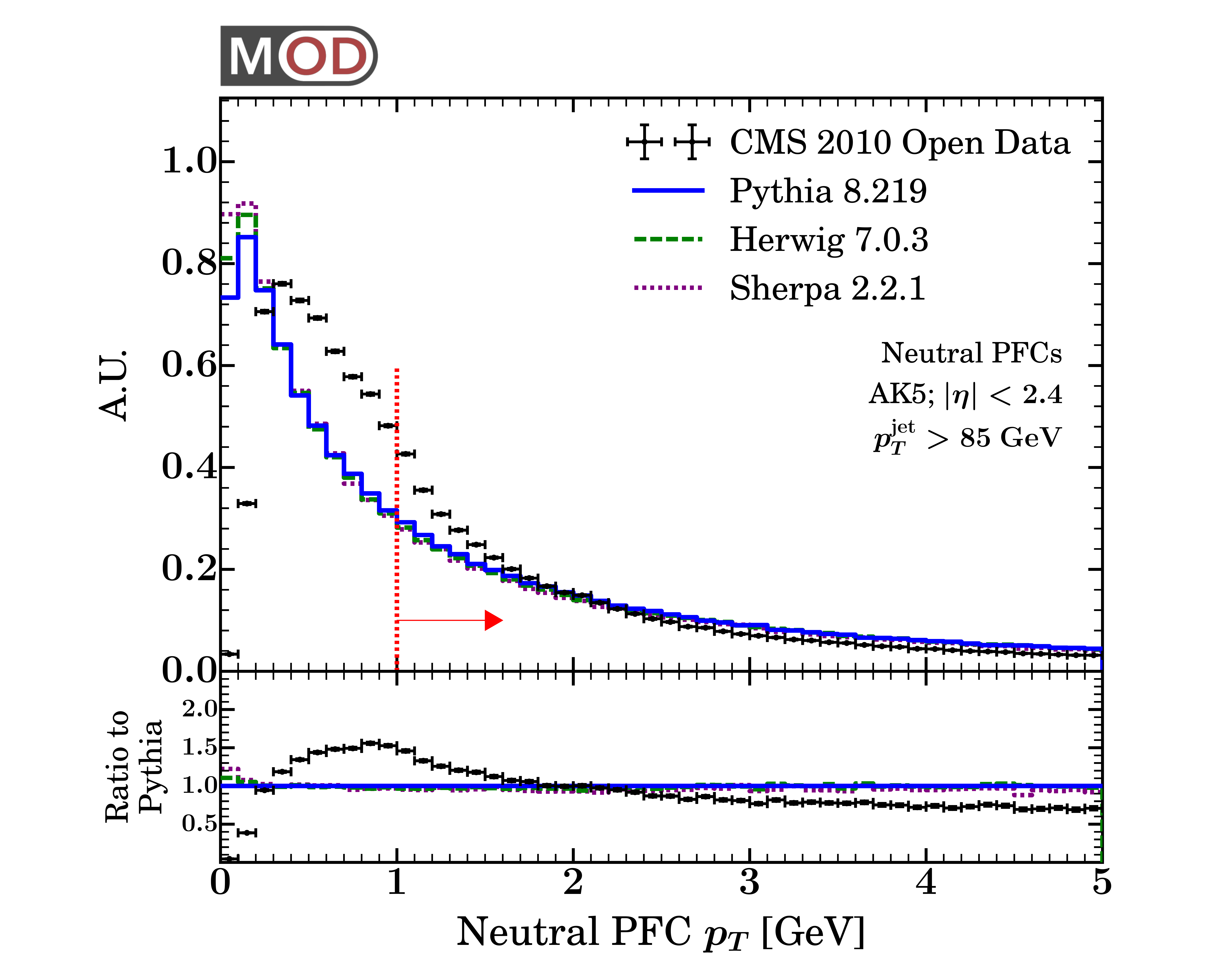}
}
\subfloat[]{
\label{fig:PFC_charged}
\includegraphics[width=\columnwidth, page=1]{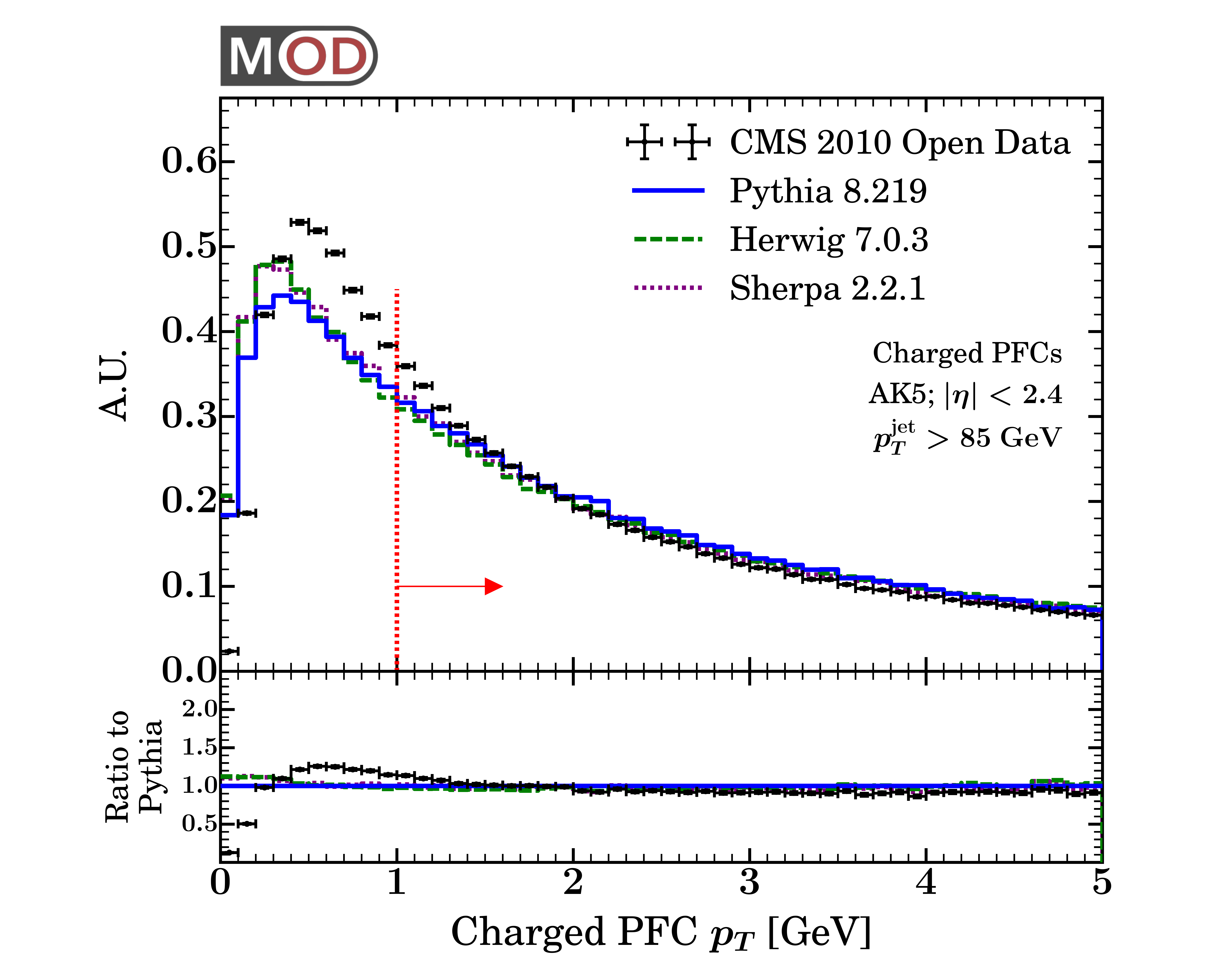}
}
\caption{Transverse momentum spectrum of raw PFCs, for (a) neutral candidates and (b) charged candidates.  These histograms are populated only with PFCs from the hardest jet in the stated jet $p_T$ range, comparing the CMS Open Data to parton shower generators.  The cuts used in our jet substructure studies are $p_T^{\rm min} = 1.0~\GeV$, applied to both neutral and charged PFCs.  For this and all remaining plots in this paper, one must keep in mind that the detector-level CMS Open Data and the particle-level parton showers are not directly comparable.  See \Fig{fig:pfc_extended} in the appendix for a version of this figure with an extended $p_T$ range.}
\label{fig:PFC}
\end{figure*}

Within \textsc{MODAnalyzer}, we have a few ways to speed up the workflow.  A large fraction of MOD events are unsuitable for analysis, mostly because the hardest jet was below the $85~\GeV$ minimum $p_T$ threshold set in \Tab{tab:trigger_table}.  We can therefore perform event skimming, where we read in each MOD file and generate a new MOD file with only events where the assigned trigger fired.\footnote{For the trigger and luminosity studies in \Sec{subsection:jet_primary_dataset}, we of course had to use the unskimmed MOD files.}  Similarly, because our analysis is only based on the hardest jet in the event, we can output MOD files with a \texttt{Hardest\_Jet\_Selection} header, where only the PFC constituents of the hardest jet are stored, and the minimally required \texttt{Trig}, \texttt{Cond}, and \texttt{AK5} information is consolidated under the \texttt{1JET} keyword.

After \texttt{gzip} compression, the \texttt{Hardest\_Jet\_Selection} MOD files only take 725 megabytes, which is small enough that we plan to make the files publicly available ourselves through \textsc{DSpace@MIT}.\footnote{The CMS Open Data is released under the Creative Commons CC0 waiver \cite{CC0}.  If you use the \texttt{Hardest\_Jet\_Selection} MOD files as part of an analysis, please cite the CMS Jet Primary Dataset \cite{CMS:JetPrimary} as well as this paper.}  This reduced MOD file can be used directly with \textsc{MODAnalyzer}, or one could build an alternative MOD analysis framework.

\subsection{Parton Shower Generators}
\label{sec:parton_shower}

For the initial 2010 CMS Open Data release, no simulated Monte Carlo datasets were provided.  In order to compare jet substructure results from open data with theoretical predictions, we use three parton shower generators: \pythia~\cite{Sjostrand:2007gs}, \herwig~\cite{Bellm:2015jjp}, and \sherpa~\cite{Gleisberg:2008ta}.  For each generator, we use the default settings for dijet production, since this is the process that dominates the single-jet triggers.  To efficiently populate the full phase space, we use a $p_T$-weighted event generation strategy, which is highly efficient for jet production with $p_T > 85~\GeV$, allowing us to use a single parton shower run to probe multiple $p_T$ ranges.  Our analyses are based on the raw output of the parton shower generators, without any detector simulation.\footnote{As mentioned in the introduction, we attempted to use the fast detector simulation tool \textsc{Delphes 3.3.2} \cite{deFavereau:2013fsa}, but the default CMS-like detector settings were intended to be used for jet studies, not jet substructure studies.  In the future, since \textsc{Delphes} does have a rudimentary version of particle flow reconstruction, it should be possible to tune \textsc{Delphes} to match published CMS jet substructure results.}

Each generator outputs to \textsc{HepMC} format \cite{Dobbs:2001ck}, which we then convert to the same MOD file format used for the open data, suitably modified to eliminate CMS-specific information like triggers, luminosity, and JEC factors. After event skimming and applying \texttt{Hardest\_Jet\_Selection}, the MOD files from the open data and the parton shower generators look essentially identical, such that the same workflow can be used for all sources.

Because the parton showers do not include detector effects by default, we have to be careful in drawing conclusions about agreement or disagreement with the open data.  For example, depending on the kinematics, the CMS particle flow reconstruction can sometimes reconstruct $\pi^0 \to \gamma \gamma$ as a single ``photon'' instead of two photons, which can affect jet substructure observables like constituent multiplicity.\footnote{One could partially mitigate this effect by forcing the $\pi^0$ to be stable within the generators, but this is not a replacement for a real detector simulation.}

To partially account for the finite energy resolution of the CMS detector, we impose a restriction of $p_T^{\rm min} = 1.0~\GeV$ on each PFC (or truth-level particle in the case of the parton showers).  This cut is motivated by \Fig{fig:PFC}, which suggests that PFCs below 1 GeV are subject to inefficiencies and mismeasurements.  Crucially, this $p_T^{\rm min}$ restriction is only imposed for substructure observables; the original jet kinematics are given by all PFCs with the CMS-provided JEC factors.  This universal $p_T^{\rm min}$ cut is similar in spirit to the \textsc{SoftKiller} approach to pileup mitigation \cite{Cacciari:2014gra}.

Comparing \Fig{fig:PFC_neutral} for neutral PFCs to \Fig{fig:PFC_charged} for charged PFCs, we see comparatively smaller differences between the CMS Open Data and the parton shower for charged PFCs; this will also be reflected in the substructure studies below.  For this reason, we always perform cross checks with track-based variants to address the finite granularity of the CMS calorimeter.  Since the particle flow algorithm uses information from both the tracker and the calorimeter, the angular resolution of charged particles is much better than for neutral particles.  This allows us to test whether there are large distortions to jet substructure observables from finite calorimeter cell size, especially for soft-dropped observables which probe the collinear core of the jet.\footnote{We also tried pre-clustering the jet into small subjets as a way to mimic finite angular resolution, but this simply led to increased smearing without improved agreement between data and parton showers.}  These track-based variants exploit the excellent track resolution of the CMS detector at the expense of losing neutral particle information, but since almost all of our substructure observables we study are dimensionless, the impact of switching to track-based variants is mild (see also \cite{Chang:2013rca,Chang:2013iba}).

\begin{figure*}
\subfloat[]{
\label{fig:jet_pt_spectrum}
\includegraphics[width=\columnwidth, page=1]{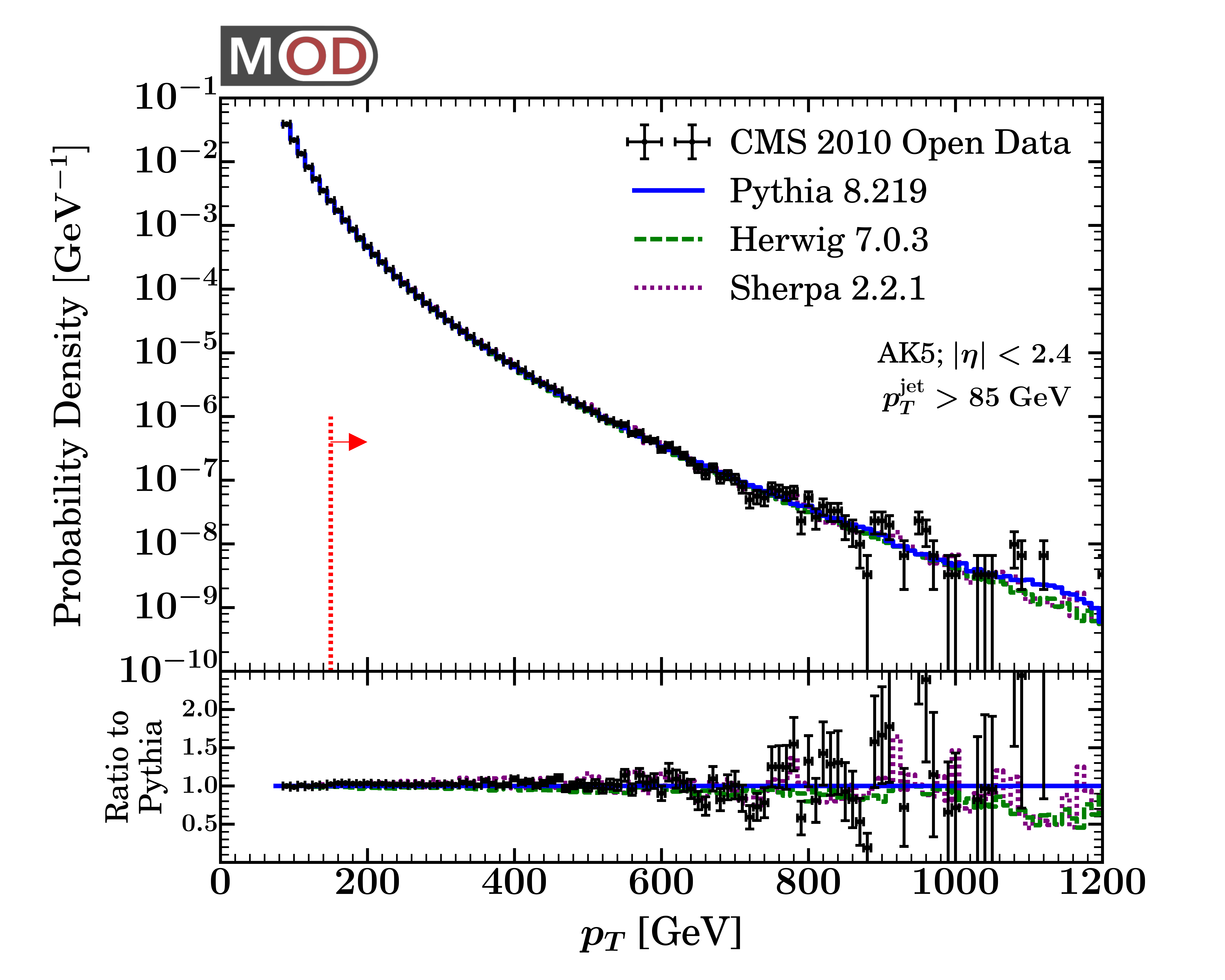}
}
\subfloat[]{
\label{fig:pT_corrected_vs_uncorrected}
\includegraphics[width=\columnwidth, page=1]{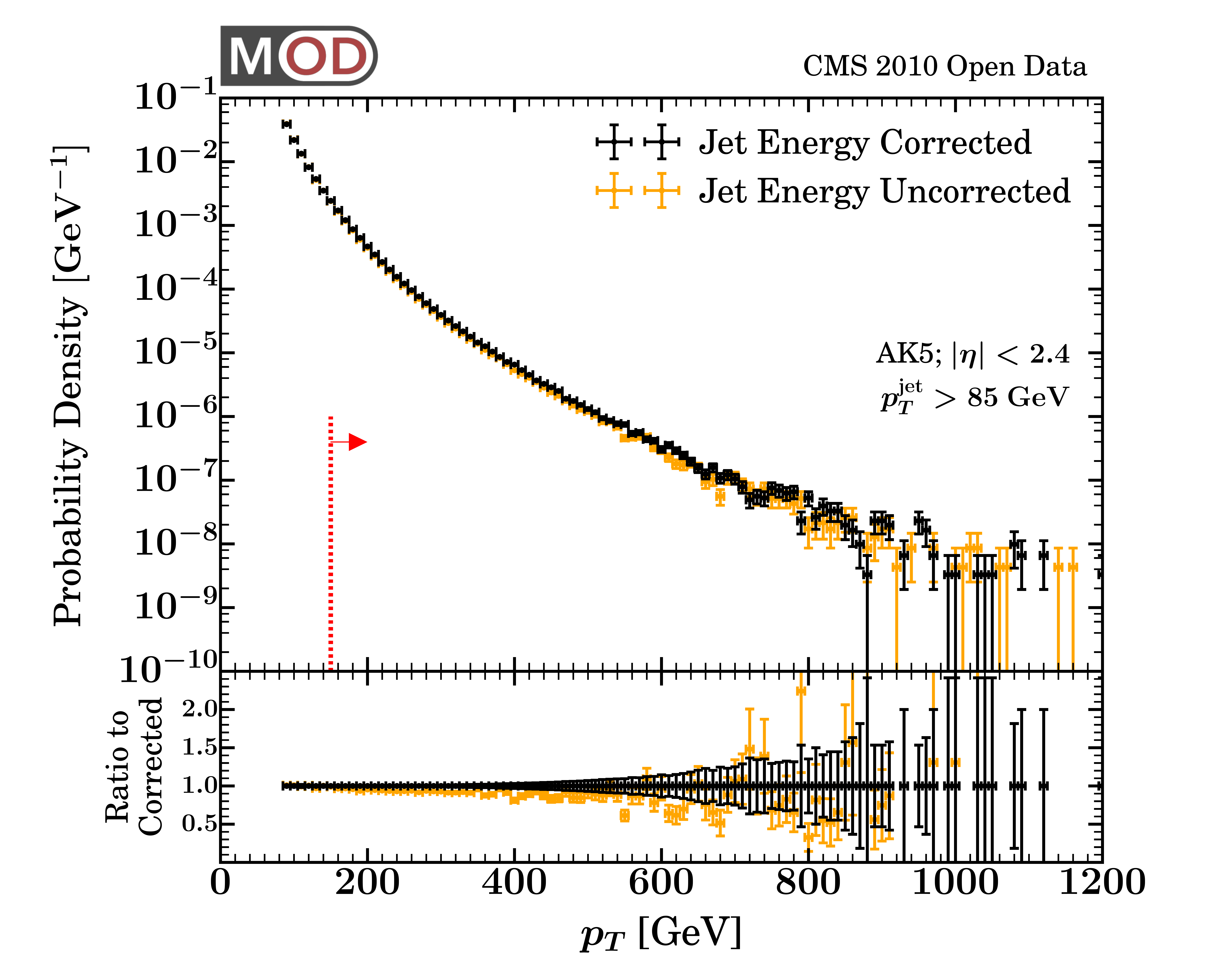}
}
\caption{(a) Hardest jet $p_T$ spectrum, comparing the CMS Open Data with \pythia, \textsc{Herwig 7.0.3}, and \textsc{Sherpa 2.2.1}.  The maximum jet $p_T$ in the Jet Primary Dataset is $1277~\mathrm{GeV}$.  (b) Hardest jet $p_T$ before and after applying the appropriate JEC factors.  Because these are normalized histograms with the same $p_T > 85~\GeV$ cut, the mismatch in JEC values is only apparent at high $p_T$. }
\end{figure*}

\begin{figure*}
\subfloat[]{
\label{fig:jet_azimuthal_spectrum}
\includegraphics[width=\columnwidth, page=1]{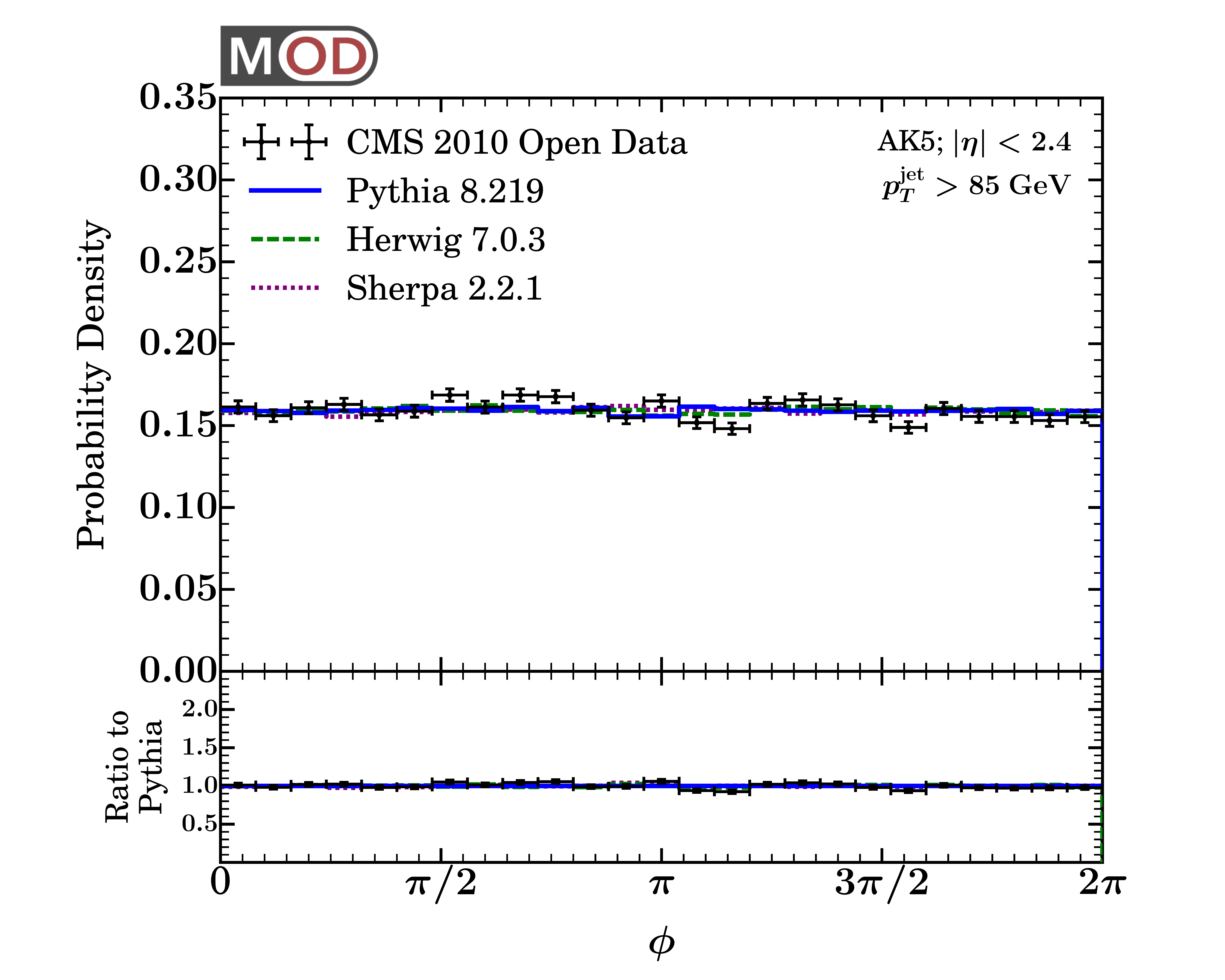}
}
\subfloat[]{
\label{fig:jet_rapidity_spectrum}
\includegraphics[width=\columnwidth, page=1]{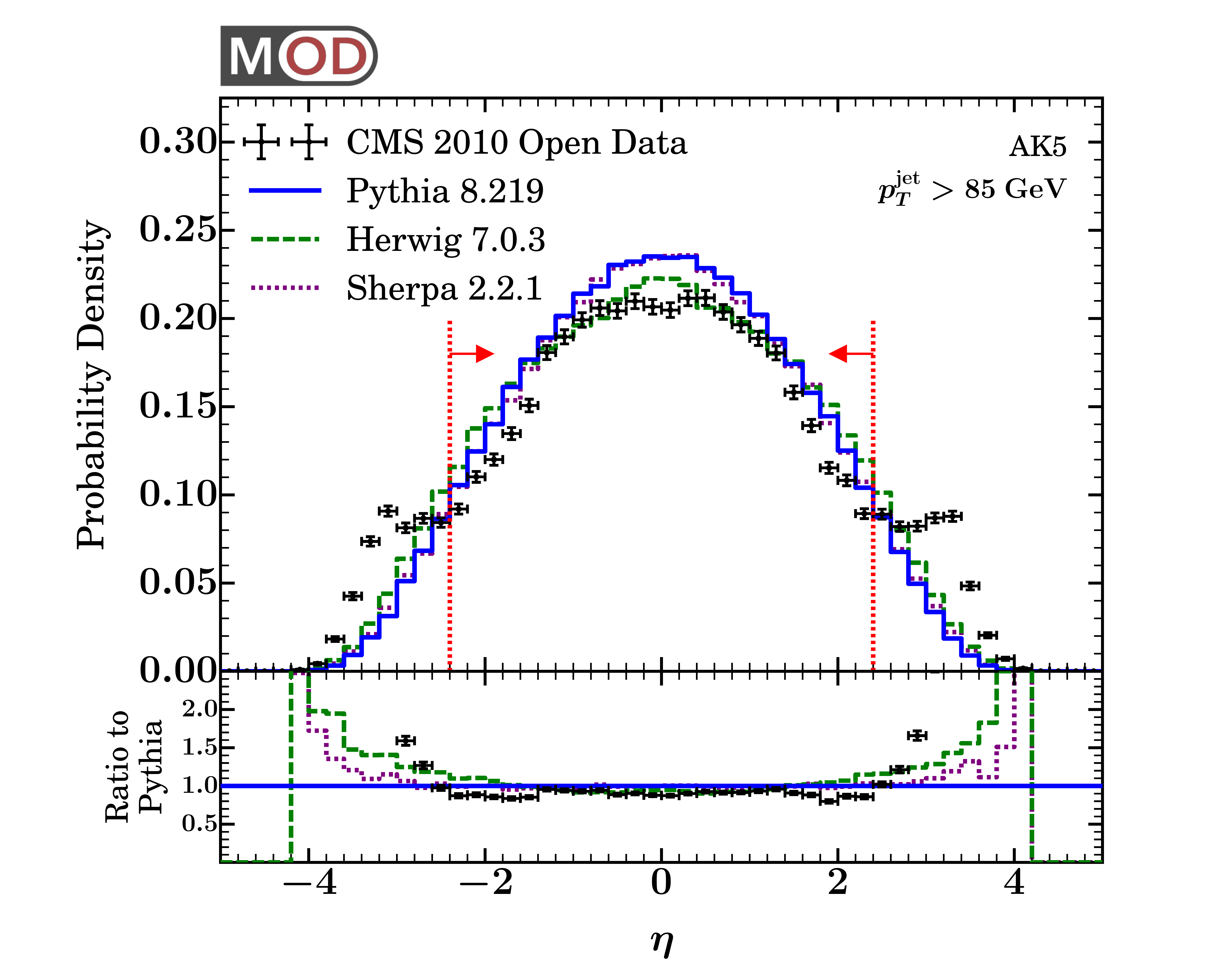}
}
\caption{(a) Azimuthal angle of the hardest jet, which is flat as desired.  (b)  Pseudorapidity spectrum for the hardest jet.  Note the population of anomalous jets at $|\eta| > 2.4$, coming from the edge of tracking acceptance, which is why we enforce $|\eta| < 2.4$ in our analysis.}
\end{figure*}

\section{Hardest Jet Properties}
\label{sec:hardest_jet}

We now present basic kinematic and substructure observables for the hardest $p_T$ jet in an event, comparing CMS Open Data to parton shower generators.  Unless otherwise stated, the jet $p_T$ values always include the appropriate JEC factors, and we restrict our attention to jets with $|\eta| < 2.4$ and $p_T > 85~\GeV$.  Following the 2010 CMS default, the anti-$k_t$ jet radius is always $R = 0.5$.  In the text, we primarily show distribution for $p_T > 150~\GeV$ in order to avoid the large prescale values associated with the \texttt{HLT\_Jet15U/Jet30U} triggers.  In the \texttt{arXiv} source for this paper, each figure corresponds to a multipage file that has distributions for the full $p_T > 85~\GeV$ range, as well as for each of the $p_T$ ranges defined in \Tab{tab:trigger_table}.

\subsection{Jet Kinematics}

The $p_T$ spectrum for the hardest jet is shown in \Fig{fig:jet_pt_spectrum}, going down to the $85~\GeV$ threshold set by the lowest trigger in \Tab{tab:trigger_table}. We see excellent agreement with parton shower predictions.  As shown in \Fig{fig:pT_corrected_vs_uncorrected}, this good agreement is only possible because proper JEC factors were applied.  Because we plot normalized histograms and because the $p_T$ spectrum is steeply falling, the impact of the JEC factors is not so apparent at low $p_T$, but becomes increasingly visible going to higher $p_T$.

Turning to angular information, we show the jet azimuthal spectrum in \Fig{fig:jet_azimuthal_spectrum}, which is flat as expected.  For the jet pseudorapidity distribution in \Fig{fig:jet_rapidity_spectrum}, central jets with $|\eta| < 2.4$ match parton shower expectations within uncertainties.  We see, however, a population of jets at $|\eta| > 2.4$ above parton shower expectations.  These are most likely jet fakes that are able to erroneously pass the jet quality criteria due to the lack of tracking information at forward rapidities.  For this reason, we restrict our attention to jets with $|\eta| < 2.4$ in our substructure studies.\footnote{Even if the jet axis satisfies $|\eta| < 2.4$, the jet constituents can extend to higher $\eta$ values where the tracking degrades quickly.  We explicitly checked that none of the jet substructure distributions studied below is substantially modified by taking the more conservative restriction of $|\eta| < 1.9$ (i.e.\ $2.4$ minus the $R = 0.5$ jet radius).  We further checked that there were no obvious pathologies for jets with $1.9 < |\eta| < 2.4$, even for observables like track multiplicity.  For substructure studies, this tracking issue is subdominant to the choice of $p_T^{\rm min}$ in \Fig{fig:PFC_charged}, in part because the jet cross section is falling with increasing $|\eta|$, so any tracking pathologies affect only a small portion of phase space.  The CMS jet mass study in \Ref{Chatrchyan:2013vbb} considers $|y| < 2.5$ despite similar potential tracking issues.}

\subsection{Basic Substructure Observables}
\label{subsec:basic_substructure}

\begin{figure*}
\subfloat[]{
\label{fig:constituent_multiplicity}
\includegraphics[width=0.9\columnwidth, page=1]{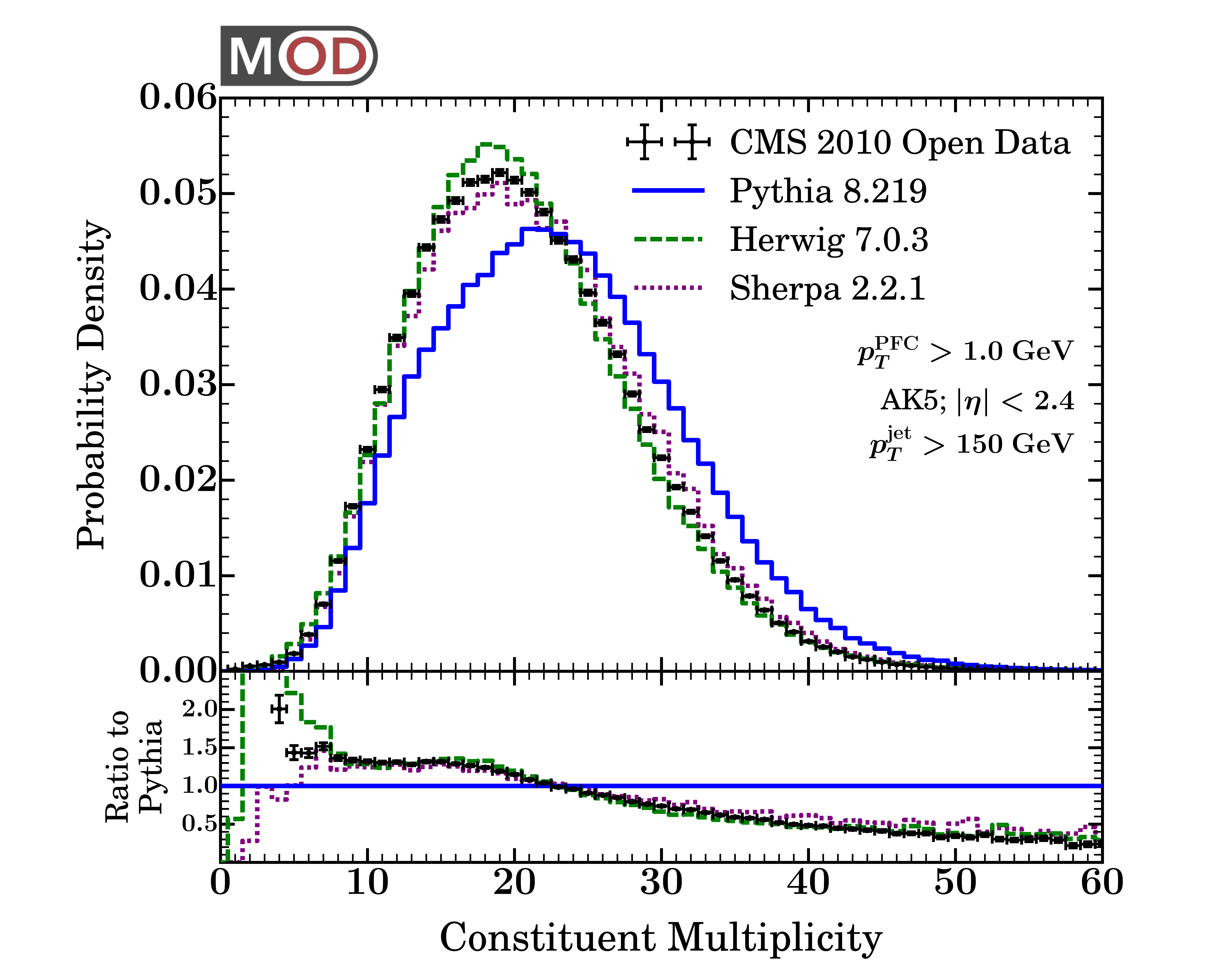}
}
\subfloat[]{
\label{fig:constituent_multiplicity_track}
\includegraphics[width=0.9\columnwidth, page=1]{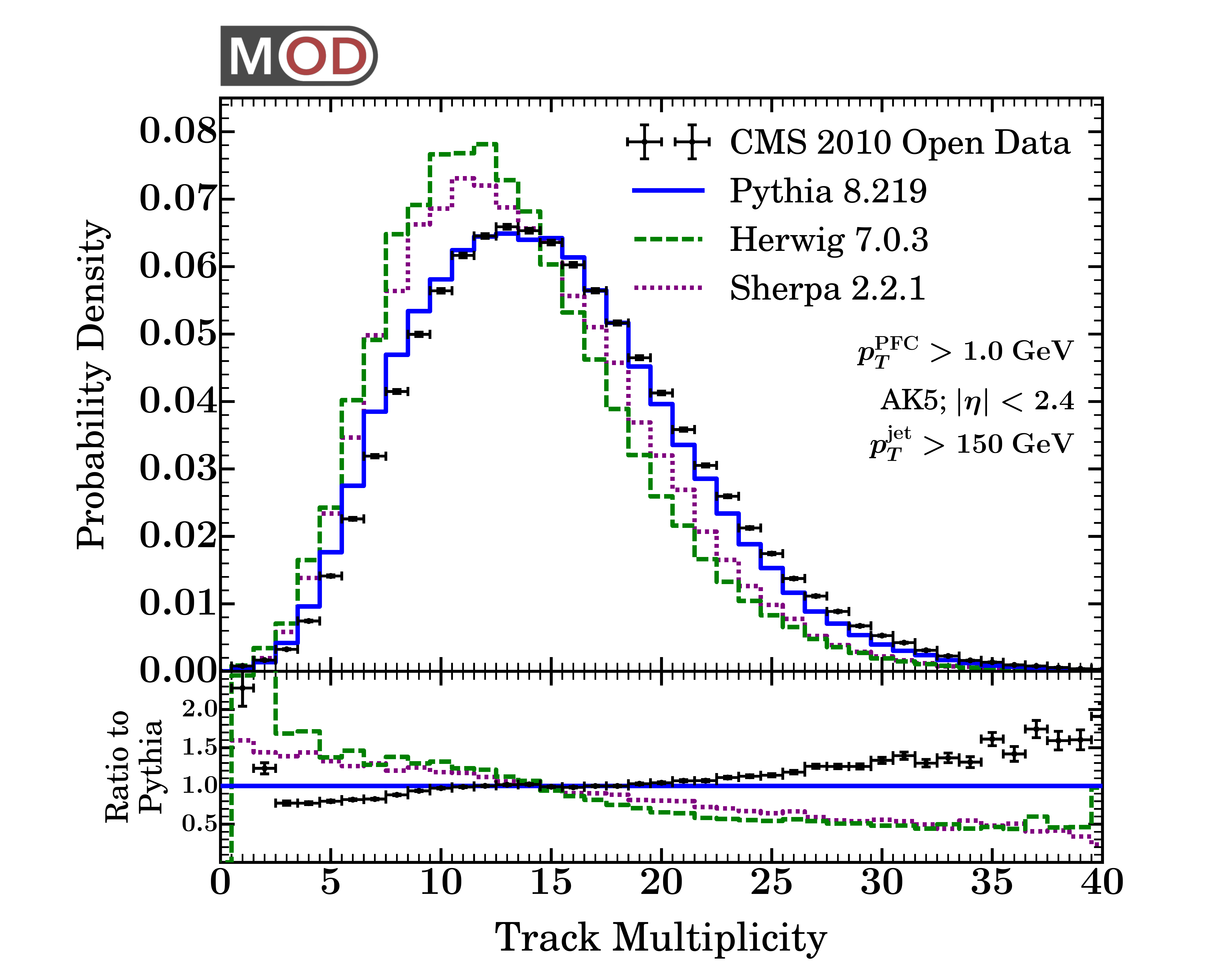}
}

\subfloat[]{
\label{fig:pTD}
\includegraphics[width=0.9\columnwidth, page=1]{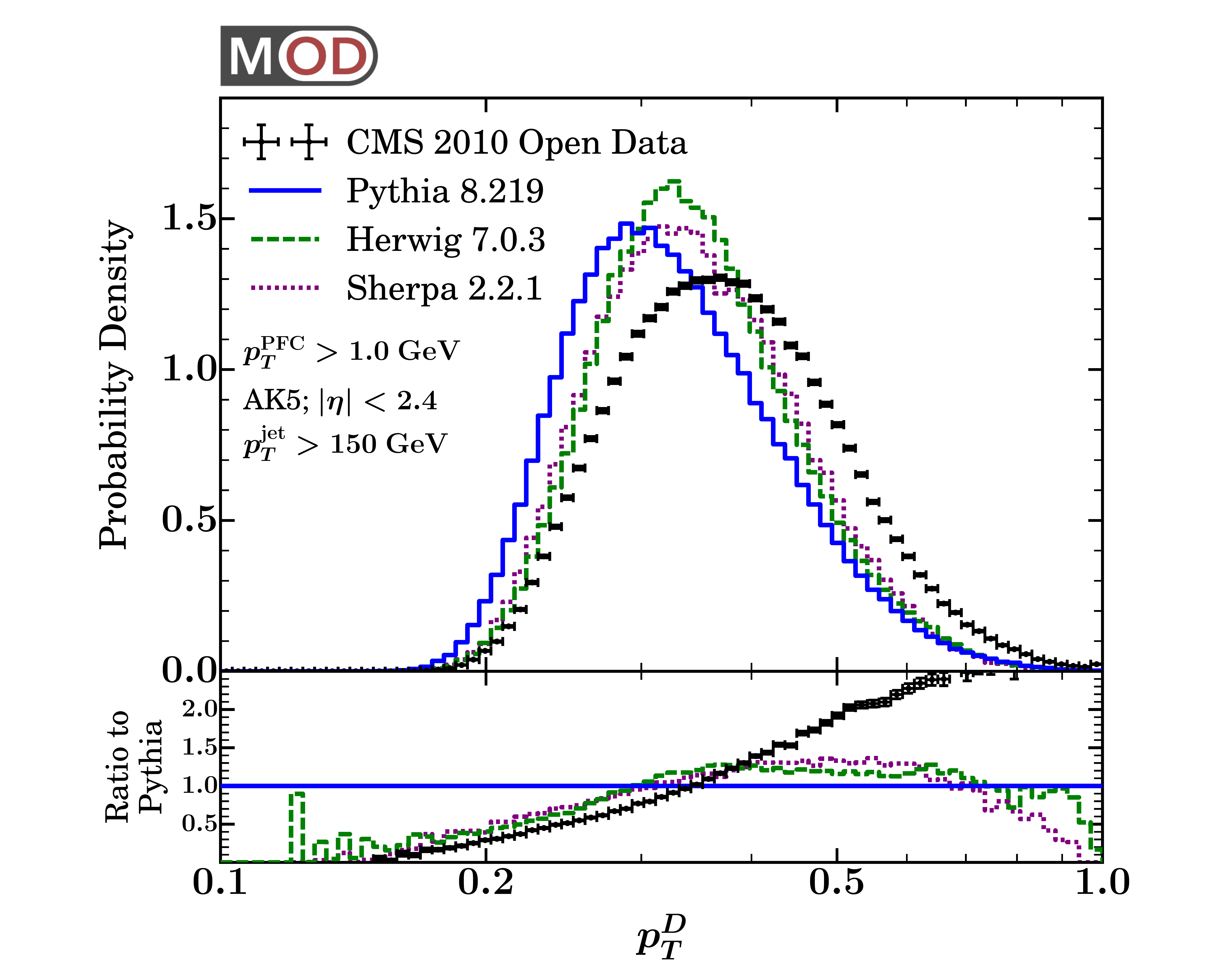}
}
\subfloat[]{
\label{fig:pTD_track}
\includegraphics[width=0.9\columnwidth, page=1]{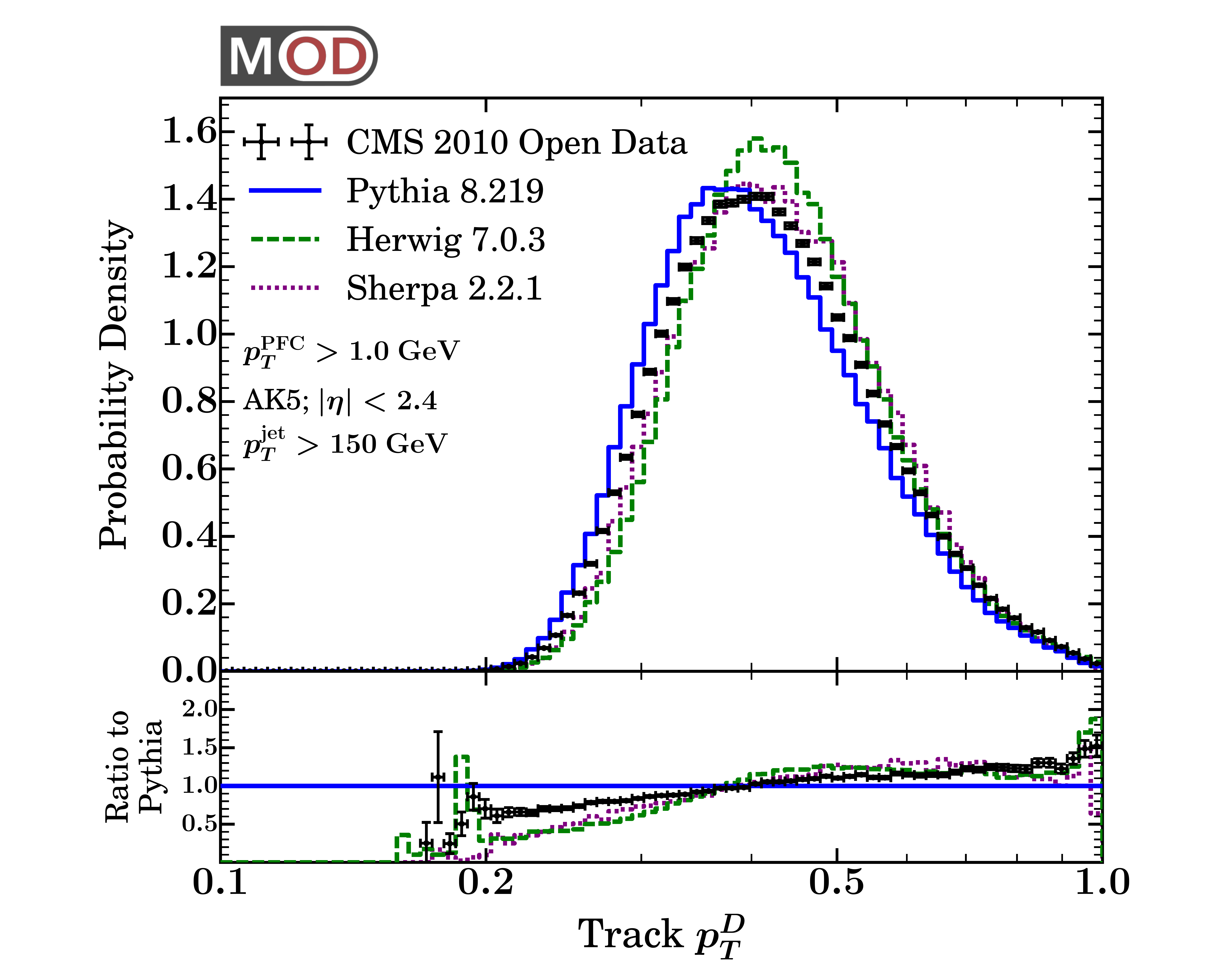}
}

\subfloat[]{
\label{fig:jet_mass}
\includegraphics[width=0.9\columnwidth, page=1]{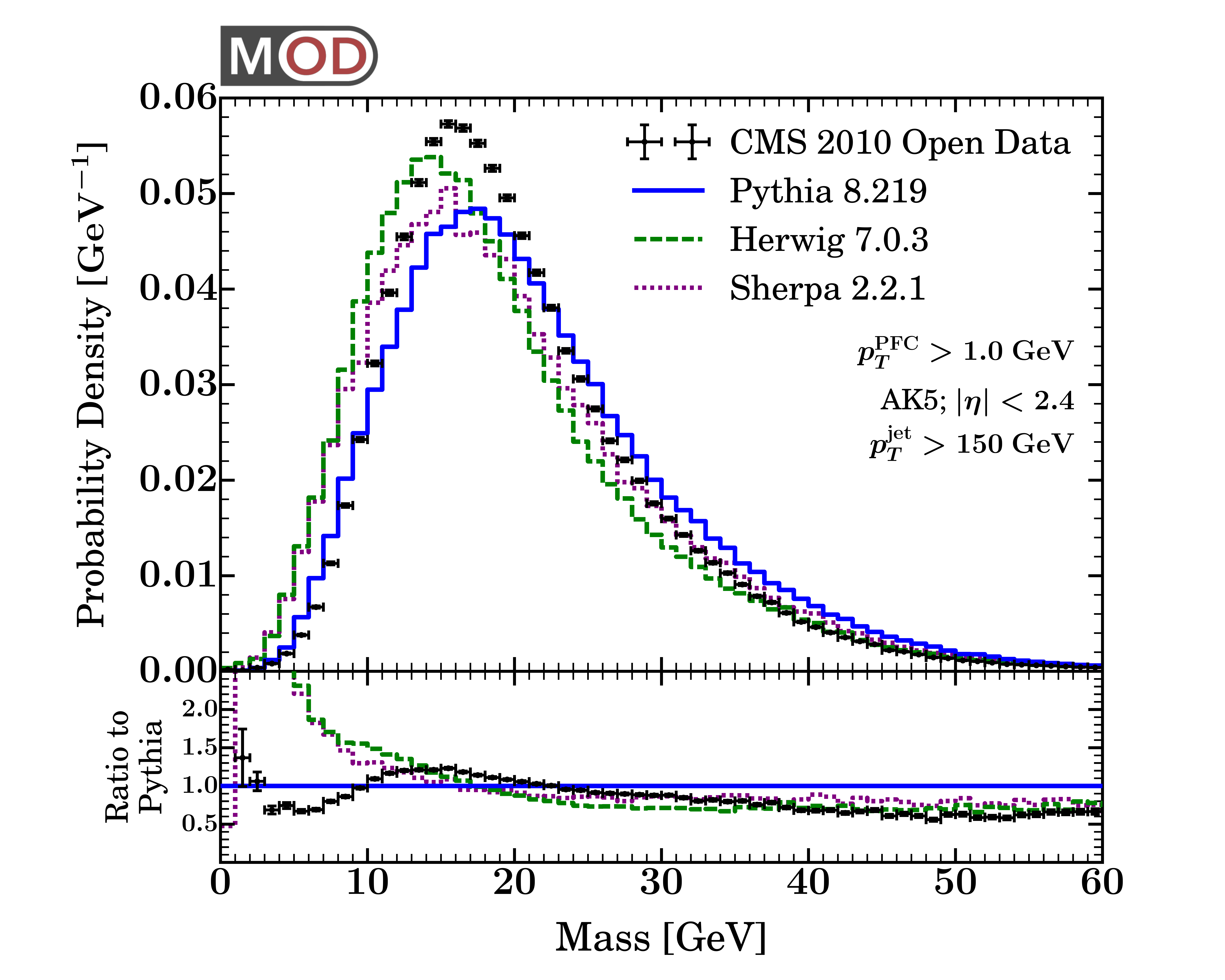}
}
\subfloat[]{
\label{fig:jet_mass_track}
\includegraphics[width=0.9\columnwidth, page=1]{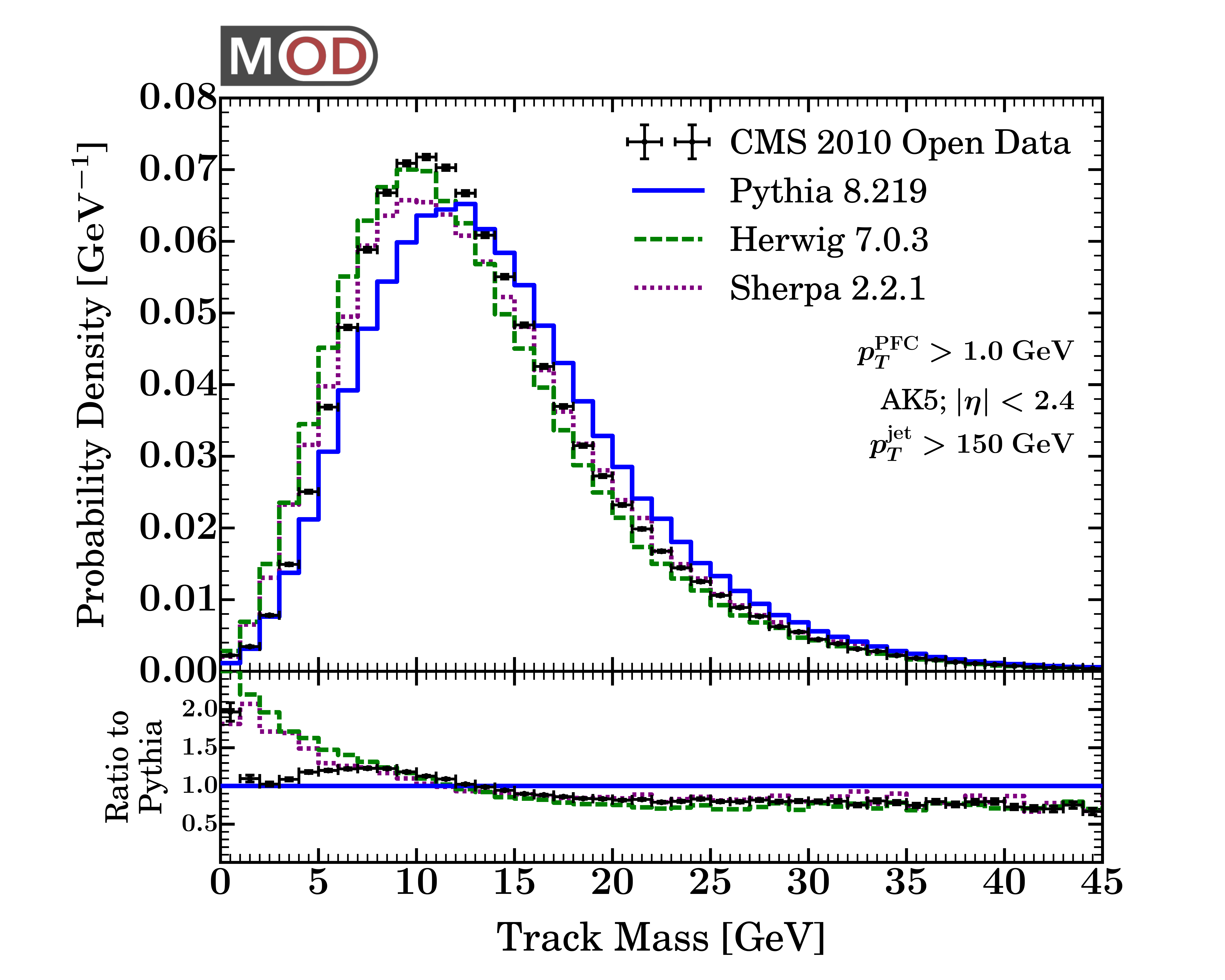}
}

\caption{Basic substructure observables for the hardest jet, using (left column) all PFCs and (right column) only charged PFCs, in both cases imposing a PFC cut of $p_T^{\rm min} = 1.0~\GeV$.  The observables are (top row) constituent multiplicity, (middle row) $p_T^D$ on a logarithmic scale and (bottom row) jet mass.  We emphasize that in this and all subsequent figures, the distributions are not directly comparable, since the CMS Open Data has not been unfolded to account for detector effects and the parton shower generators have not been folded with detector effects.  Similarly, only statistical uncertainties are shown for the open data.}
\label{fig:basic_substructure}
\end{figure*}

The most basic jet substructure observable is the multiplicity of jet constituents, though this is very sensitive to the details of CMS's particle flow reconstruction.   As mentioned in \Sec{sec:parton_shower}, we impose a cut of $p_T^{\rm min} = 1.0~\GeV$ on each PFC to avoid counting very soft particles that might not be efficiently reconstructed.  That said, CMS cannot resolve arbitrarily small angles and therefore particles can be merged by the particle flow algorithm, especially for $\pi_0 \to \gamma \gamma$.  For this reason, without a proper detector model, one has to be careful drawing conclusions from these substructure distributions.  With that caveat in mind, we proceed to overlay the detector-level CMS Open Data with the particle-level parton shower generators.  

In \Fig{fig:constituent_multiplicity}, we show the CMS Open Data constituent multiplicity distribution, which matches rather well to \textsc{Herwig} and \textsc{Sherpa}.  Once one restricts to charged particles in \Fig{fig:constituent_multiplicity_track}, however, the open data distribution shifts to lie closer to the \textsc{Pythia} distribution.  We therefore conclude that finite resolution of the calorimeter is an important detector effect that impacts jet substructure studies.  Without a detector model, though, we cannot meaningfully comment on the correspondence between the open data and the parton showers, especially for distributions like multiplicity that are infrared and collinear (IRC) unsafe.  The large differences between parton shower generators for charged particle multiplicity has been previously noted in e.g.~\Ref{Aad:2016oit}, indicating that unfolded measurement of multiplicity should be used in parton shower tuning.

We can see the same sensitivity to detector effects for the observable $p_T^D$ \cite{Pandolfi:1480598,Chatrchyan:2012sn}, defined as
\be
p_T^D = \frac{\sqrt{\sum_{i \in \text{jet}} p_{Ti}^2}}{\sum_{i \in \text{jet}} p_{Ti}}.
\ee
This observable is soft safe but collinear unsafe and used in CMS's quark/gluon discrimination studies \cite{CMS:2013kfa}.  Using a logarithmic scale to emphasize the shape, we see in \Fig{fig:pTD} that the CMS Open Data is at systematically at higher values of $p_T^D$ compared to parton shower predictions, again indicative of particle merging by the particle flow algorithm.  Testing the track-based variant in \Fig{fig:pTD_track}, we see much better agreement between the open data and the parton shower generators, where the differences between detector-level and particle-level  are comparable to the differences seen between generators.

\begin{figure*}
\subfloat[]{
\label{fig:jet_LHA}
\includegraphics[width=0.9\columnwidth, page=1]{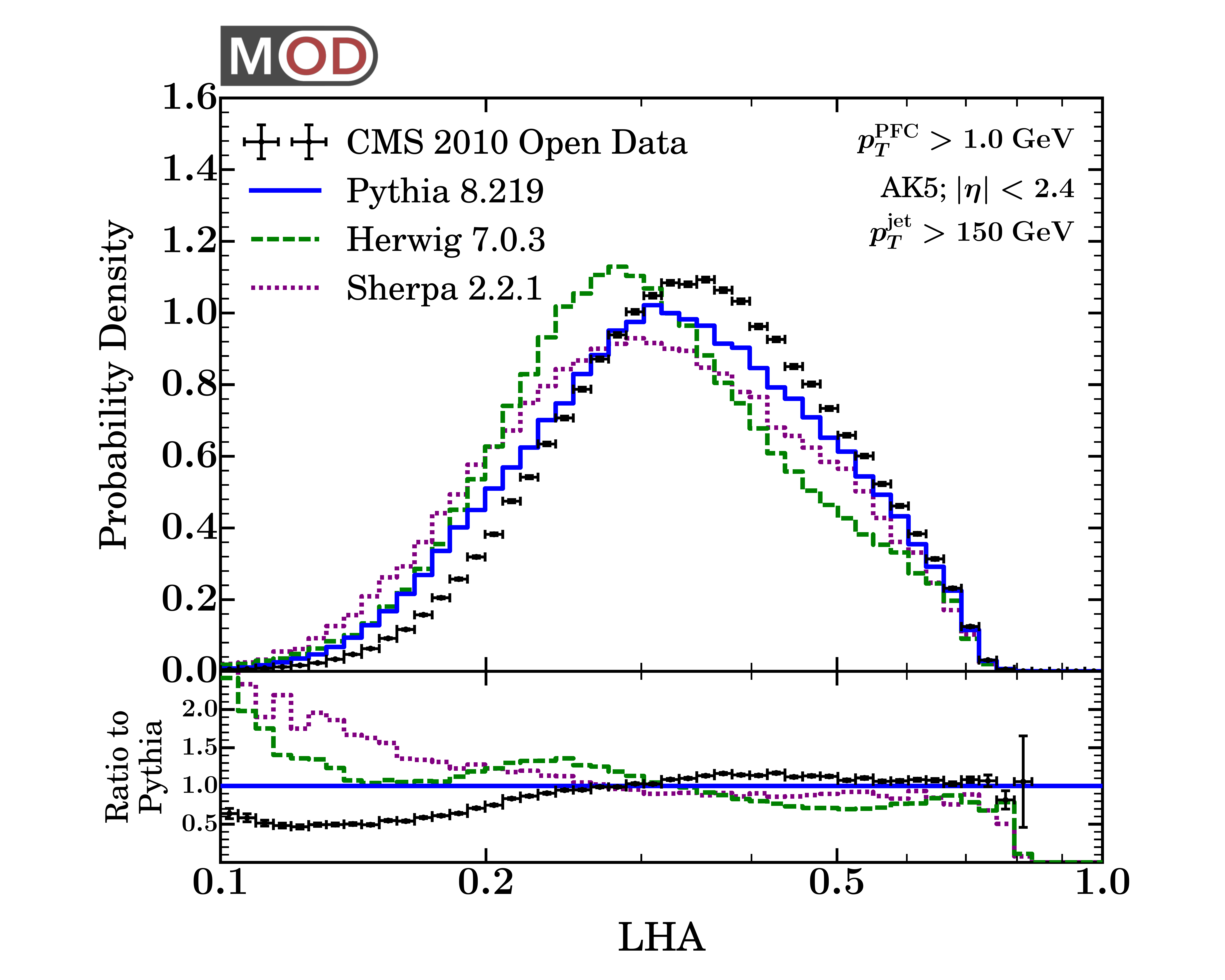}
}
\subfloat[]{
\label{fig:jet_LHA_track}
\includegraphics[width=0.9\columnwidth, page=1]{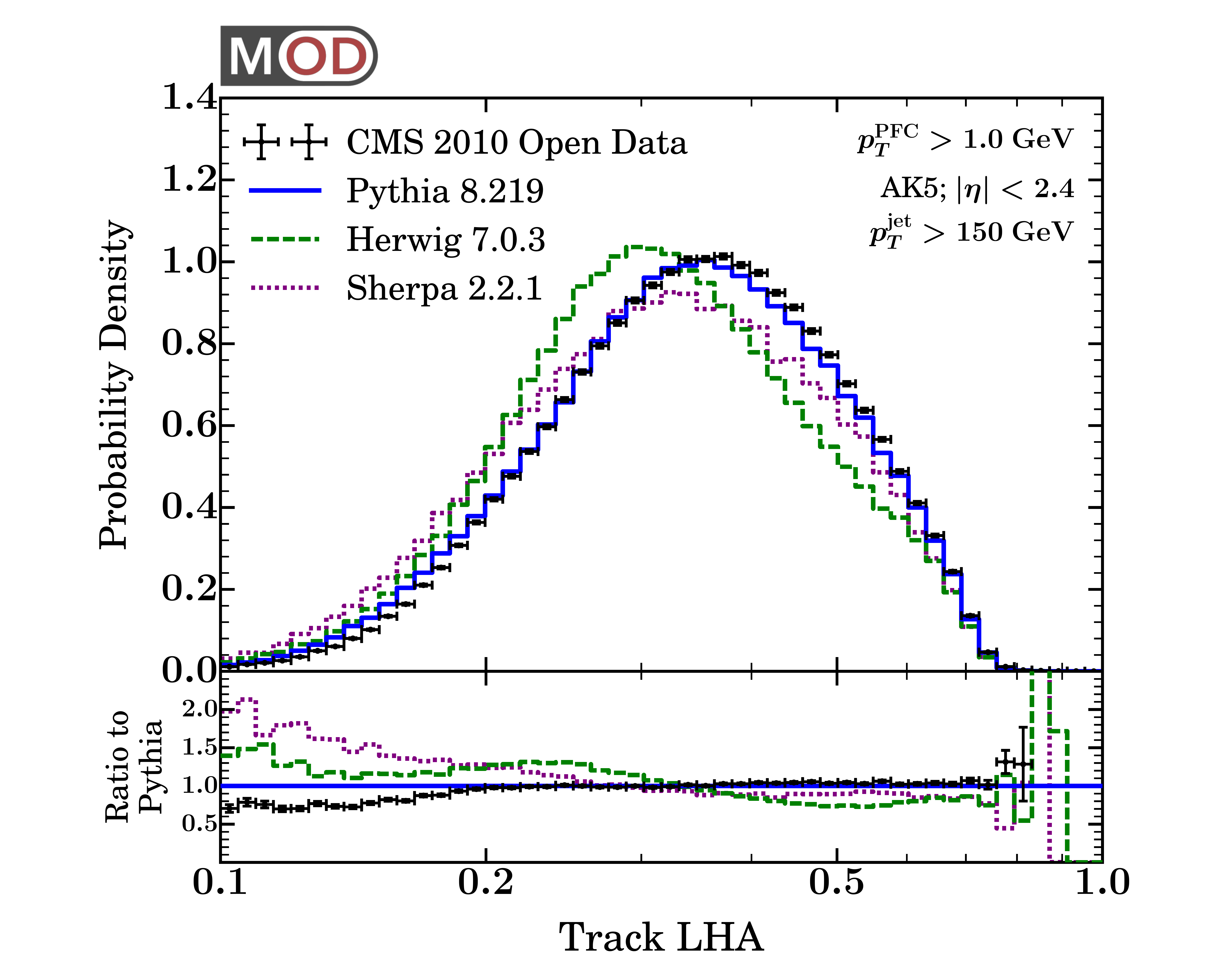}
}

\subfloat[]{
\label{fig:jet_width}
\includegraphics[width=0.9\columnwidth, page=1]{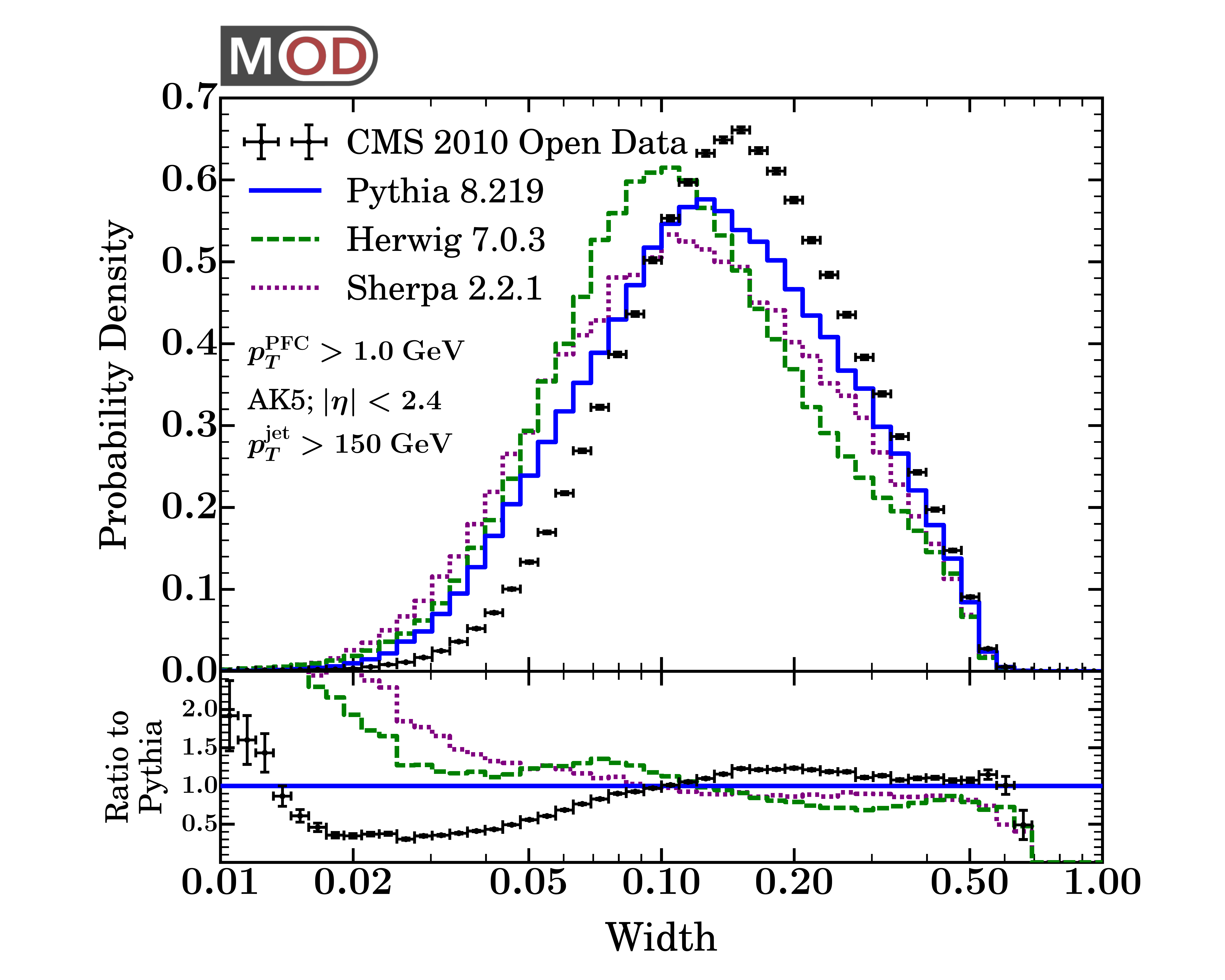}
}
\subfloat[]{
\label{fig:jet_width_track}
\includegraphics[width=0.9\columnwidth, page=1]{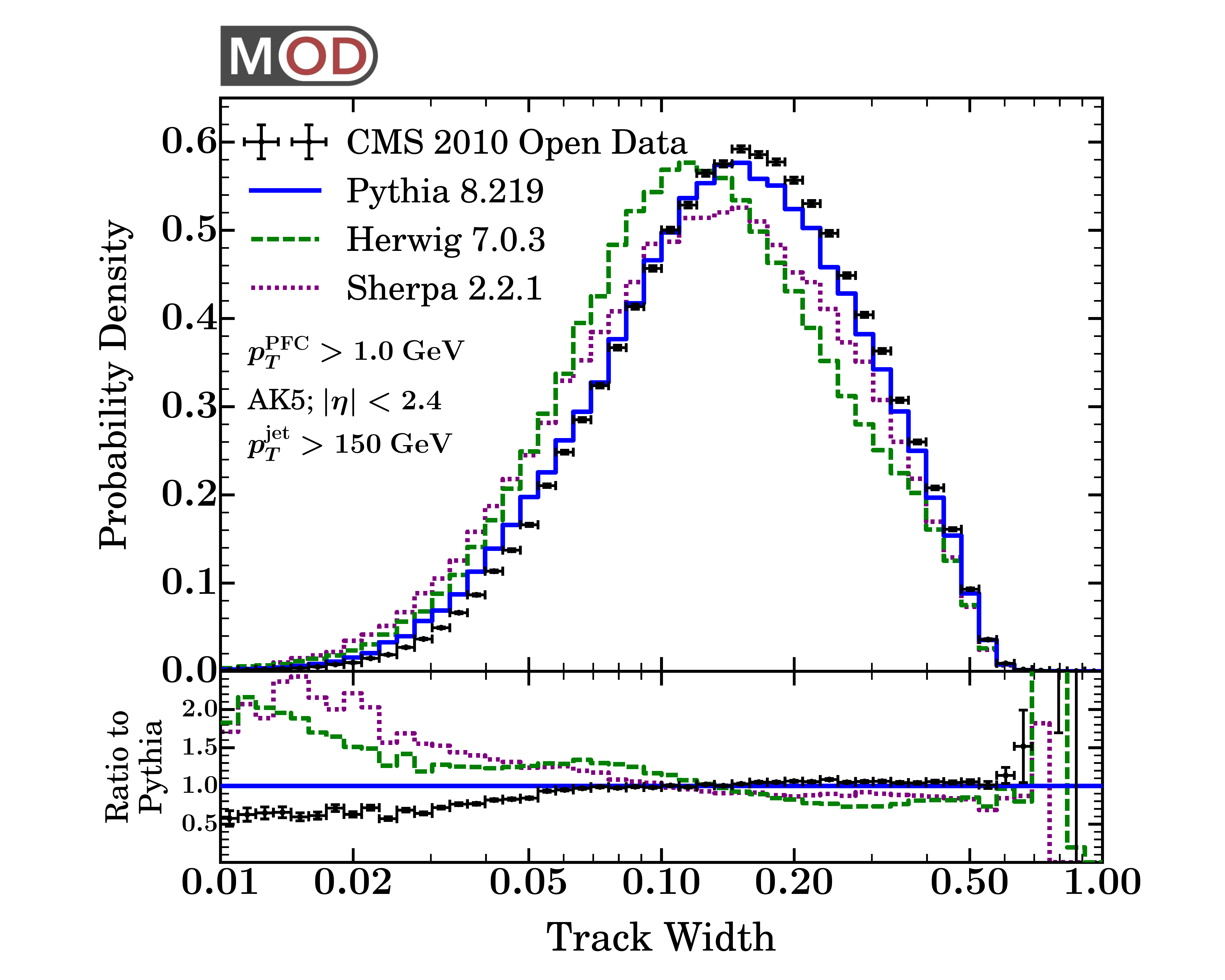}
}

\subfloat[]{
\label{fig:jet_thrust}
\includegraphics[width=0.9\columnwidth, page=1]{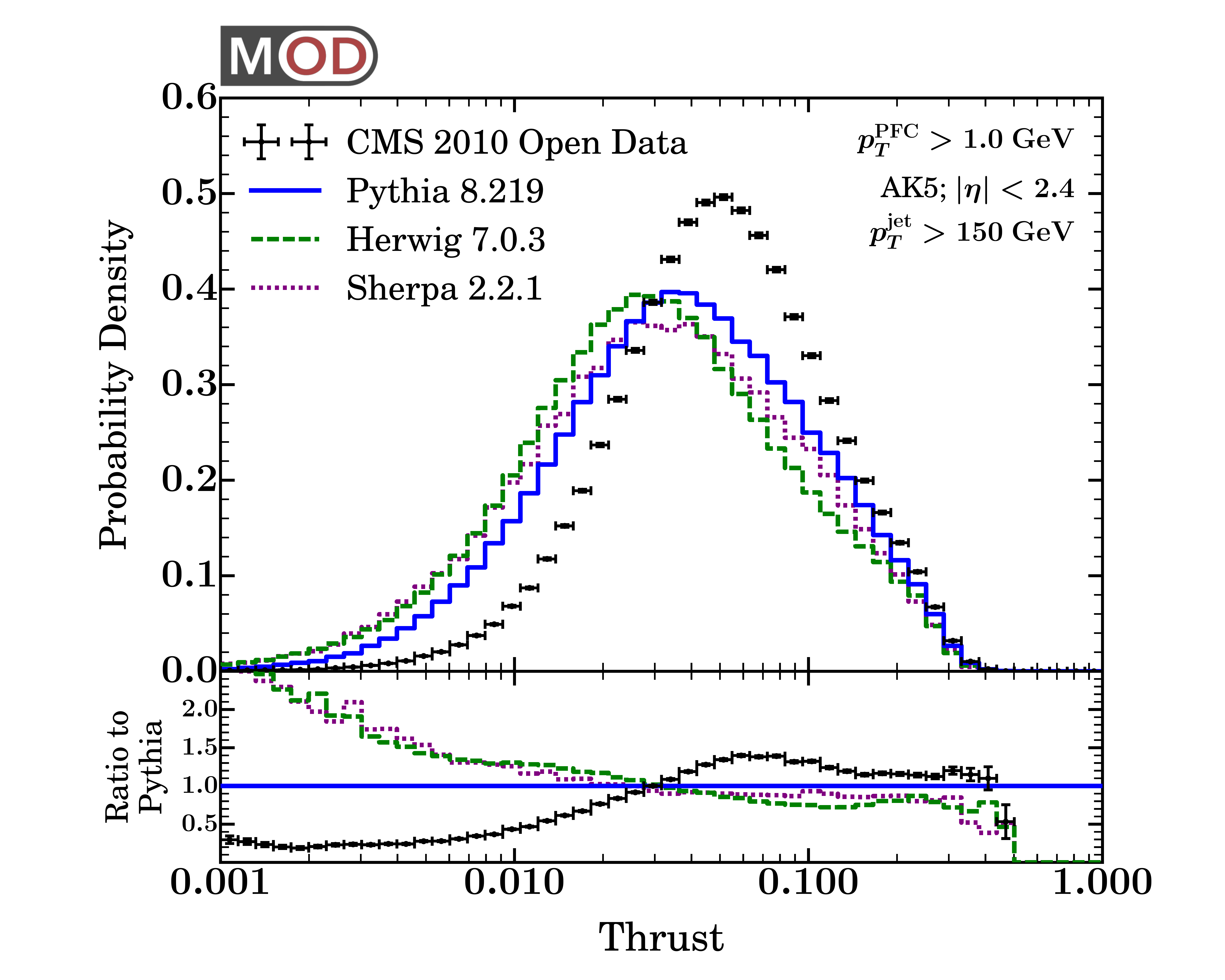}
}
\subfloat[]{
\label{fig:jet_thrust_track}
\includegraphics[width=0.9\columnwidth, page=1]{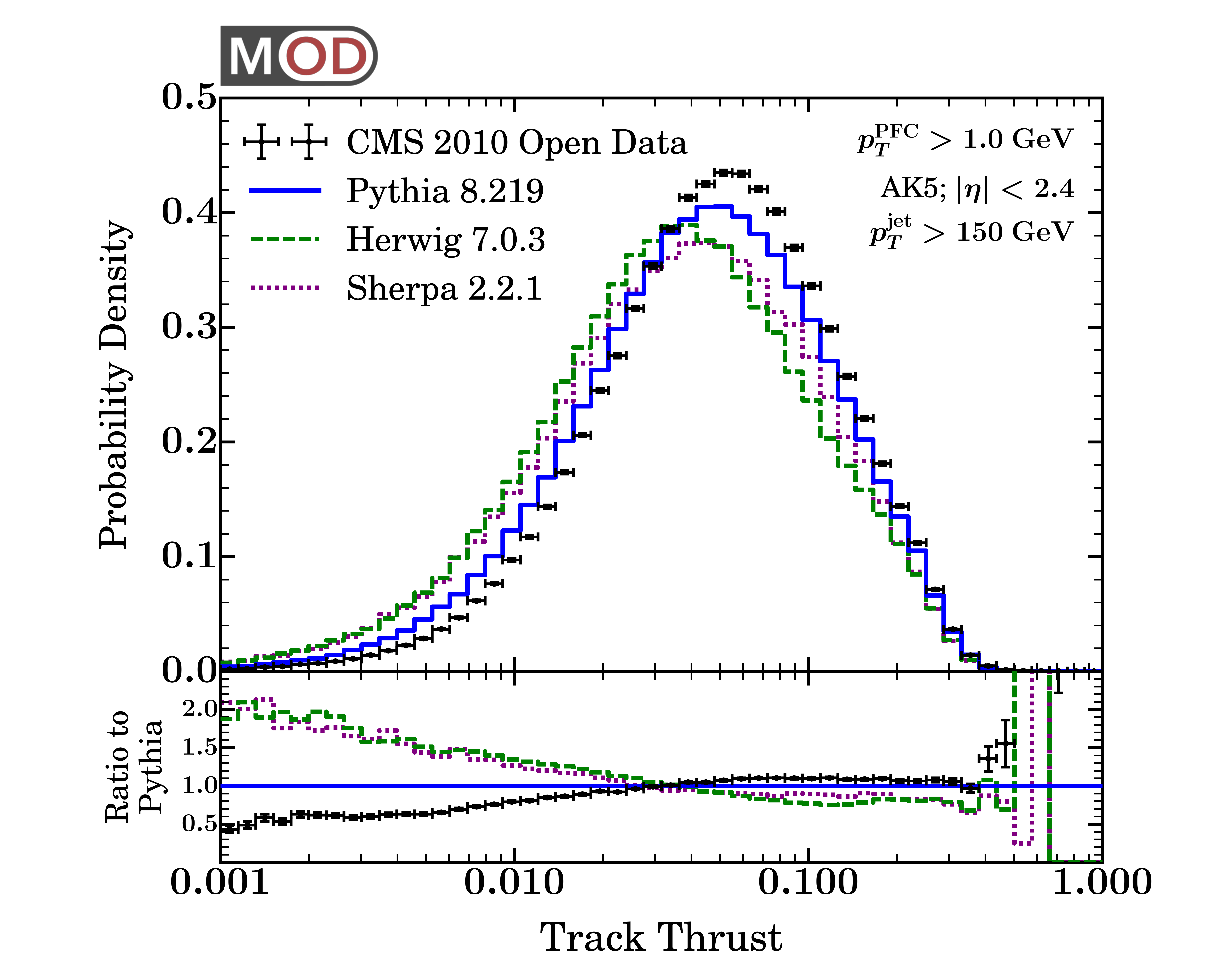}
}
\caption{Same as \Fig{fig:basic_substructure} but for the IRC-safe recoil-free jet angularities:  (top row) LHA with $\alpha = 1/2$, (middle row) jet width with $\alpha = 1$, and (bottom row) jet thrust with $\alpha = 2$.  Once again we compare (left column) all particle distributions to (right column) track-only variants.  Note the logarithmic scale of the distributions.}
\label{fig:jet_angularities}
\end{figure*}

For IRC-safe observables, we expect the impact of finite angular and energy resolution of the CMS detector to be less pronounced.  In \Fig{fig:jet_mass}, the jet mass distribution agrees rather well between CMS Open Data and the parton showers, with differences again comparable to the differences between generators.   Here, we have not applied the JEC factor to the mass distribution, since these are obtained after the  PFC cut of $p_T^{\rm min} = 1.0~\GeV$.  In \Fig{fig:jet_mass_track}, we show the track-based variant (which is not corrected for the charged energy fraction), which shows similar agreement between the open data and the parton showers.  While the lack of a detector model means that we cannot use the CMS Open Data to make quantitative statements about the jet mass distribution, we can say that the overall CMS detector performance is sufficient to draw qualitative conclusions about jet substructure distributions.

\subsection{Jet Angularities}
\label{subsec:angularities}

A powerful way to study the radiation pattern of quark and gluon jets is to use jet angularities \cite{Berger:2003iw,Almeida:2008yp,Ellis:2010rwa,Larkoski:2014uqa,Larkoski:2014pca}.  These are IRC-safe observables, defined as
\be
\label{eq:angularities}
e^{(\alpha)} = \sum_{i\in \text{jet}} z_i \theta_i^\alpha,
\ee
where
\be
\quad z_i = \frac{p_{Ti}}{\sum_{j \in \text{jet}} p_{Tj}}, \quad \theta_i = \frac{R_i}{R},
\ee
and $R_i$ is the rapidity/azimuth distance to a recoil-free axis.  Because the jet axis itself is sensitive to recoil \cite{Catani:1992jc,Dokshitzer:1998kz,Banfi:2004yd,Larkoski:2013eya,Larkoski:2014uqa}, we use the winner-take-all axis \cite{Larkoski:2014uqa,Bertolini:2013iqa,Salam:WTAUnpublished} defined from Cambridge/Aachen (C/A) clustering~\cite{Wobisch:1998wt,Dokshitzer:1997in}.  

By adjusting the value of $\alpha$ one can test radiation patterns mainly in the core ($\alpha < 1$) or periphery ($\alpha > 1$) of the jet.  Three commonly used benchmarks are the Les Houches Angularity (LHA, $\alpha = 1/2$) \cite{Badger:2016bpw,Gras:2017jty}, jet width ($\alpha = 1$) \cite{Catani:1992jc,Rakow:1981qn,Ellis:1986ig}, and jet thrust ($\alpha = 2$) \cite{Farhi:1977sg}.  The corresponding distributions are shown in \Fig{fig:jet_angularities}, plotted on a logarithmic scale to emphasize the behavior in the soft and collinear limit (i.e.~small values of the angularities).  Even though these are IRC-safe observables, we continue to place a cut of $p_T^{\rm min} = 1.0~\GeV$ on both the detector-level and particle-level constituents. 

At large values of the angularities, the agreement between the CMS Open Data and the parton showers is rather good.  At small values of the angularities where energy and angular resolution play an important role, the CMS Open Data is shifted to systematically higher values than the parton shower.  Since the shift is less pronounced for the track-based variants, we suspect that the finite angular resolution of neutral PFCs is driving the bulk of the disagreement.  For this reason, in the soft drop study presented next, we have to be mindful of the challenge of resolving small angular scales using neutral particles.  

\section{Two-Prong Jet Substructure}
\label{sec:2prong}

We now test the 2-prong substructure of the hardest jet using soft drop declustering \cite{Larkoski:2014wba}.  This method has been used in both ATLAS \cite{Aad:2015rpa} and CMS \cite{CMS:2016dmr,CMS:2016ldu,CMS:2016ude,CMS:2016flr, CMS:2016hxa,CMS:2016knm,CMS:2016pwo,
CMS:2016mjh,CMS:2016jog,CMS:2016jys,CMS:2016usi, CMS:2016rtp, CMS:2016pnc,CMS:2016bja, CMS:2016ehh, CMS:2016ccy, CMS:2016ete, CMS:2016kkf, CMS:2016zte, CMS:2016tvk, CMS:2014joa, CMS:2014ata} jet studies, including a recent CMS heavy ion result \cite{CMS:2016jys}.  There are also proposals to use soft drop to study the deadcone effect in top quarks \cite{Maltoni:2016ays} and gluon splitting to heavy flavor \cite{Ilten:2017rbd}.  Here, we exploit the fact that  soft drop is amenable to first-principle QCD calculations \cite{Larkoski:2014wba,Larkoski:2015lea,Frye:2016okc,Frye:2016aiz,Marzani:2017mva}.  While there are a variety of different 2-prong observables one could test on the CMS Open Data (e.g.~$N$-subjettiness \cite{Thaler:2010tr,Thaler:2011gf}, energy correlation functions \cite{Larkoski:2013eya,Larkoski:2014gra}, and Qjet volatility \cite{Ellis:2012sn,Ellis:2014eya}), soft drop has the advantage that it removes soft contamination from a jet, making it relatively robust to potential pileup and detector effects associated with soft particles.

As in the basic substructure analysis in \Secs{subsec:basic_substructure}{subsec:angularities}, we impose a restriction of $p_T^{\rm min} = 1.0~\GeV$ on all PFCs before passing them to the soft drop algorithm.  We again perform cross checks with track-based variants which use only charged PFCs, which are expected to better resolve the small angular scales probed by soft drop.

\subsection{Soft Drop Declustering}

\begin{figure}
\includegraphics[width=\columnwidth]{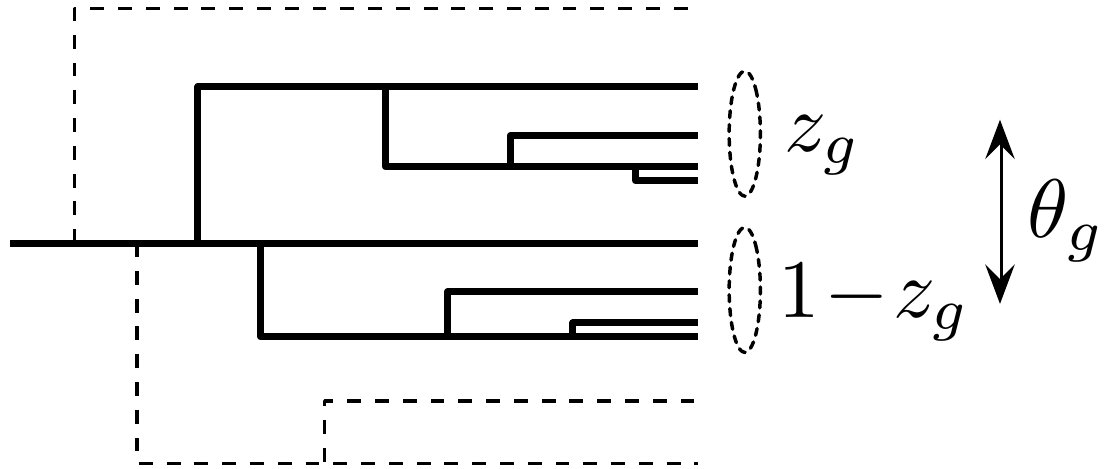}
\caption{Schematic of the soft drop algorithm, which recursively removes branches from the C/A clustering tree if the momentum fraction $z$ fails to satisfy $z > z_\text{cut} \theta^\beta$.  The $g$ subscript indicates the final groomed kinematics.}
\label{fig:soft_drop_diagram}
\end{figure}

The soft drop algorithm reclusters the constituents of a jet using the C/A algorithm \cite{Wobisch:1998wt,Dokshitzer:1997in} to create an angular-ordered clustering tree.   As shown in \Fig{fig:soft_drop_diagram}, soft drop then declusters the jet starting from the top of the tree, removing the softer $p_T$ branch until a $1 \to 2$ branching is found that satisfies
\be
z > z_{\rm cut} \theta^\beta.
\ee
Here, $z_{\rm cut}$ is an energy fraction cut, $\beta$ is an adjustable angular exponent, and the $1 \to  2$ kinematics are defined by 
\be
z = \frac{\min[p_{T1}, p_{T2}]}{p_{T1} + p_{T2}}, \qquad \theta \equiv \frac{R_{12}}{R}.
\ee
For the branching that passes the soft drop condition, we denote the resulting kinematic observables by $z_g$ and $\theta_g$, which characterize the hard 2-prong substructure of the jet.  The $g$ subscript is a reminder that these are groomed observables, subject to the soft drop condition.

In effect, soft drop simultaneously performs three tasks.  First, it removes wide-angle soft contamination from jets, which helps mitigate the effect of jet contamination from ISR, UE, and pileup.  Second, it dynamically changes the effective jet radius to match the size of the hard jet core.  Third, it provides the 2-prong kinematic observables $z_g$ and $\theta_g$, which can be used to perform foundational tests of QCD \cite{Larkoski:2015lea,Larkoski:2017bvj,CMS:2016jys,StarTalk,AlicePoster,Ilten:2017rbd} as well discriminate boosted $W$, $Z$, and Higgs bosons from ordinary quark/gluon jets \cite{Butterworth:2008iy,Aad:2015owa}. In general, groomers like soft drop have an interesting interplay with discrimination variables  \cite{Dasgupta:2016ktv,Moult:2016cvt,Salam:2016yht}.

In our study, we focus on the soft drop parameters
\be
z_{\rm cut} = 0.1, \qquad \beta = 0.
\ee
For this choice of $\beta$, the soft drop condition reduces to $z > z_{\rm cut}$ and becomes independent of angular information.  This then matches the behavior of the mMDT with $\mu = 1$  \cite{Dasgupta:2013ihk,Dasgupta:2013via}.  Without any explicit cut on $\theta_g$, this enables us to probe rather small angular scales within the jet, though we need to be cognizant of the finite angular resolution of the CMS detector.  We show distributions for five observables derived from soft drop:
\be
\label{eq:bigfive}
z_g, \quad \theta_g, \quad e_g^{(1/2)}, \quad e_g^{(1)}, \quad e_g^{(2)},
\ee
where
\be
e_g^{(\alpha)}= z_g \theta_g^\alpha
\ee
is a single-emission groomed variant of the angularities introduced in \Eq{eq:angularities}.

\subsection{MLL Analytic Predictions}
\label{subsec:MLL}

In addition to parton shower predictions, we compare the CMS Open Data to first-principles QCD theory distributions made using the techniques of \Refs{Larkoski:2014wba,Larkoski:2013paa,Larkoski:2015lea}, working to modified leading-logarithmic (MLL) accuracy.

For the observable $\theta_g$, it is convenient to express the probability distribution as
\be
\label{eq:thetagprob}
\frac{1}{\sigma} \frac{\df \sigma}{\df \theta_g} \equiv p(\theta_g) = \frac{\df}{\df \theta_g} \Sigma(\theta_g),
\ee
where the cumulative probability distribution $\Sigma(\theta_g)$ was calculated to MLL accuracy in \Ref{Larkoski:2014wba}.  For $\beta=0$, the result for a parton of flavor $i$ is
\be
\label{eq:MLLthetag}
\begin{split}
&\Sigma_i^{\rm MLL}(\theta_g; \mu_\theta) \\
&\quad = \exp \left[-\frac{2  C_i}{\pi} \int_{\theta_g}^1\frac{\df \theta}{\theta}\int_{\zcut}^{1/2} \!  \df z  \,\bar{\alpha}_s(z \, \theta \, \mu_\theta) \, \overline{P}_i(z) \right],
\end{split}
\ee
where $C_i$ is the Casimir factor ($C_F = 4/3$ for quarks and $C_A = 3$ for gluons).  At lowest non-trivial order, the QCD splitting functions are 
\begin{align}
P_q(z)&=\frac{1+(1-z)^2}{2z},  \label{splittings_q} \\
P_g(z)&= \frac{1-z}{z}+\frac{z(1-z)}{2} +\frac{n_f T_R}{2C_A}\left[z^2+(1-z)^2 \right], \label{splittings_g}
\end{align}
with $n_f = 5$ and $T_R = 1/2$; these appear in \Eq{eq:MLLthetag} in a symmetrized form,
\be
\label{eq:splittingcombo}
\overline{P}_i(z)=P_i(z)+P_i(1-z).
\ee
The one-loop QCD running coupling is $\bar{\alpha}_s$, where the bar indicates that we have frozen the running below the IR scale $\mu_\text{NP}\sim1.0$~GeV,
\be
\label{asfreezing}
\bar{\alpha}_s(\mu) = \as(\mu)\Theta \left(\mu-\mu_\text{NP}\right)+\as(\mu_\text{NP})\Theta \left(\mu_\text{NP}-\mu\right).
\ee
The running coupling is evaluated at the canonical renormalization group scale
\be
\mu_\theta = p_T R,
\ee
and we estimate uncertainties by varying both this scale and $\mu_\text{NP}$ up and down by a factor of two.  

To get a physical distribution for $\theta_g$, we need to determine the relative fraction of quark and gluon jets with our selection, such that the final cumulative distribution is
\be
\Sigma^{\rm MLL} = f_q \Sigma_q^{\rm MLL} + f_g \Sigma_g^{\rm MLL}.
\ee
To determine the fractions $f_q$ and $f_g$, we generate a leading-order sample of dijets using \textsc{MadGraph5\_aMC@NLO} 2.4.0 \cite{Alwall:2014hca} with parton distribution functions (PDFs) given by NNPDF2.3 LO \cite{Ball:2012cx}, extracting the average flavor composition from both jets.  We set the renormalization and factorization scales to the total transverse momentum of the dijet event,
\be
\mu_h = p_{T1} + p_{T2},
\ee
and vary this up and down by a factor of 2 to estimate uncertainties.  Note that the renormalization scale does not affect the relative quark and gluon composition since it only rescales the total cross section by changing $\alpha_s$.  By contrast, the factorization scale does affect the flavor composition through the PDFs.  

Strictly speaking, the above method for determining the quark/gluon fraction of the hardest jet is not IRC safe, since the flavor composition of the hardest jet at NLO is no longer the same as the average flavor composition at LO.  In practice, though, the hardest jet at NLO is more or less randomly determined from the two degenerate jets at LO, so the strategy used in this paper is sufficient for the current level of theoretical accuracy.  There are various ways we could improve this procedure in a future analysis.  Arguably the easiest method would be to study the inclusive jet spectrum instead of focusing on just the hardest jet in the event.  While conceptually straightforward, it is technically more involved, since for dijet events close to a trigger boundary, the same event can have different assigned triggers for the two different jets.  If we only wanted to study a single jet per event, we could use a dijet trigger for event selection but then only analyze the more central of the two jets, since that is a well-defined selection at LO.  

To predict the probability distributions for $z_g$ and $e_g^{(\alpha)}$, we use the strategy of \Ref{Larkoski:2015lea}.  Since $z_g = e_g^{(0)}$ (i.e.~$\alpha = 0$), we can use the same method to calculate the remaining four observables in \Eq{eq:bigfive}.  We express the full probability distribution for $e_g^{(\alpha)}$ and $\theta_g$,
\be
p(e_g^{(\alpha)}, \theta_g) \equiv \frac{1}{\sigma} \frac{\df^2 \sigma}{\df e_g^{(\alpha)} \, \df \theta_g},
\ee
in terms of the probability for $\theta_g$ from \Eq{eq:thetagprob} multiplied by the conditional probability for $e_g^{(\alpha)}$ given $\theta_g$,
\be
p(e_g^{(\alpha)}, \theta_g) = p(\theta_g) \, p(e_g^{(\alpha)} | \theta_g).
\ee
To obtain the probability for $e_g^{(\alpha)}$ alone, we simply integrate over all values of $\theta_g$,
\be
p(e_g^{(\alpha)}) = \int \df \theta_g \,  p(\theta_g) \, p(e_g^{(\alpha)} | \theta_g).
\ee
To leading fixed order in the collinear limit, the conditional probability distribution is 
\begin{align}\label{cond-fo}
p^\text{LO-c}(e_g^{(\alpha)} | \theta_g; \mu_z) &= \frac{\bar{\alpha}_s(e_g^{(\alpha)} \theta_g^{1-\alpha} \mu_z)\, \theta_g^{-\alpha} \, \overline{P}_i\left(e_g^{(\alpha)}  \theta_g^{-\alpha}\right)}{ \int_{\zcut}^{1/2} \df z \, \bar{\alpha}_s(z \theta_g \mu_z) \overline{P}_i(z) }\nonumber\\
&
\hspace{-1cm}
\times \Theta\left( \theta_g^{\alpha}-2 e_g^{(\alpha)}  \right)\,  \Theta\left( e_g^{(\alpha)}- \zcut \theta_g^{\alpha}  \right).
\end{align}
We note the dependence on a (in principle different) renormalization group scale,
\be
\mu_z = p_T R,
\ee
which can be varied up and down by a factor of two.  

In summary, these theory distributions depend on four different scales
\be
\mu_{\rm NP}, \quad \mu_\theta, \quad \mu_h, \quad \mu_z,
\ee
which can be varied to estimate theoretical uncertainties. As established, these variations do yield properly normalized distributions.  To estimate perturbative uncertainties, we take the envelope of all scale variations, noting that the envelope will not, in general, be normalized.

There are two known effects which are not included in our theoretical uncertainty estimates.  The first is genuine nonperturbative corrections.  The above distributions are calculated perturbatively, with only the frozen coupling in \Eq{asfreezing} acknowledging the impact of nonperturbative physics.  When $z_g$ or $\theta_g$ are dominated by nonperturbative dynamics, though, these perturbative distributions can no longer be trusted.  For double-differential distributions, this occurs when 
\be
z_g \theta_g \lesssim \frac{\Lambda}{p_T R},
\ee
where $\Lambda\sim \mathcal{O}(\text{GeV})$ and $p_T$ is the lowest value in the plotted range.  Projecting to the single observables, nonperturbative dynamics becomes relevant when:
\be
\theta_g\lesssim \frac{\Lambda}{\zcut p_T R}, \quad
e_g^{(\alpha)} \lesssim \max\{1, z_{\rm cut}^{1-\alpha}\} \left(\frac{\Lambda}{p_T R}\right)^{\alpha}.
\ee
To indicate this in the plots below, we change the theory curves to a dashed style when nonperturbative modes dominate, using $\Lambda  = 2~\GeV$ for concreteness.  Note that $z_g$ itself ($\alpha = 0$) is a collinear unsafe observable, so strictly speaking it is always sensitive to nonperturbative dynamics.  Because $z_g$ is a Sudakov safe \cite{Larkoski:2013paa,Larkoski:2015lea} observable, though, the collinear singularity is regulated by the Sudakov form factor for $\theta_g$.  Also note that the theory calculations do not include the $p_T^{\min} = 1.0~\GeV$ cut, which can be considered as part of the nonperturbative uncertainty.

The second missing effect is matching to fixed-order matrix elements.  This is expected to have a small impact because the jet radius is reasonably small and we are mostly focused on the $e_g^{(\alpha)}\ll 1$ limit.  Nevertheless, there will be important fixed-order corrections to our theory predictions above the characteristic scale of $e_g^{(\alpha)} \simeq \zcut$, though we have not indicated that scale explicitly on the plots below.  Indeed, there is noticeable disagreement between our theory predictions and the open data/parton showers in the fixed-order regime, especially for $\theta_g \to 1$ (as illustrated in \Fig{fig:softdrop_rg}).   A detailed study of fixed-order corrections is beyond the scope of this paper, and would anyway require a proper IRC-safe definition of the measured jet.

\subsection{Open Data Results}

\begin{figure*}
\subfloat[]{
\label{fig:softdrop2D_data}
\includegraphics[width=0.9\columnwidth, page=1]{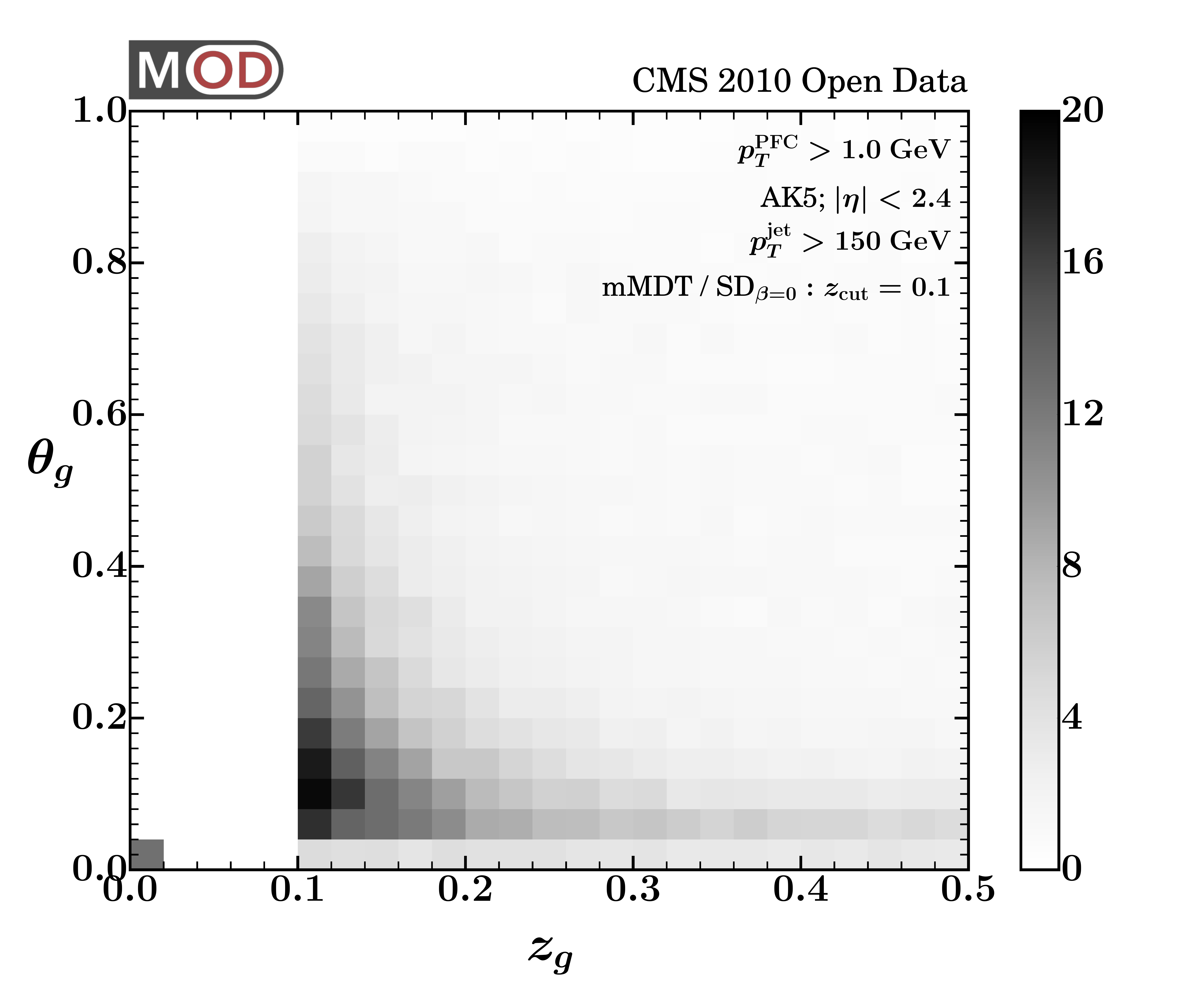}
}
\subfloat[]{
\label{fig:softdrop2D_analytic}
\includegraphics[width=0.9\columnwidth, page=1]{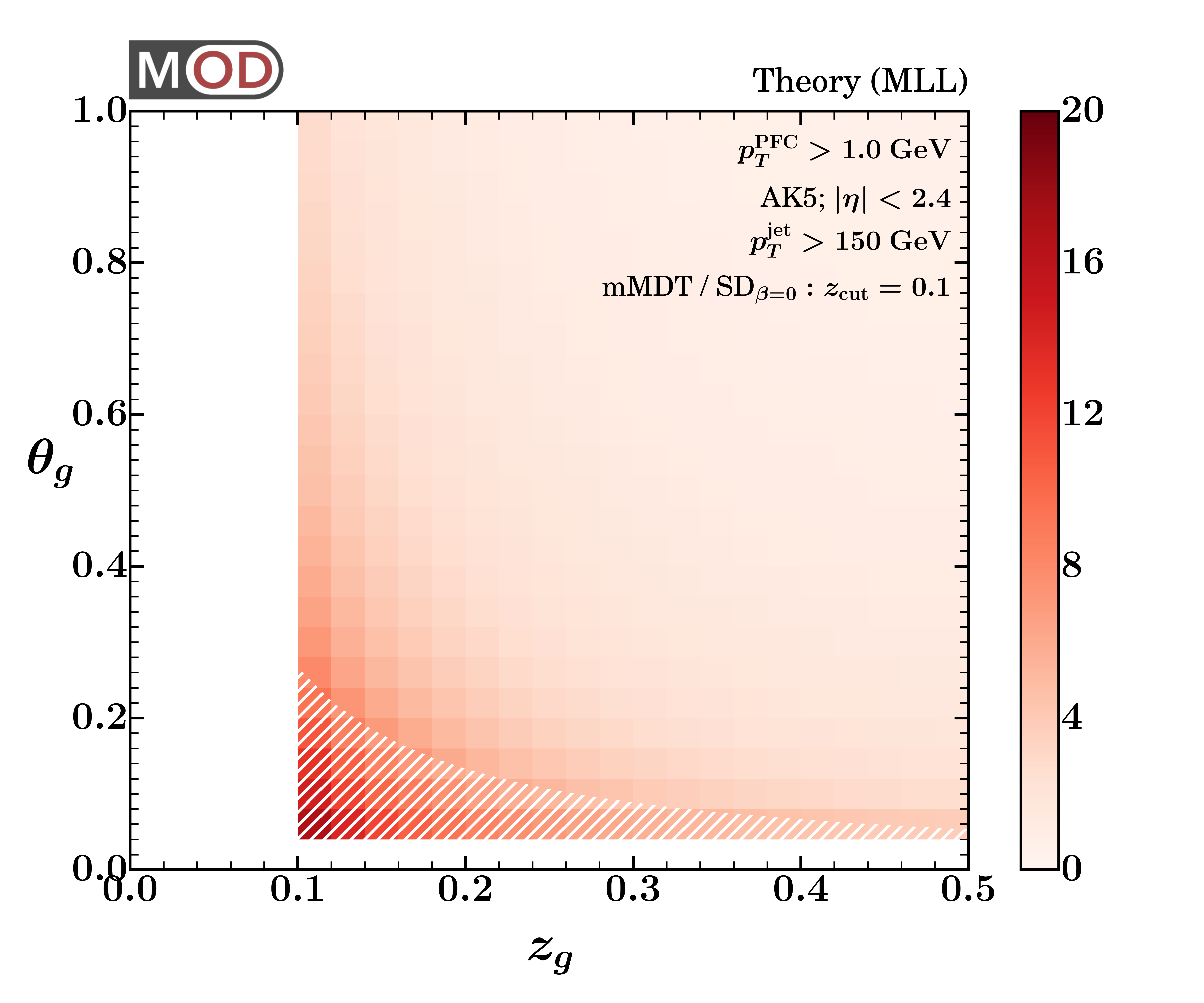}
}

\subfloat[]{
\label{fig:softdrop2D_pythia}
\includegraphics[width=0.9\columnwidth, page=1]{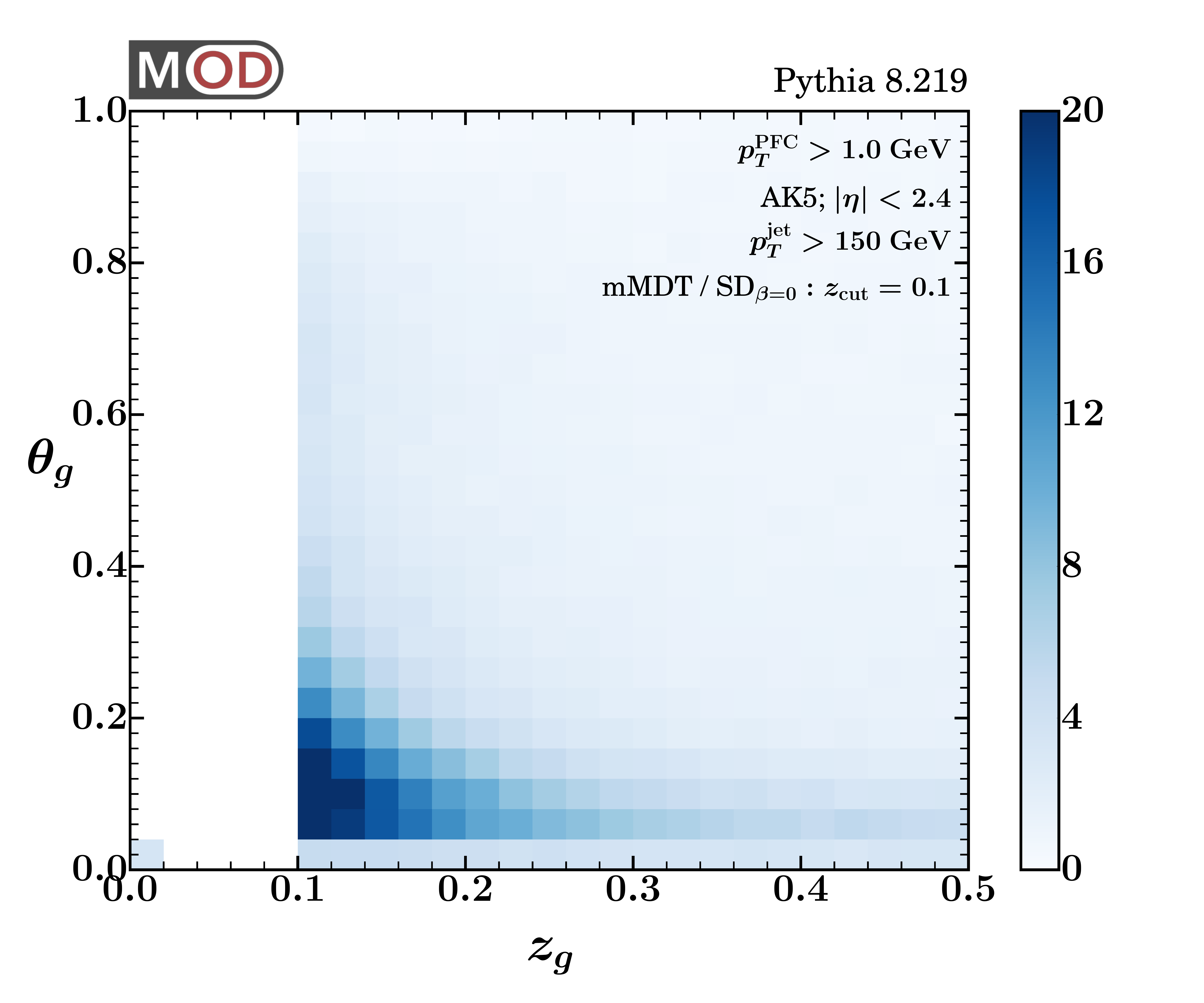}
}

\subfloat[]{
\label{fig:softdrop2D_herwig}
\includegraphics[width=0.9\columnwidth, page=1]{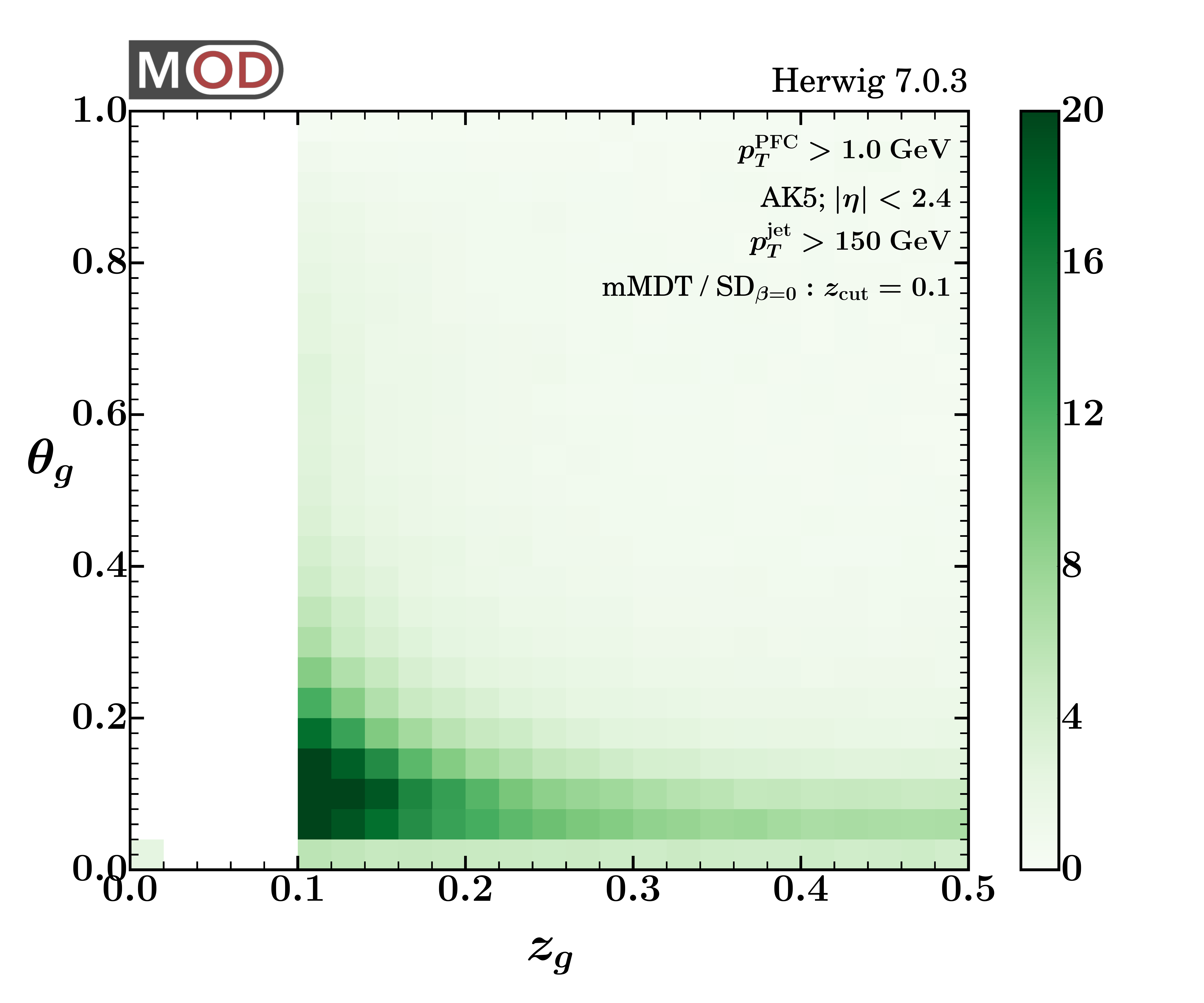}
}
\subfloat[]{
\label{fig:softdrop2D_sherpa}
\includegraphics[width=0.9\columnwidth, page=1]{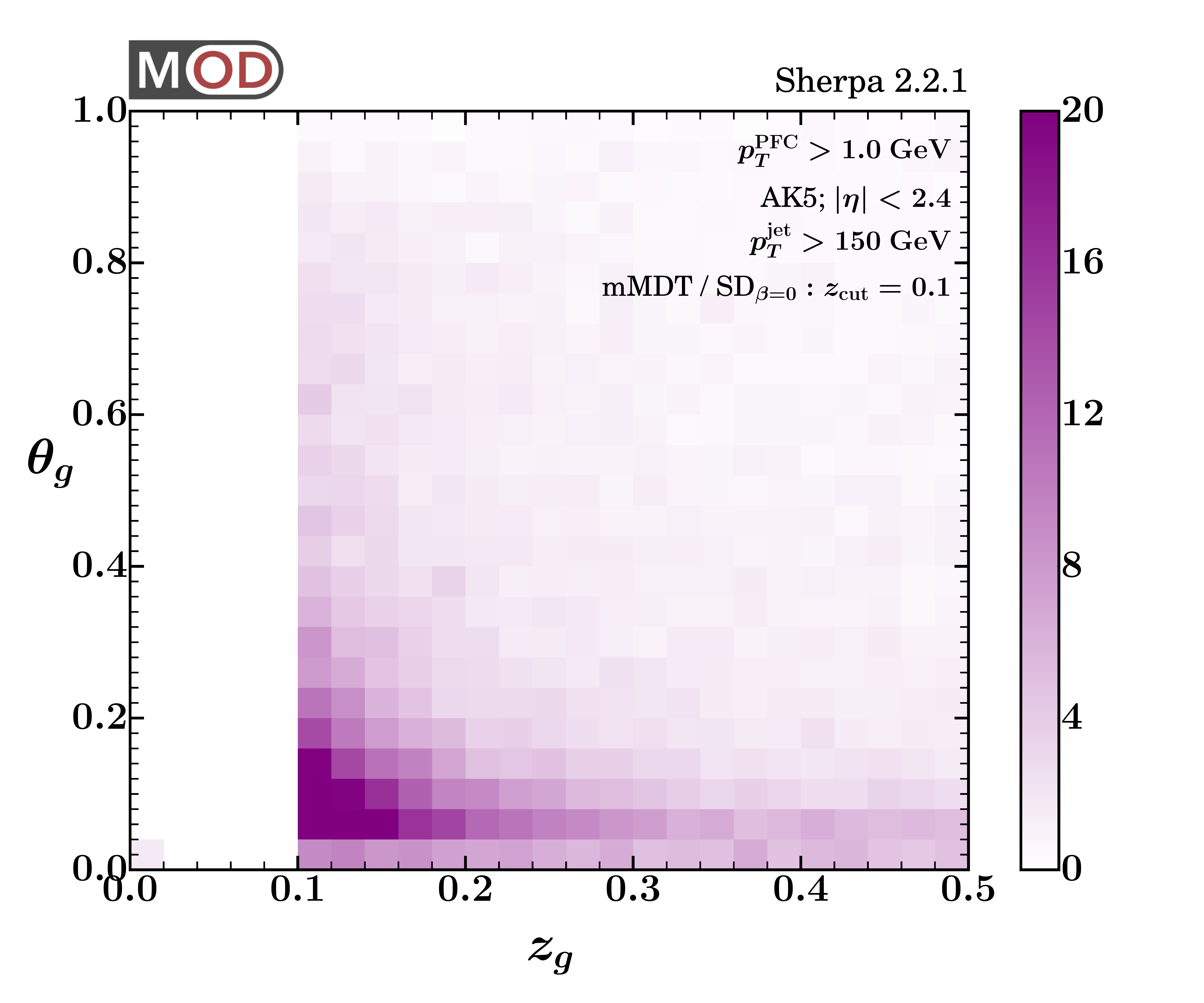}
}
\caption{Two dimensional distributions of $z_g$ versus $\theta_g$ from soft drop with $\beta = 0$ (i.e.\ mMDT with $\mu = 1$) in (a) CMS Open Data and (b) the MLL analytic prediction, compared to (c) \textsc{Pythia}, (d) \textsc{Herwig}, and (e) \textsc{Sherpa}.  Here, we are plotting the dimensionless probability density $p(z_g, \theta_g)$ whose integral is 1.  The hard vertical cut corresponds to $z_g = z_{\rm cut}$, and the $(0,0)$ entry corresponds to jets that fail the soft drop procedure (not present for the analytic calculation).  The white hashing in the MLL distribution corresponds to where nonperturbative physics dominates.}
\label{fig:softdrop2D}
\end{figure*}

\begin{figure*}
\subfloat[]{
\label{fig:softdrop2D_data_log}
\includegraphics[width=0.9\columnwidth, page=1]{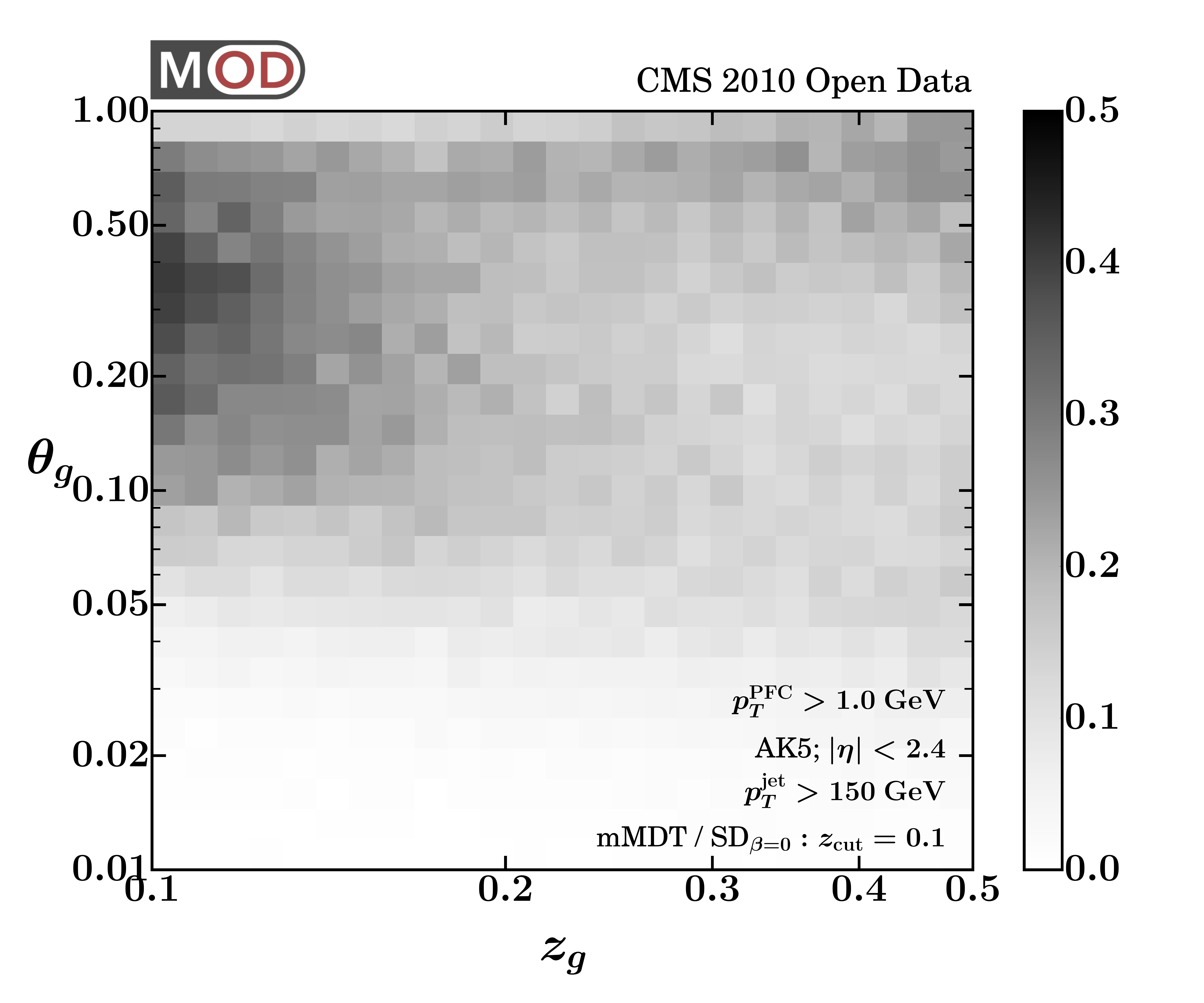}
}
\subfloat[]{
\label{fig:softdrop2D_analytic_log}
\includegraphics[width=0.9\columnwidth, page=1]{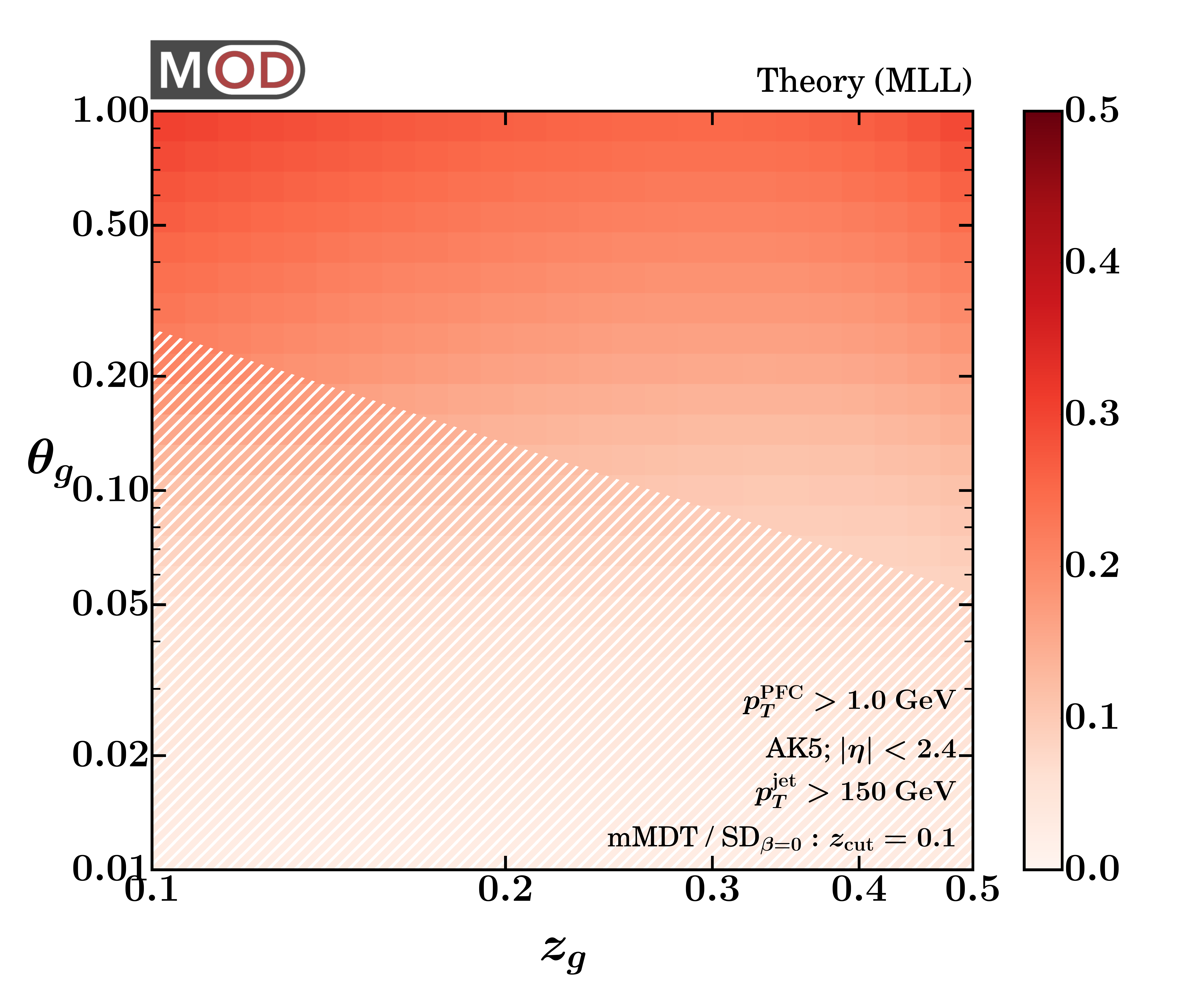}
}

\subfloat[]{
\label{fig:softdrop2D_pythia_log}
\includegraphics[width=0.9\columnwidth, page=1]{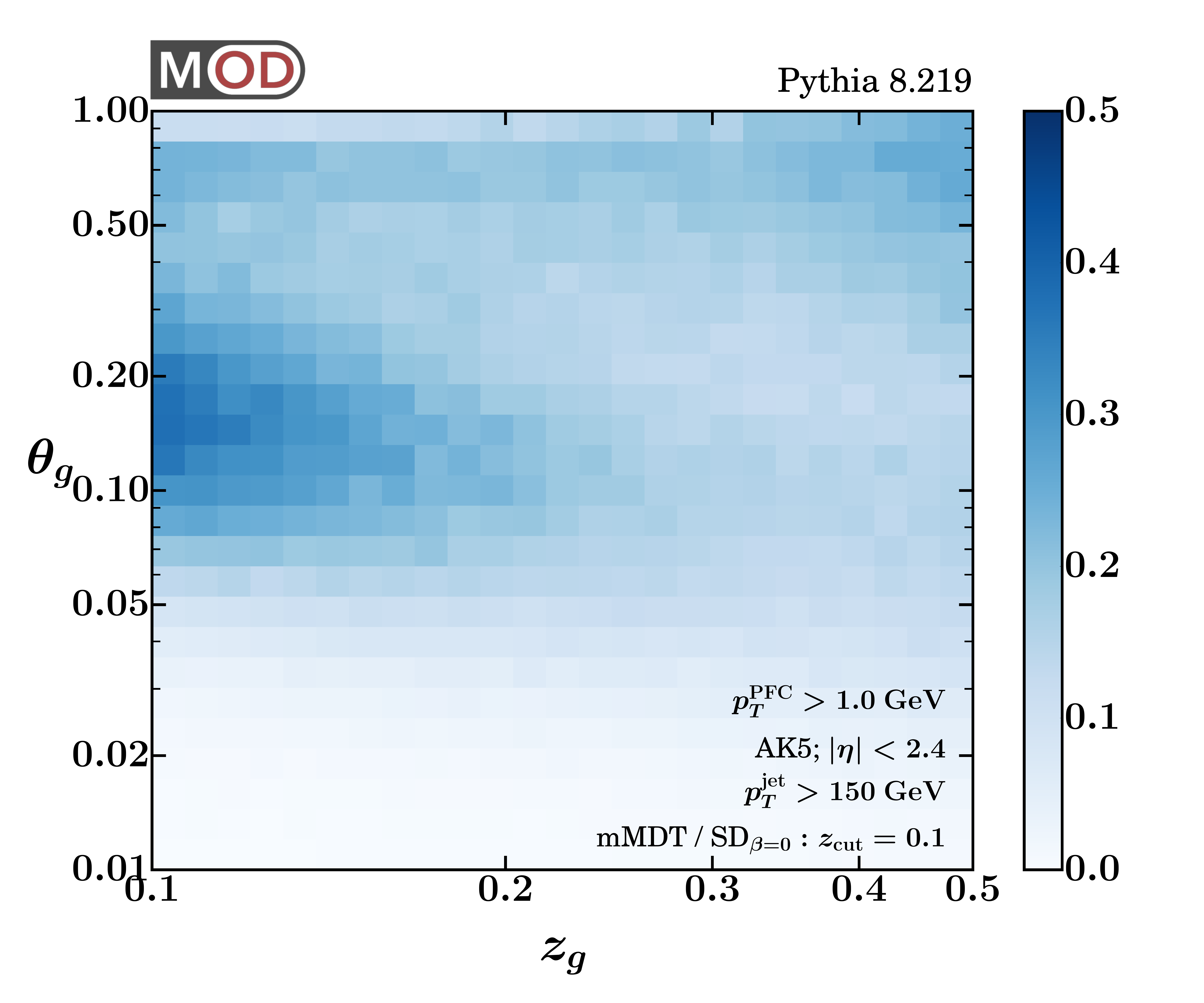}
}

\subfloat[]{
\label{fig:softdrop2D_herwig_log}
\includegraphics[width=0.9\columnwidth, page=1]{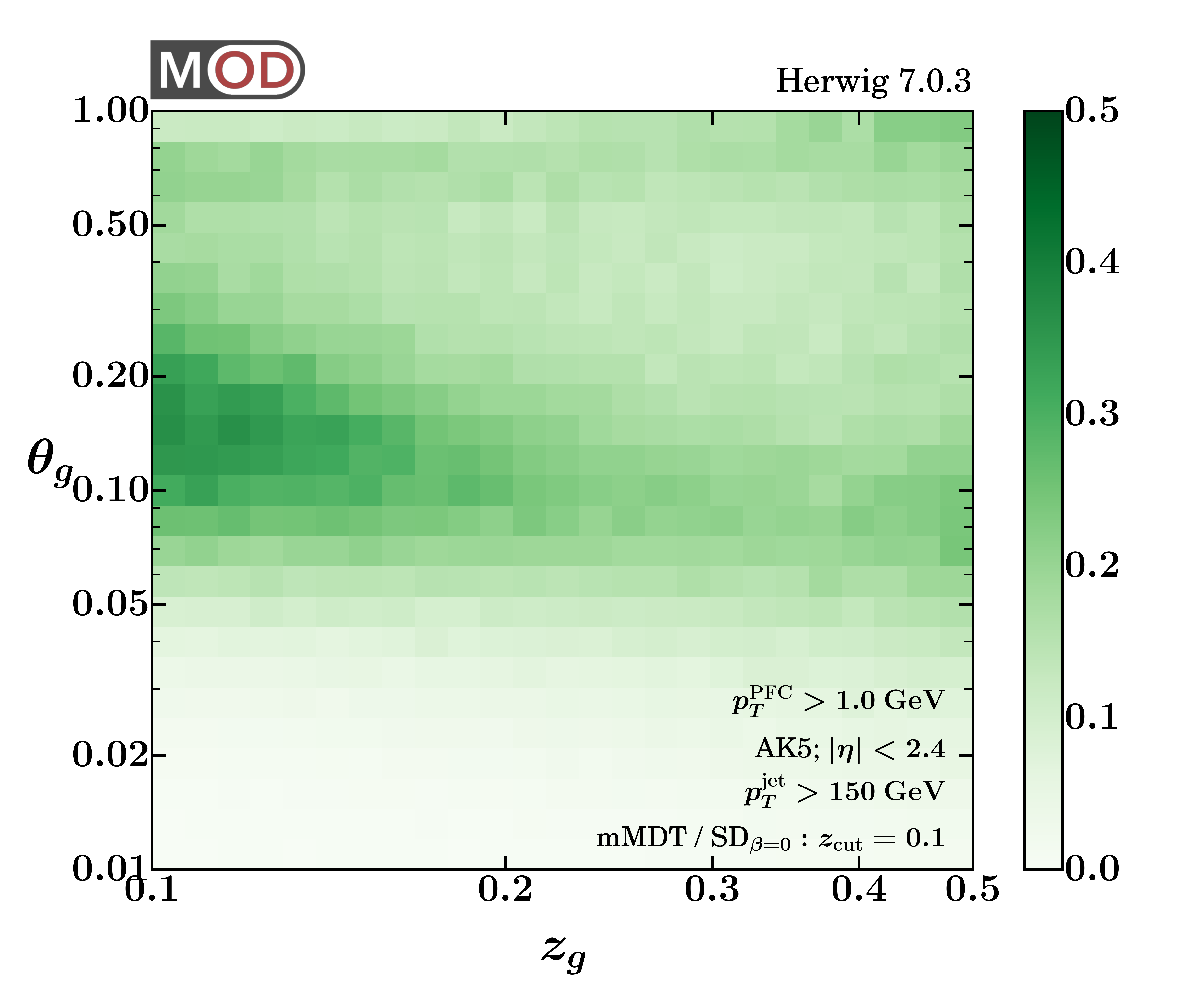}
}
\subfloat[]{
\label{fig:softdrop2D_sherpa_log}
\includegraphics[width=0.9\columnwidth, page=1]{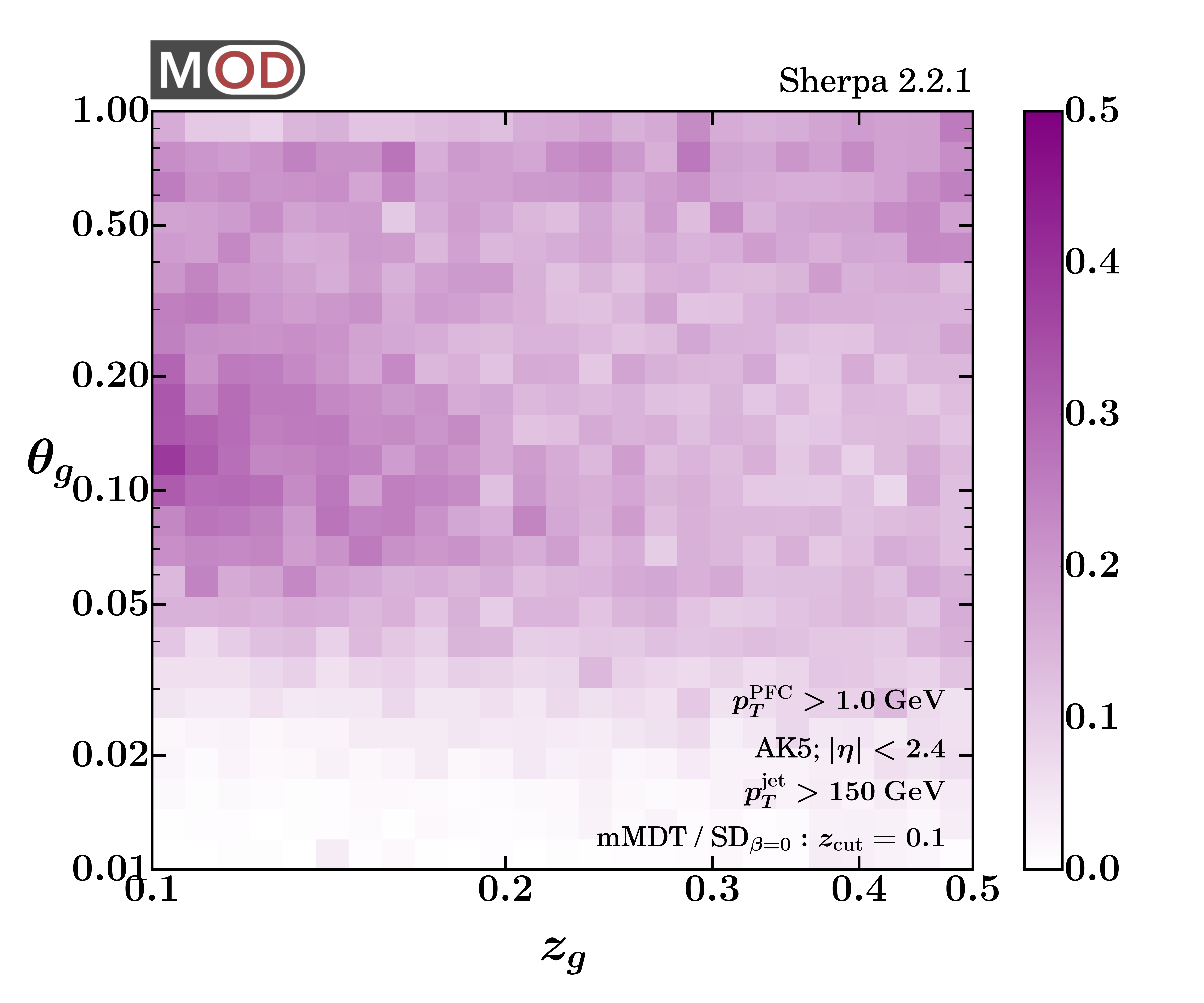}
}
\caption{Same as \Fig{fig:softdrop2D} but on a logarithmic scale to highlight the soft/collinear limit.  Here, we are plotting the dimensionless probability density $p(\log z_g, \log \theta_g) = z_g \, \theta_g \, p(z_g, \theta_g)$ whose integral is 1 in logarithmic variables.}  
\label{fig:softdrop2D_log}
\end{figure*}

\begin{figure*}
\subfloat[]{
\label{fig:softdrop_zg_linear}
\includegraphics[width=\columnwidth, page=1]{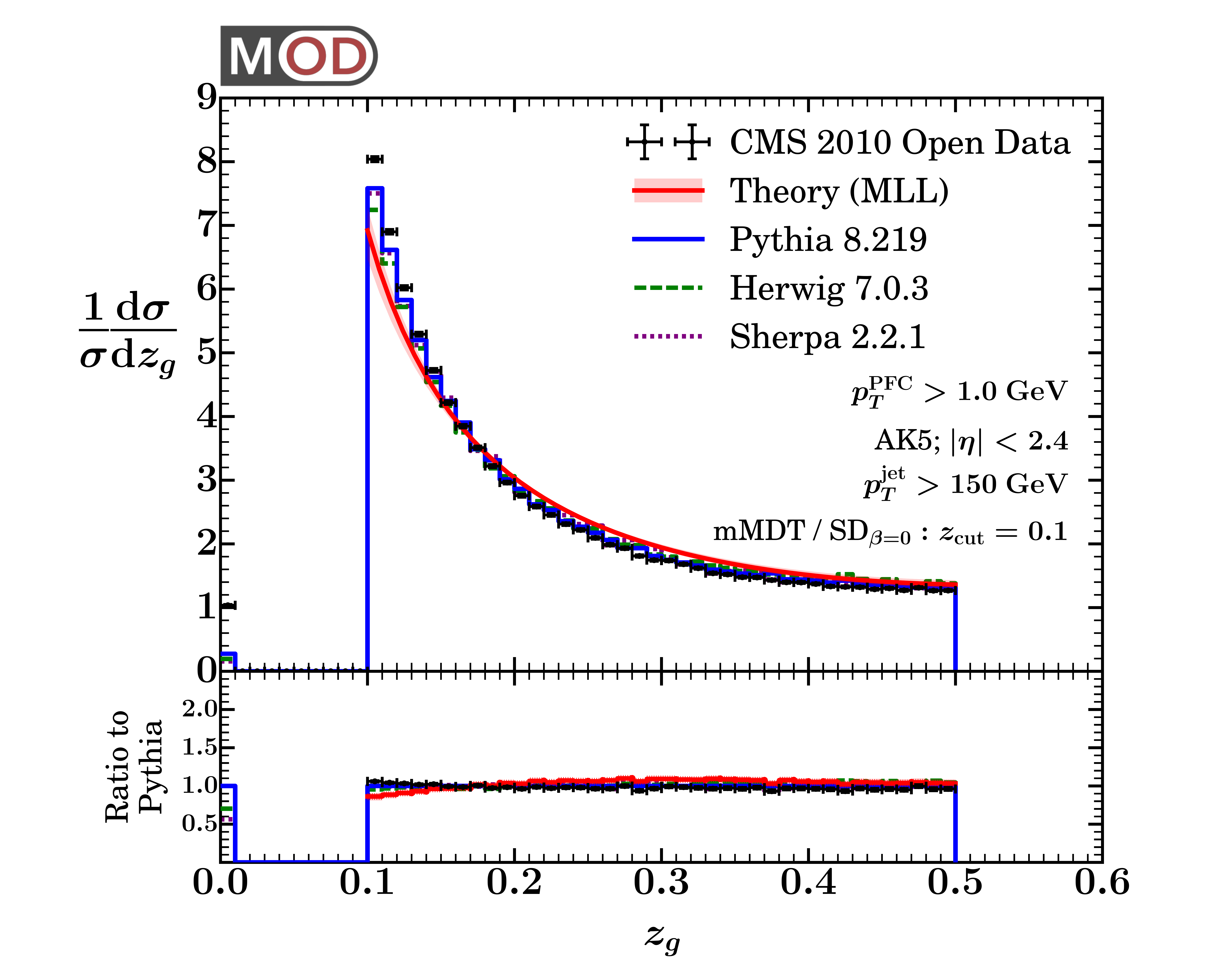}
}
\subfloat[]{
\label{fig:softdrop_zg_track_linear}
\includegraphics[width=\columnwidth, page=1]{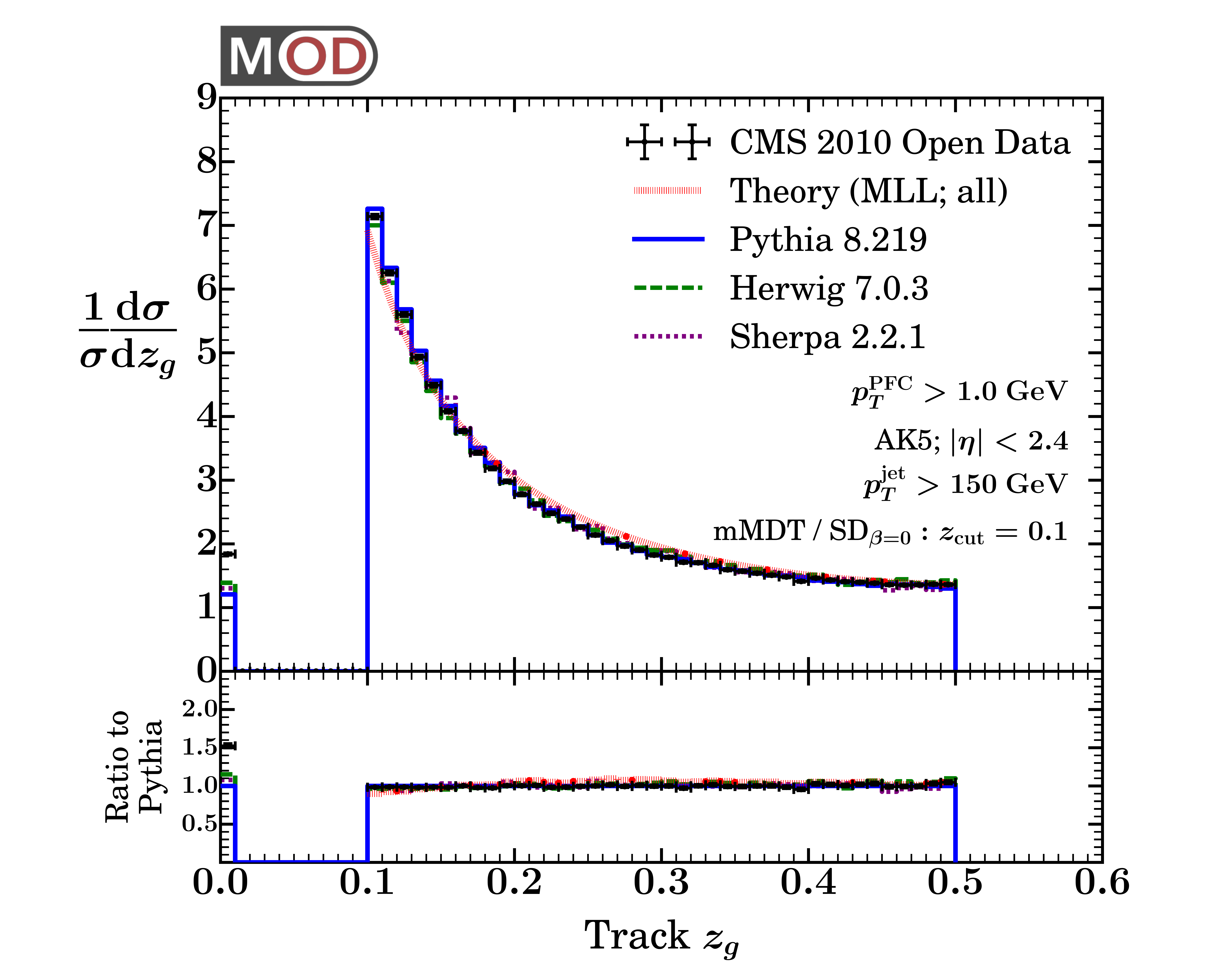}
}

\subfloat[]{
\label{fig:softdrop_zg_log}
\includegraphics[width=\columnwidth, page=1]{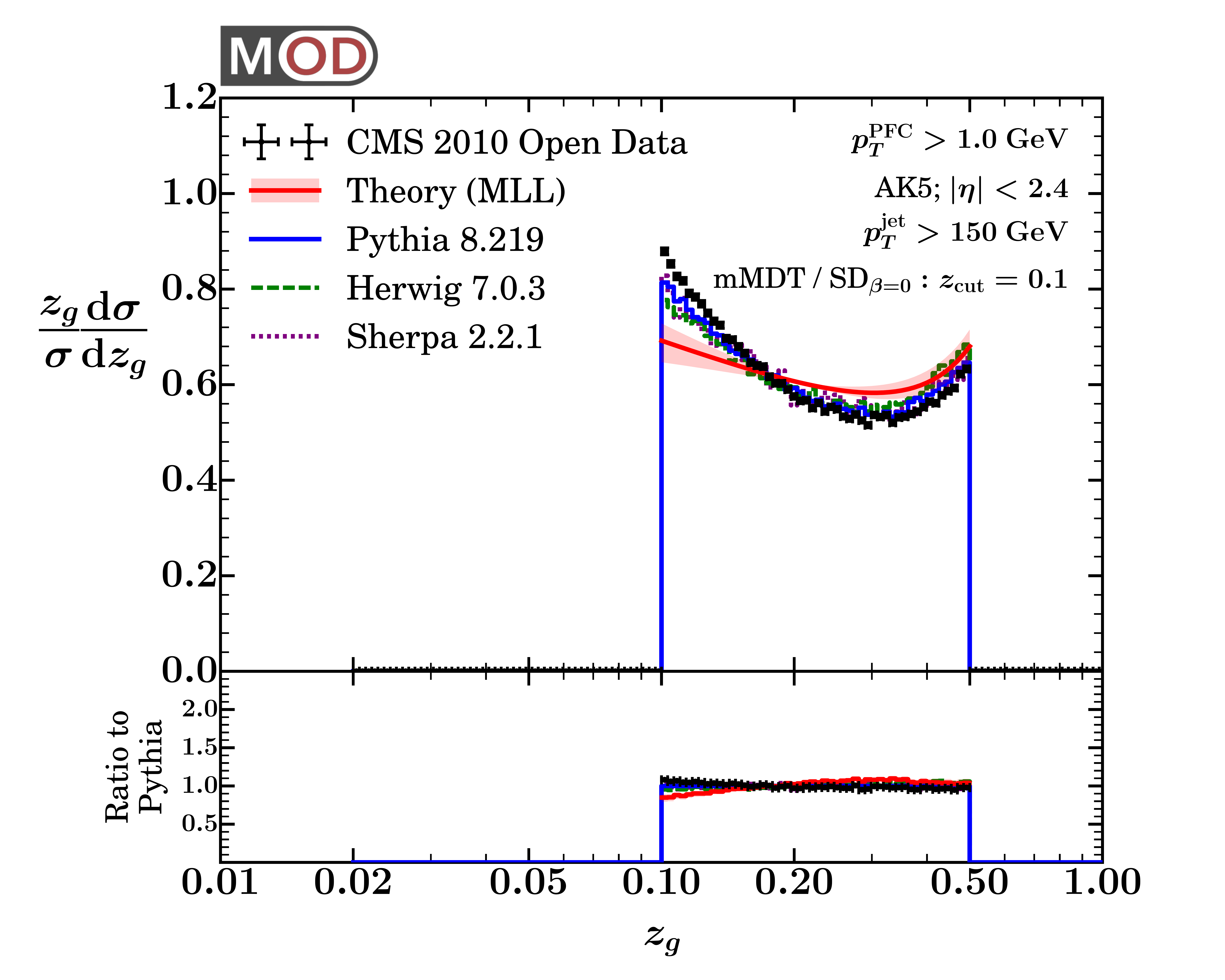}
}
\subfloat[]{
\label{fig:softdrop_zg_track_log}
\includegraphics[width=\columnwidth, page=1]{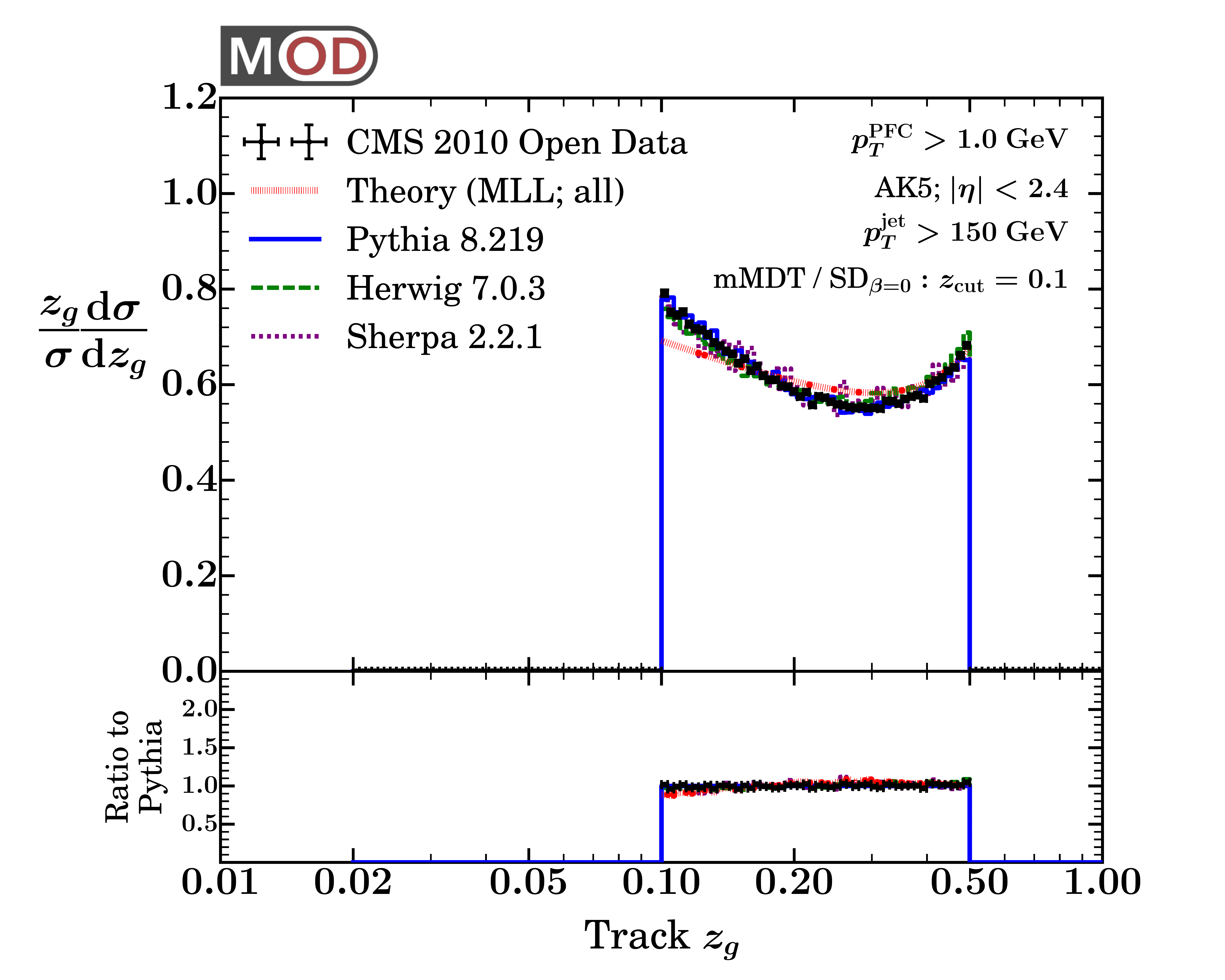}
}

\caption{Soft-dropped distributions for $z_g$ using (left column) all particles and (right column) only charged particles.  In this and subsequent plots, the MLL distributions are the same in both columns and do not account for the $p_T^{\rm min} = 1~\GeV$ cut on PFCs or the switch to charged particles (hence the dashed version on the right).  The top row shows the linear distributions while the bottom row shows the logarithmic distributions. }
\label{fig:softdrop_zg}
\end{figure*}

\begin{figure*}
\subfloat[]{
\label{fig:softdrop_rg_linear}
\includegraphics[width=\columnwidth, page=1]{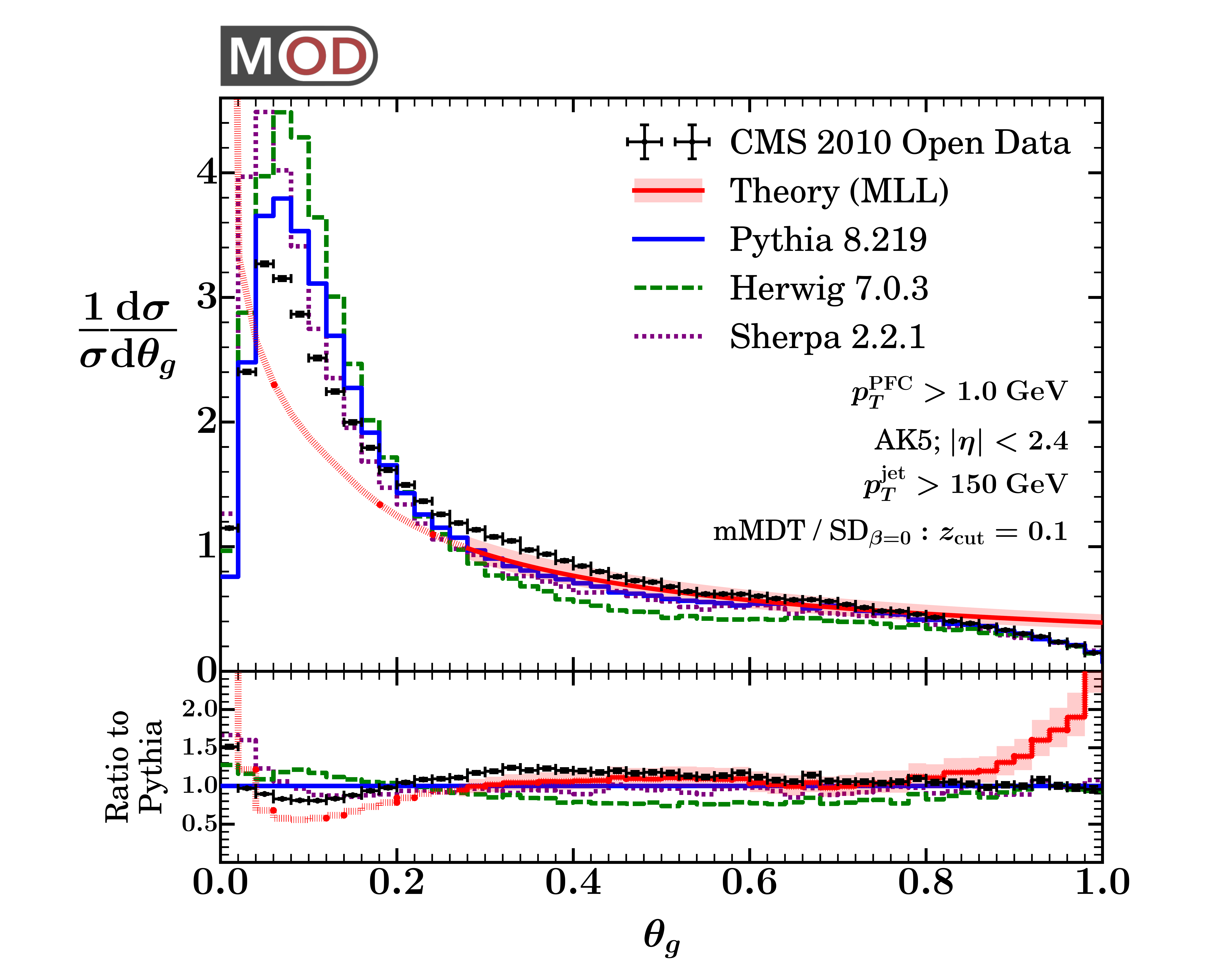}
}
\subfloat[]{
\label{fig:softdrop_rg_track_linear}
\includegraphics[width=\columnwidth, page=1]{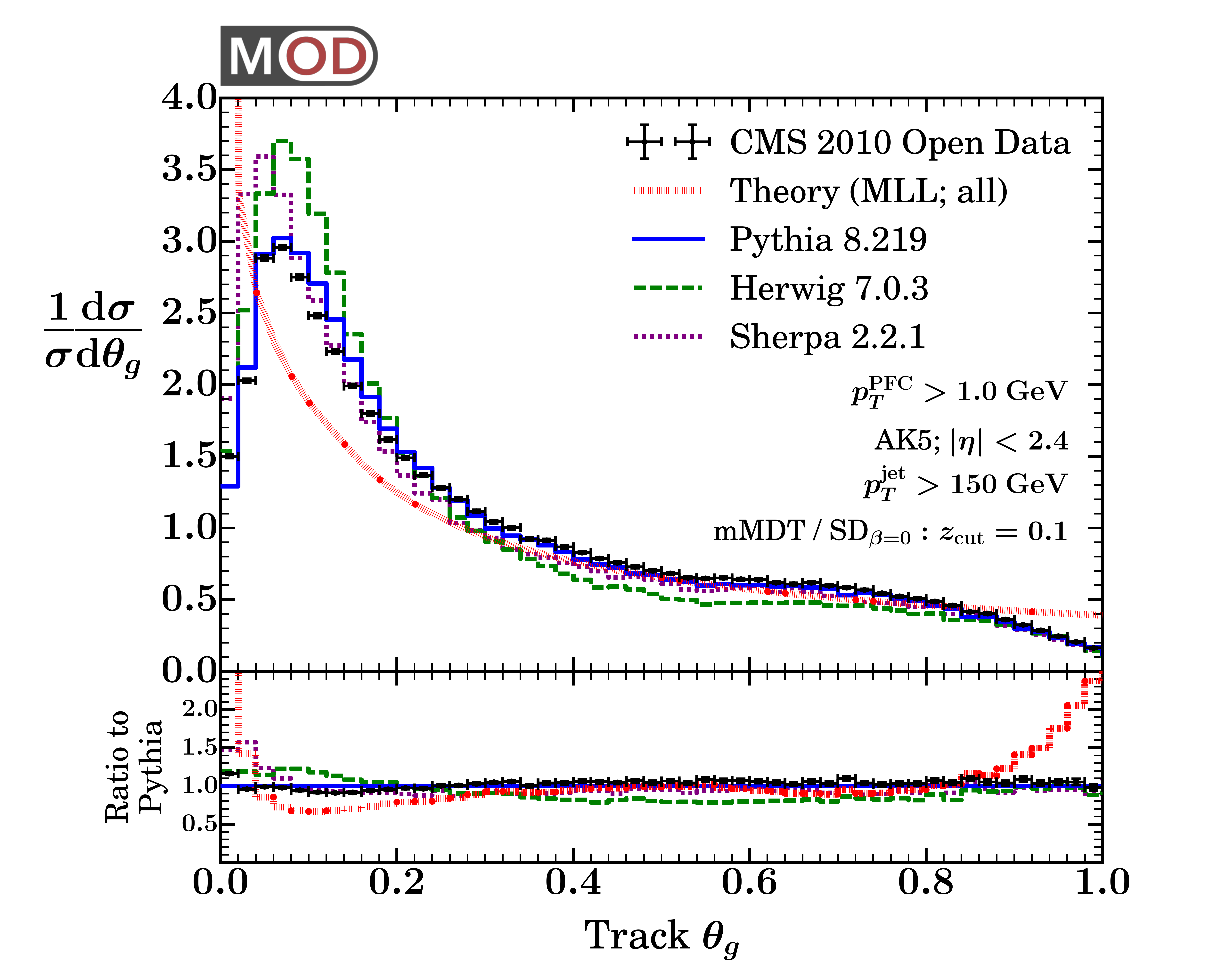}
}

\subfloat[]{
\label{fig:softdrop_rg_log}
\includegraphics[width=\columnwidth, page=1]{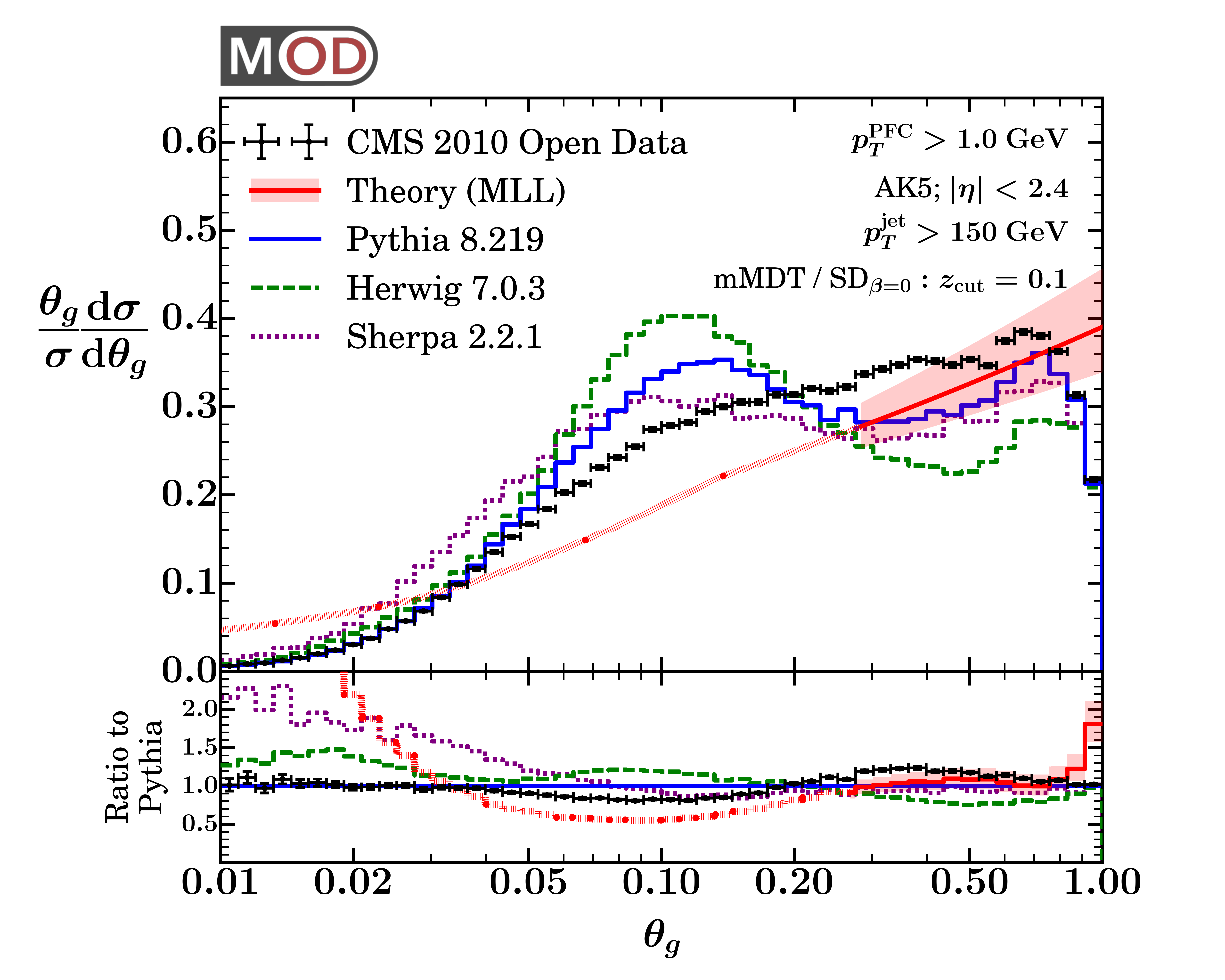}
}
\subfloat[]{
\label{fig:softdrop_rg_track_log}
\includegraphics[width=\columnwidth, page=1]{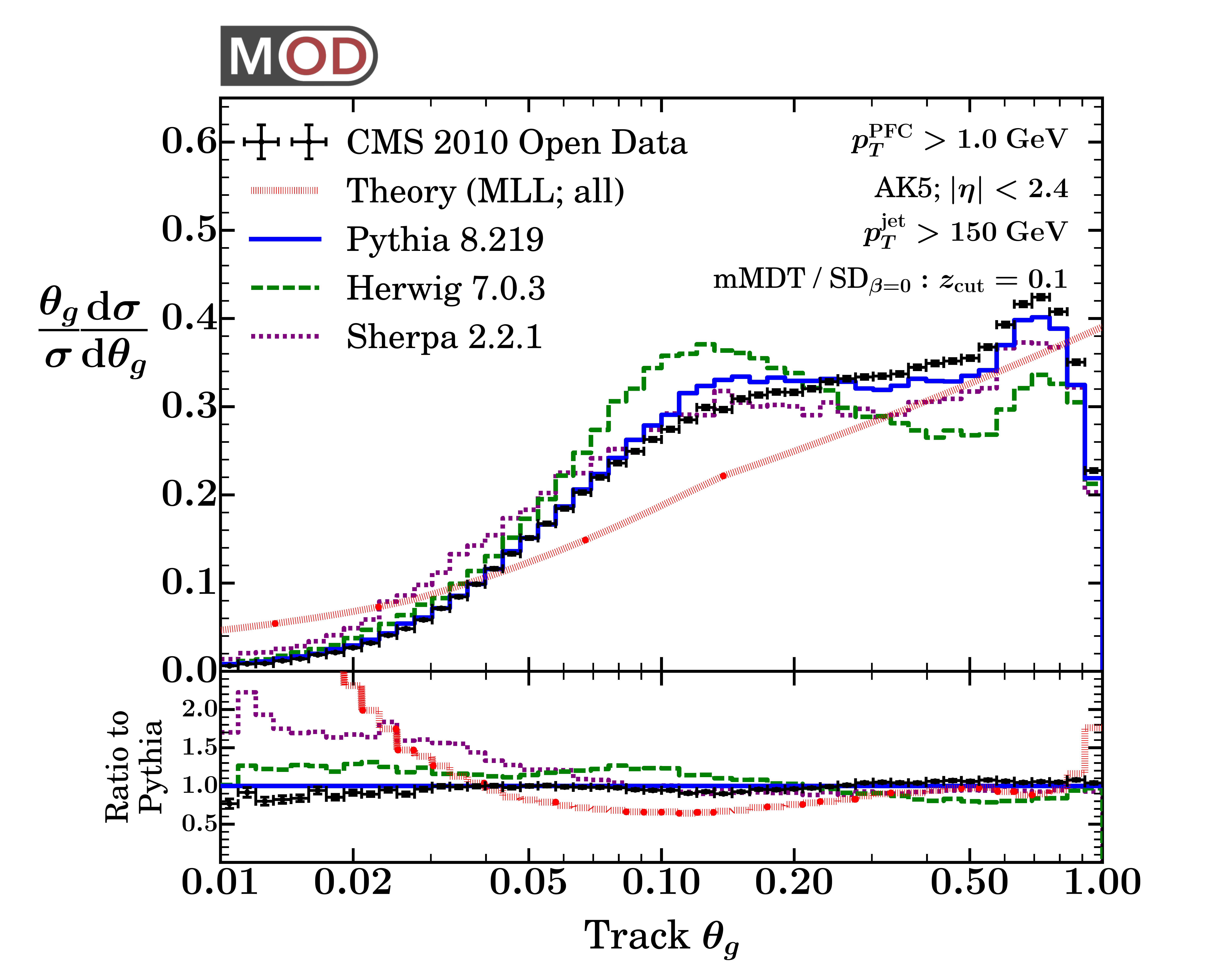}
}
\caption{Same as \Fig{fig:softdrop_zg} but for $\theta_g$.  For the MLL distributions, the region where nonperturbative dynamics matters is indicated by the use of dashing.  We do not indicate the regime where fixed-order corrections matter, since we have no first-principles estimate for the transition point.}
\label{fig:softdrop_rg}
\end{figure*}

\begin{figure*}
\subfloat[]{
\label{fig:softdrop_e05_log}
\includegraphics[width=0.9\columnwidth, page=1]{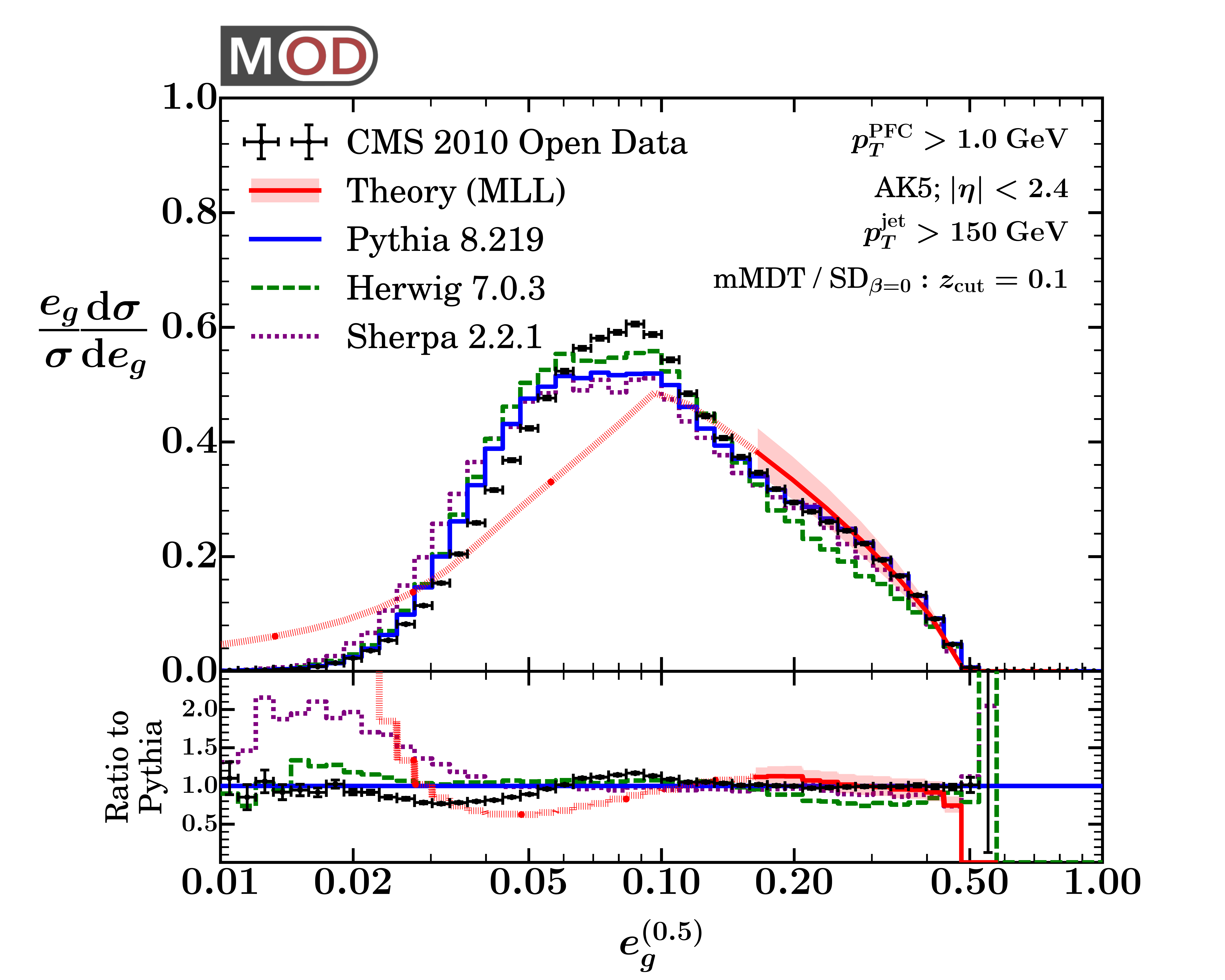}
}
\subfloat[]{
\label{fig:softdrop_e05_track_log}
\includegraphics[width=0.9\columnwidth, page=1]{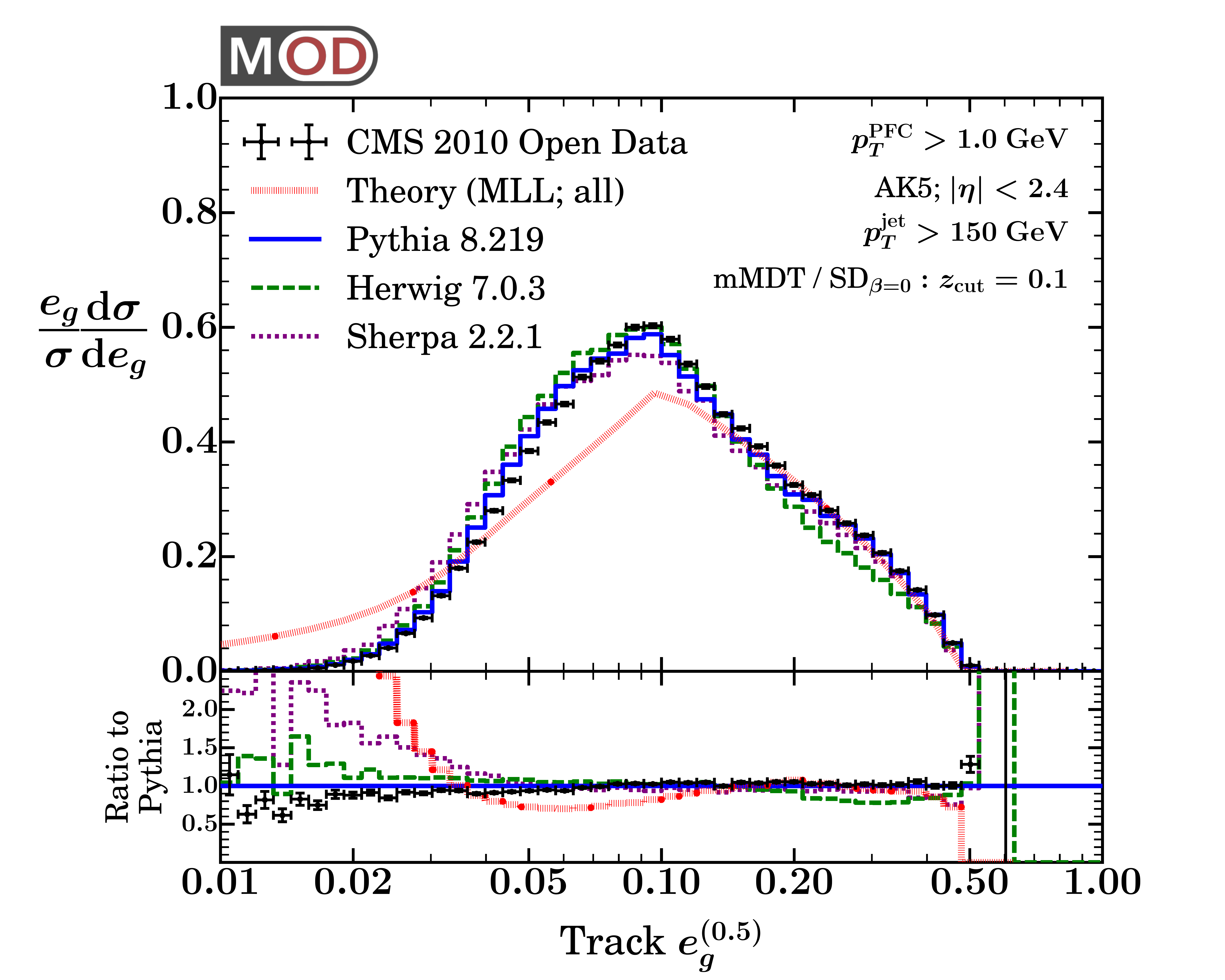}
}

\subfloat[]{
\label{fig:softdrop_e1_log}
\includegraphics[width=0.9\columnwidth, page=1]{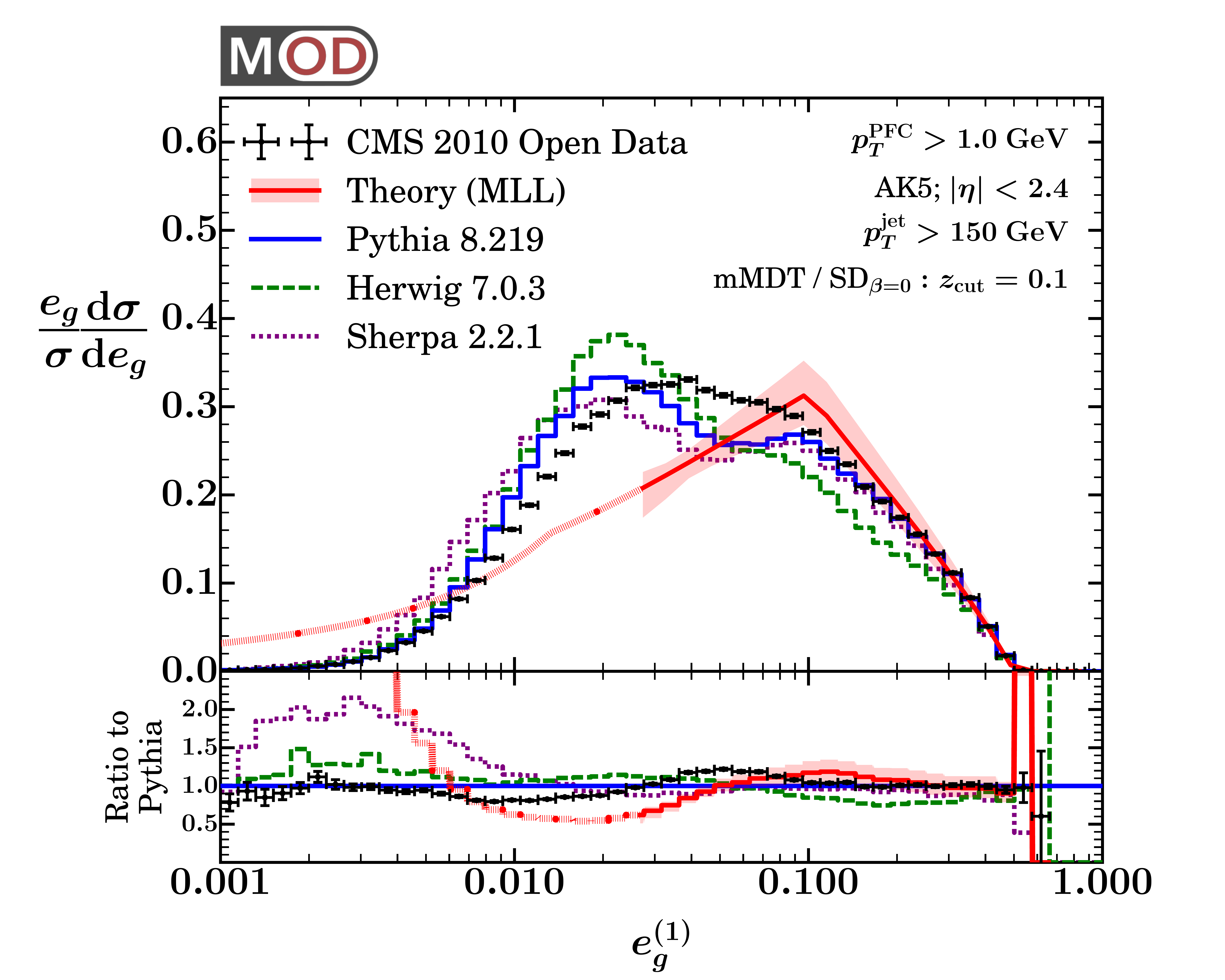}
}
\subfloat[]{
\label{fig:softdrop_e1_track_log}
\includegraphics[width=0.9\columnwidth, page=1]{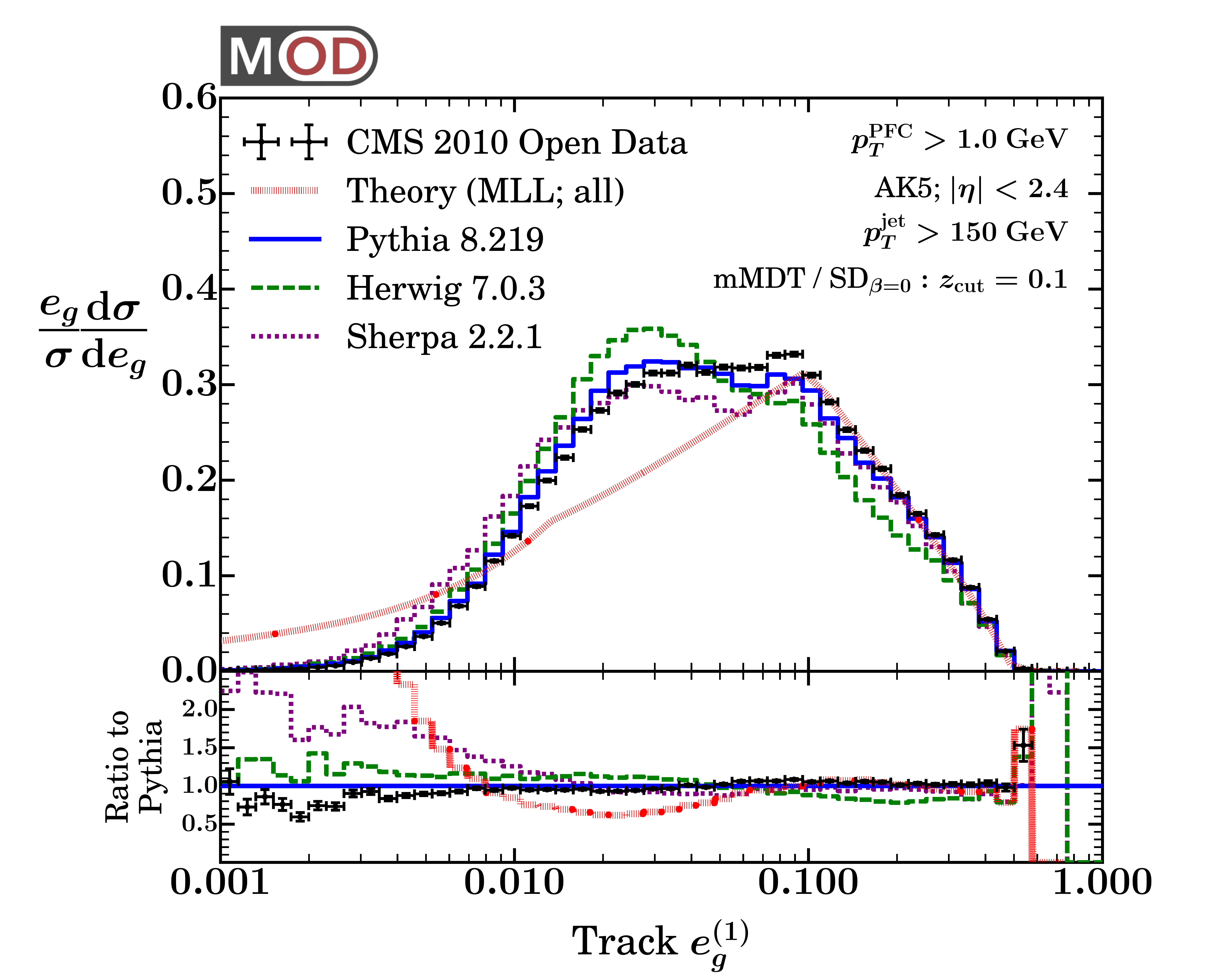}
}

\subfloat[]{
\label{fig:softdrop_e2_log}
\includegraphics[width=0.9\columnwidth, page=1]{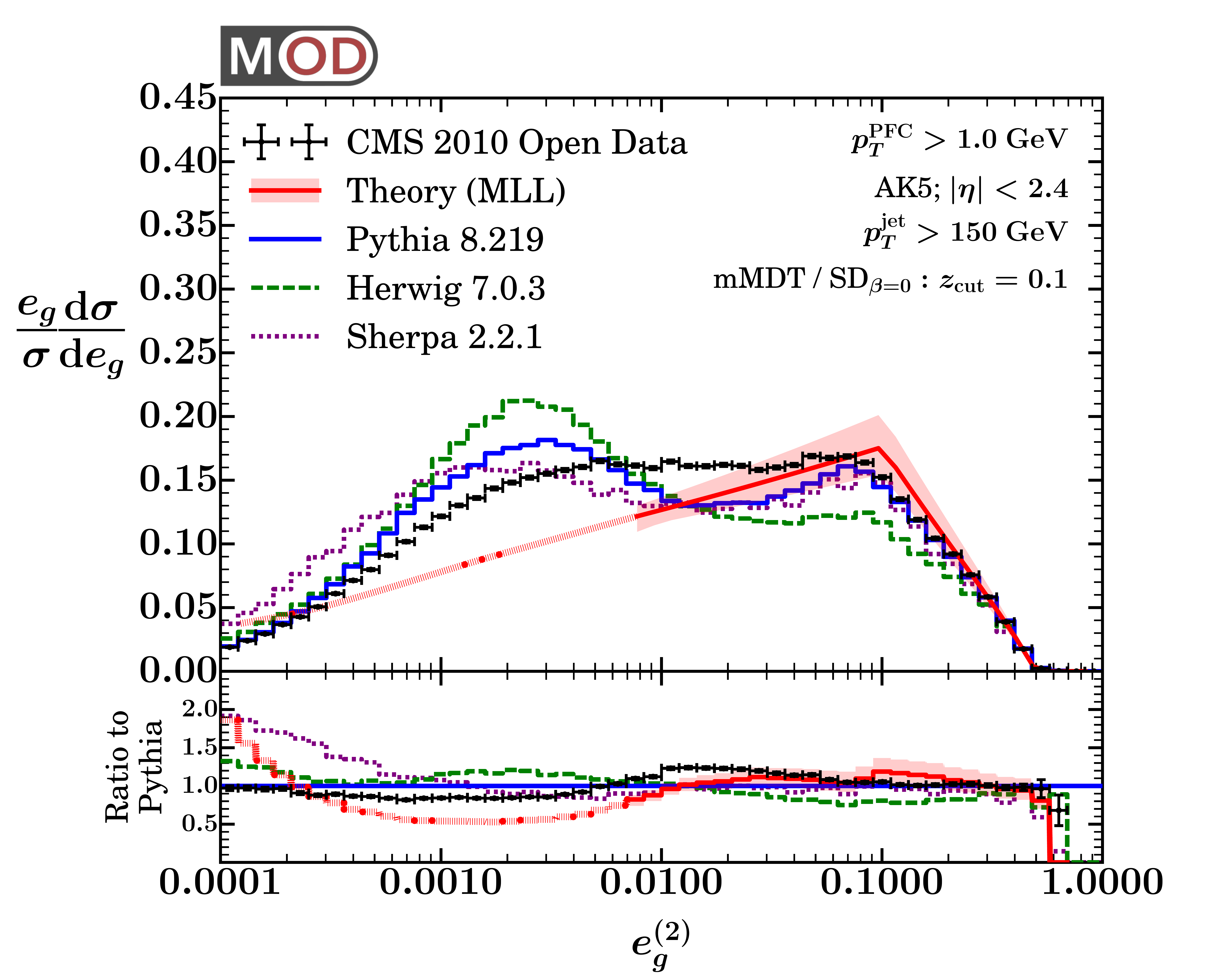}
}
\subfloat[]{
\label{fig:softdrop_e2_track_log}
\includegraphics[width=0.9\columnwidth, page=1]{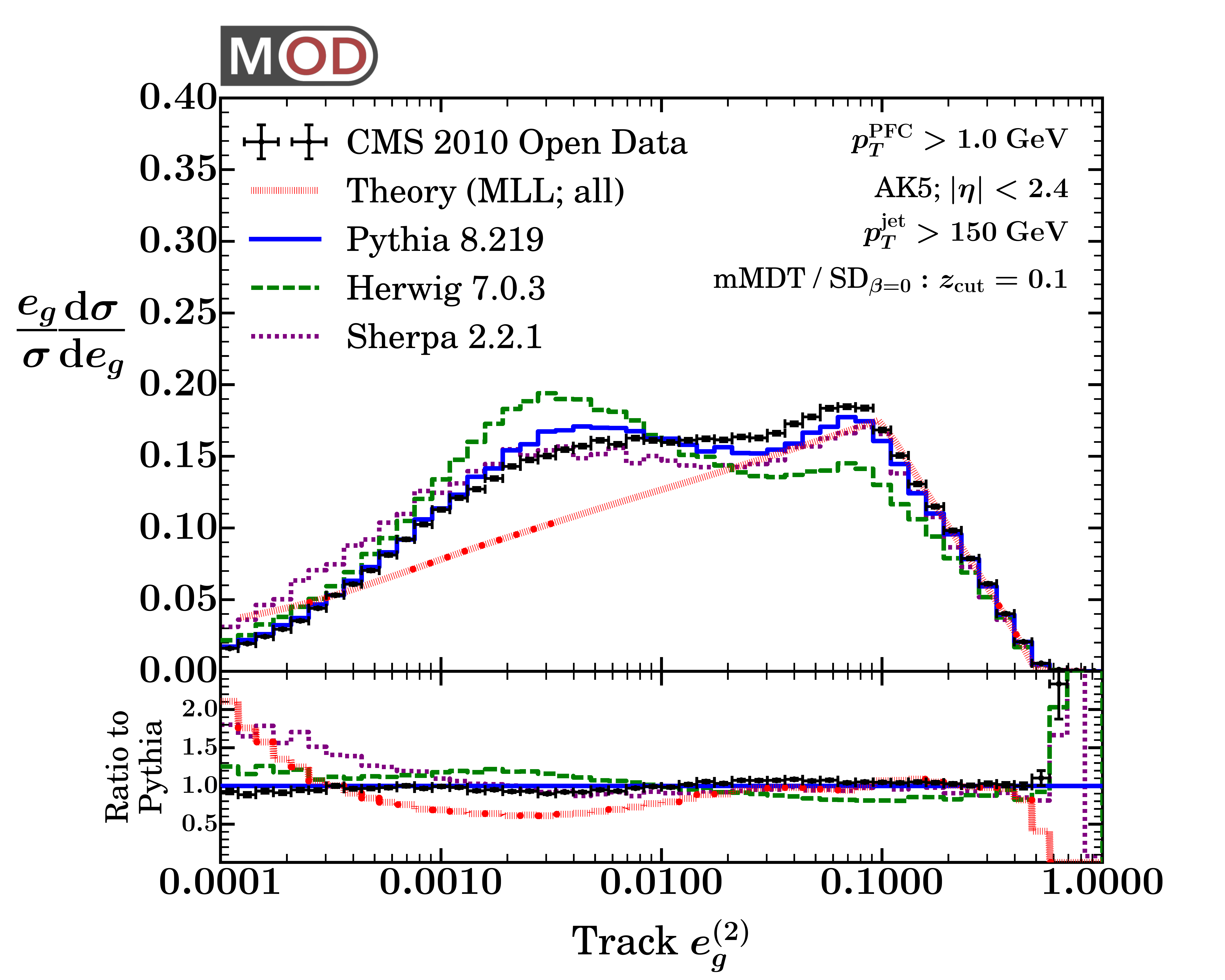}
}
\caption{Logarithmic distributions for (top row) $e_g^{(1/2)} = z_g \sqrt{\theta_g}$, (middle row) $e_g^{(1)} = z_g \theta_g$, and (bottom row) $e_g^{(2)} = z_g \theta_g^2$, using (left column) all particles and (right column) only charged particles.  As in \Fig{fig:softdrop_rg}, dashing indicates where nonpeturbative physics dominates and we have not indicated the fixed-order regime.}
\label{fig:softdrop_e_log}
\end{figure*}

We start in \Fig{fig:softdrop2D} with the full two-dimensional distributions for $p(z_g,\theta_g)$ from the Open Data, compared to the MLL analytic results and the three parton showers.  All of the distributions show a peak at small values of $z_g$ and $\theta_g$, corresponding to the soft and collinear singularities of QCD.  This structure is explained in more detail in a companion paper \cite{Larkoski:2017bvj}.  In principle, the $\theta_g$ distribution could extend all the way to $\theta_g \to 0$, but it is regulated by the perturbative form factor in \Eq{eq:MLLthetag}, nonperturbative hadronization corrections, as well as the finite angular resolution of the CMS detector.  Note the expected cut at $z_g = z_{\rm cut}$ from the soft drop condition.  The $z_g = \theta_g = 0$ bin indicates jets which only have one constituent after soft drop. 

Because of the logarithmic nature of the soft/collinear singularities of QCD, it is instructive to also plot $p(z_g,\theta_g)$ on a logarithmic scale, shown in \Fig{fig:softdrop2D_log}.  The overall qualitative structure is similar between the CMS Open Data and the theory distributions, but there are visible differences especially when nonperturbative physics is important.  Specifically, in the parton shower generators there is a strong peak around $\theta_g \simeq 0.1$, which is suppressed in the CMS Open Data.  It would be interesting to know whether the parton shower is exhibiting a physical structure that is simply washed out in the open data or if there is a pathology in the parton shower generators in this kinematic regime.  Because this feature appears exactly where nonperturbative physics is expected to matter, the perturbative MLL distribution is not a useful guide to answer this question.

To better compare the open data to theory predictions, we now consider the projected observables from \Eq{eq:bigfive}.  We show both all-particle and track-only observables to highlight the impact of angular resolution.  Strictly speaking, the MLL distributions from \Sec{subsec:MLL} are only valid for all-particle observables, but we show dashed versions of same curves on the track-only plots for ease of comparison.  One could imagine using the track function formalism \cite{Chang:2013rca,Chang:2013iba} to make sensible track-based MLL predictions, but that is beyond the scope of the present work.

We start with $z_g$ in \Fig{fig:softdrop_zg}, which is also studied in \Refs{Larkoski:2017bvj,CMS:2016jys,StarTalk,AlicePoster}.  Especially for the track-only measurement, the agreement between all five distributions is remarkable.  For the all-particle distributions, there is a noticeable excess in the CMS Open Data compared to the theory distributions at $z_g \simeq z_{\rm cut}$, as well as an excess of events that failed the soft drop procedure; both of these features could be explained by the degraded angular resolution for neutral particles.  On a logarithmic scale, one can see that the $z_g$ distribution is roughly flat, as expected from the singularity structure of the splitting functions in \Eqs{splittings_q}{splittings_g}.  

We can get a better understanding of angular effects by looking at $\theta_g$ directly in \Fig{fig:softdrop_rg}.  Not surprisingly, the largest differences between the MLL distribution and the parton showers occur in the regime where nonperturbative dynamics matters.  Especially on the logarithmic scale, the feature at $\theta_g \simeq 0.1$ is prominent in the parton shower generators.  Note that the CMS heavy ion analysis in \Ref{CMS:2016jys} placed a cut of $R_g > 0.1$ ($\theta_g > 0.2$) to avoid modeling issues in the small $\theta_g$ regime.  Given the relatively good agreement between the CMS Open Data and the parton shower generators in the track-based distributions, we do not see an immediate reason to distrust small $\theta_g$ values, and measurements of $\theta_g$ could indeed be relevant for parton shower tuning.

Turning to the groomed single-emission angularities $e_g^{(\alpha)}$, in \Fig{fig:softdrop_e_log} we see reasonable agreement between the CMS Open Data and the parton shower generators, especially for the track-based observables.  The MLL distributions exhibit the expected kinks at $e_g^{(\alpha)} = z_{\rm cut}$, but the slope below this kink value differs noticeably.  For the $p_T$ range shown, though, the location of the kink is not so far from the scale where nonperturbative physics dominates, so measurements with more energetic jets are needed to test whether or not there is any tension with perturbative predictions.

The above plots are only a subset of the soft-dropped distributions we have made with the CMS Open Data.  In the \texttt{arXiv} source files, the plots in \Figss{fig:softdrop_zg}{fig:softdrop_rg}{fig:softdrop_e_log}  are part of a multipage file that not only has multiple jet $p_T$ ranges, but also $z_{\rm cut} = 0.05$ and $z_{\rm cut} = 0.2$ distributions.  We leave a study of alternative $\beta$ values to future work.  For completeness, in \App{app:yetmore} we show soft-dropped versions of all of the substructure distributions from \Sec{sec:hardest_jet}.  We also show the fractional change in the jet $p_T$ due to soft drop, which was shown in have interesting analytic properties in \Ref{Larkoski:2014wba}.  Additional soft-dropped observables can be provided to interested readers upon request (or derived using the publicly-available MOD software framework).

\clearpage

\section{Advice to the Community}
\label{sec:recommend}

From a physics perspective, our experience with the CMS Open Data was fantastic.  With PFCs, one can essentially perform the same kinds of four-vector-based analyses on real data as one would perform on collisions from parton shower generators.  Using open data has the potential to accelerate scientific progress (pun intended) by allowing scientists outside of the official detector collaborations to pursue innovative analysis techniques.  We hope that our jet substructure studies have demonstrated both the value in releasing public data and the enthusiasm of potential external users.  We encourage other members of the particle physics community to take advantage of this unique data set.

From a technical perspective, though, we encountered a number of challenges.  Some of these challenges were simply a result of our unfamiliarity with the CMSSW framework and the steep learning curve faced when trying to properly parse the AOD file format.   Some of these challenges are faced every day by LHC experimentalists, and it is perhaps unreasonable to expect external users to have an easier time than collaboration members.  Some of these challenges (particularly the issue of detector-simulated samples) have been partially addressed by the 2011A CMS Open Data release \cite{CMS:JetPrimary2011}.  That said, we suspect that some issues were not anticipated by the CMS Open Data project, and we worry that they have deterred other analysis teams who might have otherwise found interesting uses for open data.  Therefore, we think it is useful to highlight the primary challenges we faced, followed by specific recommendations for how potentially to address them. 

\subsection{Challenges}

Here are the main issues that we faced in performing the analyses in this paper.
\begin{itemize}
\item \textit{Slow development cycle}.  As CMSSW novices, we often needed to perform run-time debugging to figure out how specific functions worked.  There were two elements of the CMSSW workflow that introduced a considerable lag between starting a job and getting debugging feedback.  The first is that, when using the \textsc{XRootD} interface, one has to face the constant overhead (and inconstant network performance) of retrieving data remotely.  The second is that, as a standard part of every CMS analysis, one has to load configuration files into memory.  Loading \texttt{FrontierConditions\_GlobalTag\_cff} (which is necessary to get proper trigger prescale values) takes around 10 minutes at the start of a run.  For most users, this delay alone would be too high of a barrier for using the CMS Open Data.  By downloading the AOD files directly and building our own MOD file format, we were able to speed up the development cycle through a lightweight analysis framework.  Still, creating the \texttt{MODProducer} in the first place required a fair amount of trial, error, and frustration.
\item \textit{Scattered documentation}.  Though the CMS Open Data uses an old version of CMSSW (v4.2 compared to the latest v9.0), there is still plenty of relevant documentation available online.  The main challenge is that it is scattered in multiple places, including online \textsc{TWiki} pages, masterclass lectures, thesis presentations, and \textsc{GitHub} repositories.  Eventually, with help from CMS insiders, we were able to figure out which information was relevant to a particular question, but we would have benefitted from more centralized documentation that highlighted the most important features of the CMS Open Data.  Centralized documentation would undoubtably help CMS collaboration members as well, as would making more \textsc{TWiki} pages accessible outside of the CERN authentication wall.
\item \textit{Lack of validation examples}.  When working with public data, one would like to validate that one is doing a sensible analysis by trying to match published results.  While example files were provided, none of them (to our knowledge) involved the complications present in a real analysis, such as appropriate trigger selection, jet quality criteria, and jet energy corrections.  Initially, we had hoped to reproduce the jet $p_T$ spectrum measured by CMS on 2010 data \cite{CMS:2011ab}, but that turned out to be surprisingly difficult, since very low $p_T$ jet triggers are not contained in the Jet Primary Dataset, and we were not confident in our ability to merge information from the MinimumBias Primary Dataset.  (In addition, the published CMS result is based on inclusive jet $p_T$ spectra, while we restricted our analysis to the hardest jet in an event to simplify trigger assignment.)  Ideally, one should be able to perform event-by-event validation with the CMS Open Data, especially if there are important calibration steps that could be missed.\footnote{In the one case where we thought it would be the most straightforward to cross check results, namely the luminosity study in \Fig{fig:integrated_lumi}, it was frustrating to later learn that the AOD files contained insufficient information.}
\item \textit{Information overload}.  The AOD files contain an incredible wealth of information, such that the majority of official CMS analyses can use the AOD format directly without requiring RAW or RECO information.  While ideal for archival purposes, it is an overload of information for external users, especially because some information is effectively duplicated.  The main reason we introduced the MOD file format was to restrict our access only to information that was essential for our analysis.  This can be compared to the Mini-AOD format currently being developed by CMS to address a similar problem \cite{Petrucciani:2015gjw}.
\item \textit{Presence of superfluous data}.  As described on the Open Data Portal, one has to apply a cut to only select validated runs.  This meant that of the initial 20 million events, only 16 million were actually usable.  That said, this turns out to be a relatively small issue compared to trigger inefficiencies, which to our knowledge is not mentioned on the Open Data Portal webpage.  The Jet Primary Dataset includes any event where one of the jet-related triggers fired (see \Tab{tab:trigger_names}).  However, these triggers are not fully efficient down to the turn-on threshold, which is why we had to derive trigger efficiency curves in \Fig{fig:trigger_efficiency_curve}. Using just the triggers in \Tab{tab:trigger_table} in the regime where they were nearly 100\% efficient reduced the number of events for our analysis to less than 1 million, which is an order-of-magnitude smaller than the starting dataset.
\item \textit{No fast simulation or Monte Carlo samples}.  While it is in principle possible to run the full CMS detector simulation on events from parton shower generators, we did not have the computing resources to do so.  Without detector information, either in the form of CMS-approved fast simulation software or simulated Monte Carlo datasets, we cannot really say whether the good agreement seen between open data and parton showers is robust or merely accidental.  Fast simulation tools like \textsc{Delphes} can be used to some extent, but because they have not been optimized for jet substructure, we were not able to use them for this study.  Official CMS Monte Carlo samples would have helped us greatly to estimate the size of detector corrections (and potentially even unfold distributions back to truth level).  We are therefore encouraged by the inclusion of Monte Carlo samples in the 2011 CMS Open Data release \cite{CMS:JetPrimary2011}.
\end{itemize}
Despite these above issues, though, we were able to perform a successful jet substructure analysis, in no small part due to the help of our CMS (and ATLAS) colleagues who generously offered their time and advice.

\subsection{Recommendations}

Given our experience, we would like to make the following recommendations to CERN and CMS about the continuation of the Open Data project.  Many of these suggestions are also relevant for the 2012 ATLAS Open Data \cite{ATLASOpenDataPortal}, though that effort is aimed more at education than research.  Here are our recommendations, in rough order of priority.
\begin{itemize}
\item \textit{Continue to release research-grade public data.}   Particle physics experiments are expensive and, in many cases, unique.  It is therefore incumbent on the particle physics community to extract as much useful information from collision data as possible.  First priority for data analysis should of course go to members of the detector collaborations, especially since proper calibration can only be performed by physicists  familiar with the detection technology.\footnote{There also needs to be a strong incentive for experimentalists to join collaborations in the first place.  Outside access to (calibrated) data should not be used to bypass the stringent internal collaboration review process.}  After an appropriate lag time---four years in the case of the 2010 CMS Open Data---outside scientists can play a useful role in data analysis, especially because collaboration members might not have the time or interest to revisit old data once new data is available.  Techniques that perform well on open data can then be incorporated into the analysis strategies used internally by the collaborations, enhancing the already strong feedback cycle within the particle physics community.\footnote{There are, of course, cases where a full open data analysis is not necessary to motivate the adoption of new techniques.  Even in that context, though, it can still be valuable for the collaborations to release official Monte Carlo samples.  At minimum, hadron-truth-level samples provide a standard benchmark to validate the performance of new techniques.  More ambitiously, detector-simulated samples can be used to assess how a new technique might be affected by detector granularity, acceptance, and efficiency.}  
\item \textit{Continue to provide a unique reference event interpretation.}   A key feature of the CMS Open Data is the presence of PFCs, which provides a unique reference event interpretation with four-vector-like objects.  From our experience, this seems to be the right level of information for an outside user.  If the CMS Open Data were to consist only of high-level objects, like reconstructed jets, then we would not have been able to pursue these jet substructure studies.  On the flip side, more low-level information (or multiple versions of the same information) could overwhelm the external user and cause confusion.  Since it is unlikely that open data could support arbitrary physics studies, the aim of open data should be to facilitate particle-level studies that do not require detailed knowledge of the detector.
\item \textit{Provide validation examples.}  We mentioned above the potential value of having centralized documentation about open data.  Even more important than documentation, though, is having example analyses performed using open data.  Explicit code helps emphasize analysis steps that might be missed by novices, including trigger selection, prescale factors, jet calibration, and luminosity extraction.  Where possible, these validation examples should reproduce official published analyses.  We expect that these validation examples will become the templates for future open data analyses, and good validation examples could minimize incorrect use of the data.  We intend to make the present analysis software public, in order to guide future open data studies.
\item \textit{Provide detector response information.}  The biggest physics gap in our study was our limited ability to estimate detector corrections.  Ideally, open data should be released with corresponding detector-simulated Monte Carlo samples, matched to the triggers of interest.  Indeed, the 2011 CMS Open Data---released in April 2016---does provide these samples, which will make it possible to estimate (some) detector systematics.\footnote{The 2012 ATLAS Open Data \cite{ATLASOpenDataPortal} does provide detector-simulated samples, but not truth-level information, so it is not possible to derive detector response information.}  Eventually, if open data is used to place (unofficial) bounds on physics beyond the standard model, an external user would also need access to a recommended fast simulation framework.  While it is probably impossible for external users to assess systematic uncertainties with the same level of care as one can do within the collaboration, some understanding of detector effects is needed before concluding that an effect observed in the data is real and interesting.
\item \textit{Cull the data set.}  Within the experimental collaborations, most studies are based on well-defined trigger paths with almost 100\% trigger efficiencies and nearly constant prescale factors.  These same requirements should be imposed on the open data such that only usable data is made available publicly.  This would not only reduce the storage requirements for open data, but it would also help avoid some spurious features showing up in the data.  Similarly, most official studies do not need the full information contained in the AOD file format, and a more restricted data format would help further shrink the data file sizes and reduce user errors.  Of course, to maximize the archival value, it may still make sense to release the original AOD files for the expert users, along with the tools used to create the culled versions.
\item \textit{Speed up the development cycle.}  For archival purposes, it is valuable to have the full CMSSW framework operating in a VM environment.  For the external user, though, it would be more efficient to have a simplified software framework that can run with minimal software dependencies.\footnote{If the use of the CMSSW framework is essential, it would be helpful to have more centralized documentation for the core classes and methods of CMSSW.}  We understand that developing an external software environment requires considerable effort by collaboration members, but a relatively small investment would greatly increase the usability of the CMS Open Data.  Our \textsc{MODAnalyzer} software (based heavily on \textsc{FastJet}) might be a good starting point for such an analysis package, as would any of the existing private tools used internally by CMS analysis teams.  It may also make sense for the collaborations to appoint an official contact to answer questions from external users, possibly in the form of an open data convenership.  
\end{itemize}

While these recommendations are perhaps ambitious in their scope, we think that the enormous scientific value of particle physics data justifies this kind of investment in open data.

\section{Conclusion}
\label{sec:conclude}

As the LHC explores the frontiers of scientific knowledge, its primary legacy will be the measurements and discoveries made by the LHC detector collaborations.  But there is another potential legacy from the LHC that could be just as important:  granting future generations of physicists access to unique high-quality data sets from proton-proton collisions at 7, 8, 13, and 14 TeV.

In our view, the best way to build a legacy data set is to invest in open data initiatives right now, such that scientists outside of the LHC collaborations can stress-test archival data strategies.  This paper represents the first such analysis made with 2010 CMS Open Data from 7 TeV collisions.   We showed how to extract jet substructure observables with the help of CMS's particle flow algorithm, yielding results that are in good agreement with parton shower generators and first-principles QCD calculations.  The recent release of the 2011 CMS Open Data is particularly exciting, since it now includes detector-simulated Monte Carlo samples, allowing one to properly estimate detector systematics.  We hope our experience motivates the LHC collaborations to further their investment in public data releases and encourages the particle physics community to exploit the scientific potential of open datasets.

\begin{acknowledgments}
We applaud CERN for the historic launch of the Open Data Portal, and we congratulate the CMS collaboration for the fantastic performance of their detector and the high quality of the resulting public data set.
We thank Alexis Romero for collaboration in the early stages of this work.
We are indebted to Salvatore Rappoccio and Kati Lassila-Perini for helping us navigate the CMS software framework.
We benefitted from code and encouragement from Tim Andeen, Matt Bellis, Andy Buckley, Kyle Cranmer, Sarah Demers, Guenther Dissertori, Javier Duarte, Peter Fisher, Achim Geiser, Giacomo Govi, Phil Harris, Beate Heinemann, Harri Hirvonsalo, Markus Klute, Greg Landsberg, Yen-Jie Lee, Elliot Lipeles, Peter Loch, Marcello Maggi, David Miller, Ben Nachman, Christoph Paus, Alexx Perloff, Andreas Pfeiffer, Maurizio Pierini, Ana Rodriguez, Gunther Roland, Ariel Schwartzman, Liz Sexton-Kennedy, Maria Spiropulu, Nhan Tran, Ana Trisovic, Chris Tully, Marta Verweij, Mikko Voutilainen, and Mike Williams.
This work is supported by the MIT Charles E.\ Reed Faculty Initiatives Fund.
The work of JT, AT, and WX is supported by the U.S. Department of Energy (DOE) under grant contract numbers DE-SC-00012567 and DE-SC-00015476.
The work of AL was supported by the U.S.\ National Science Foundation, under grant PHY--1419008, the LHC Theory Initiative.
SM is supported by the U.S.\ National Science Foundation, under grants PHY--0969510 (LHC Theory Initiative) and PHY--1619867.
AT is also supported by the MIT Undergraduate Research Opportunities Program.
\end{acknowledgments}

\appendix

\section{Additional Open Data Information}
\label{app:opendatainfo}

\begin{figure}[t]
\includegraphics[width=\columnwidth]{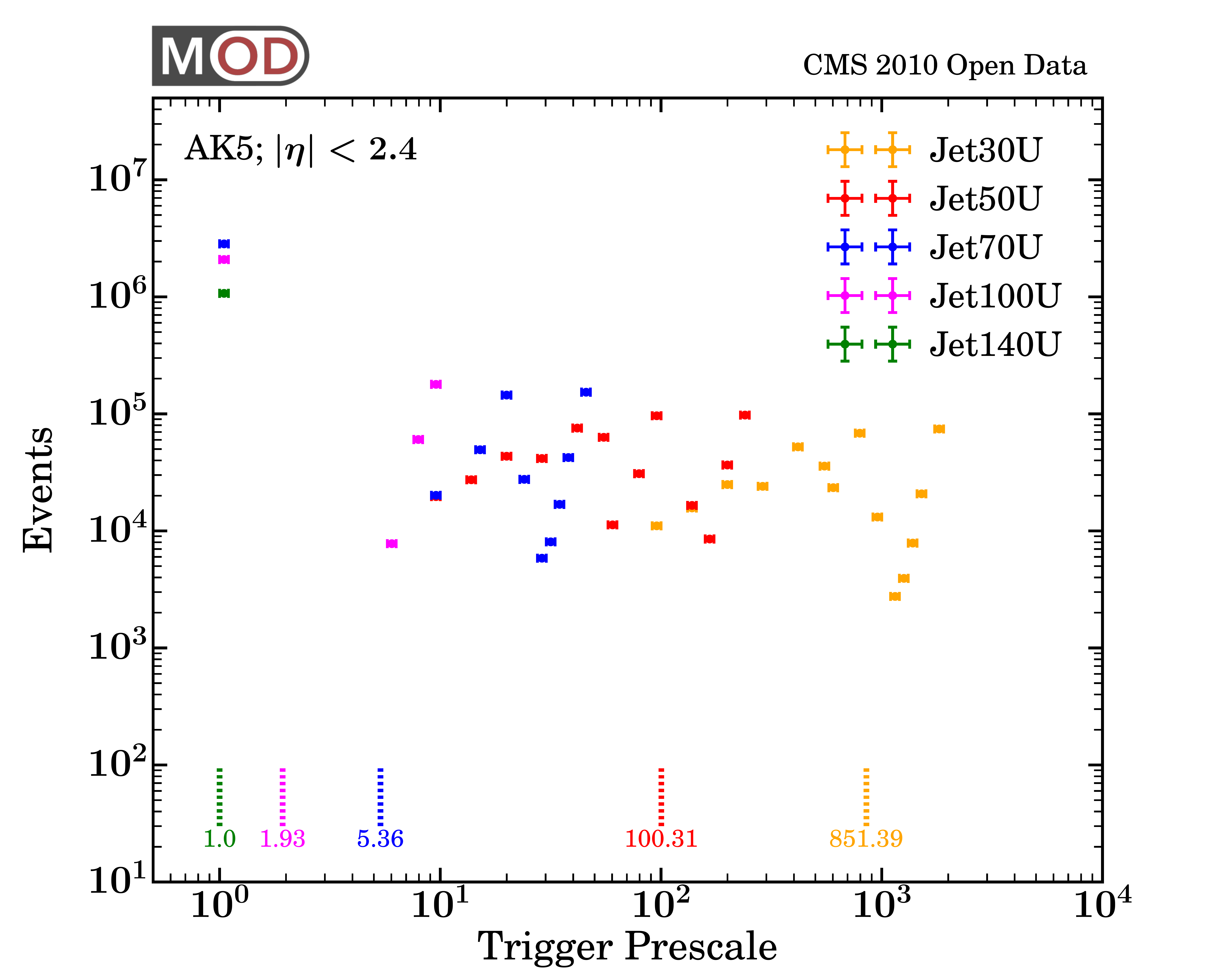}
\caption{Trigger prescale values for jets that pass the criteria in \Tab{tab:trigger_table}.   When filling histograms in this paper, we always use the average prescale values, not the individual ones.}
\label{fig:prescale_values}
\end{figure}

In this appendix, we provide additional information about the overall CMS Open Data extraction from \Sec{sec:CMSOpen}.  In \Fig{fig:prescale_values}, we show the distribution of prescale values obtained for the triggers in \Tab{tab:trigger_table}.  As expected, higher trigger thresholds have lower prescale values, but there is substantial variation in the prescale values which changed over the duration of the run.  If we were to use the given prescale factors instead of the averages, we would have seen rather large statistical uncertainties in our distributions.  Since we only ever use one trigger per $p_T$ bin, it is valid to use the average prescale value instead.

\begin{figure*}[t]
\subfloat[]{
\label{fig:JEC}
\includegraphics[width=\columnwidth, page=1]{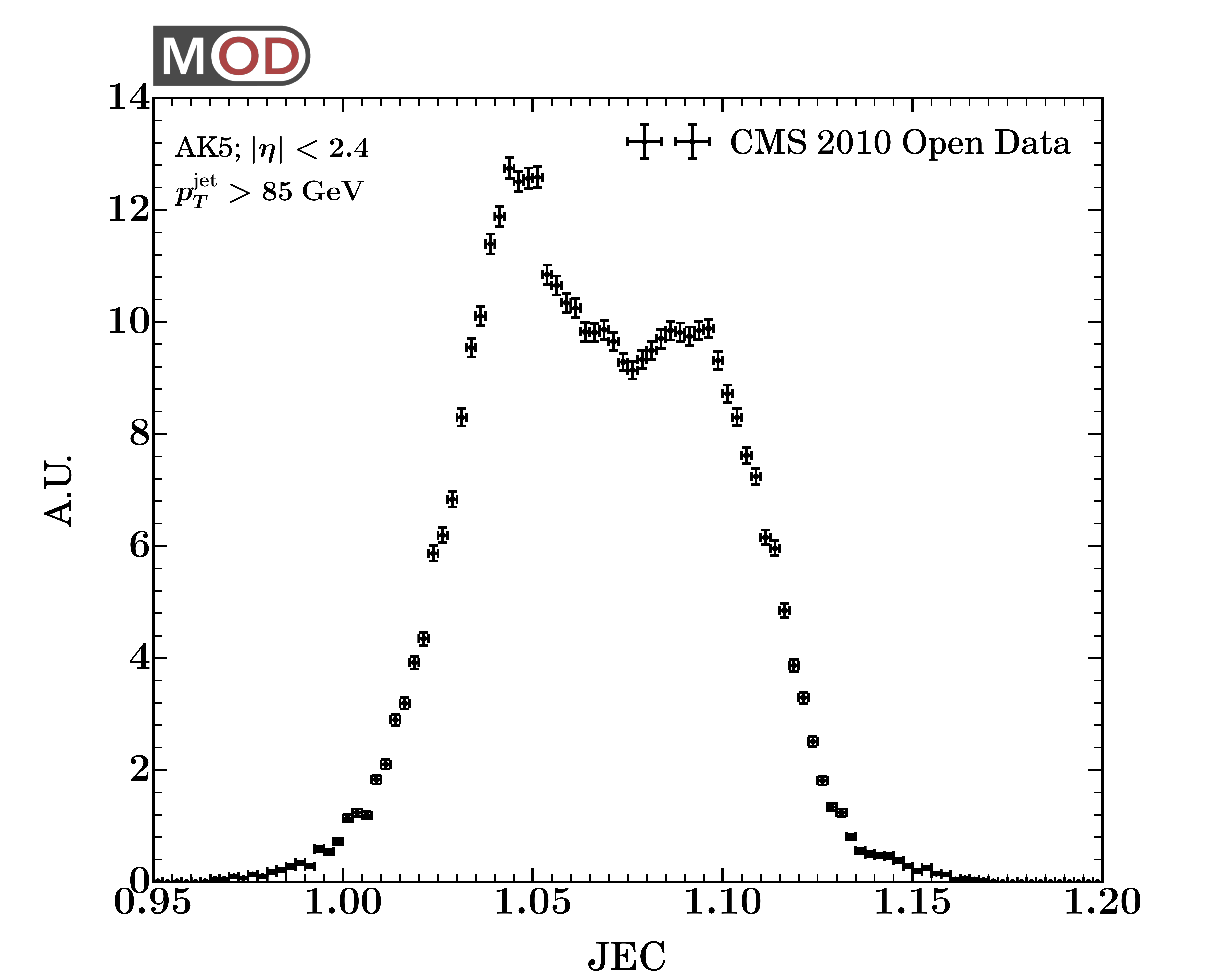}
}
\subfloat[]{
\label{fig:jet_area}
\includegraphics[width=\columnwidth, page=1]{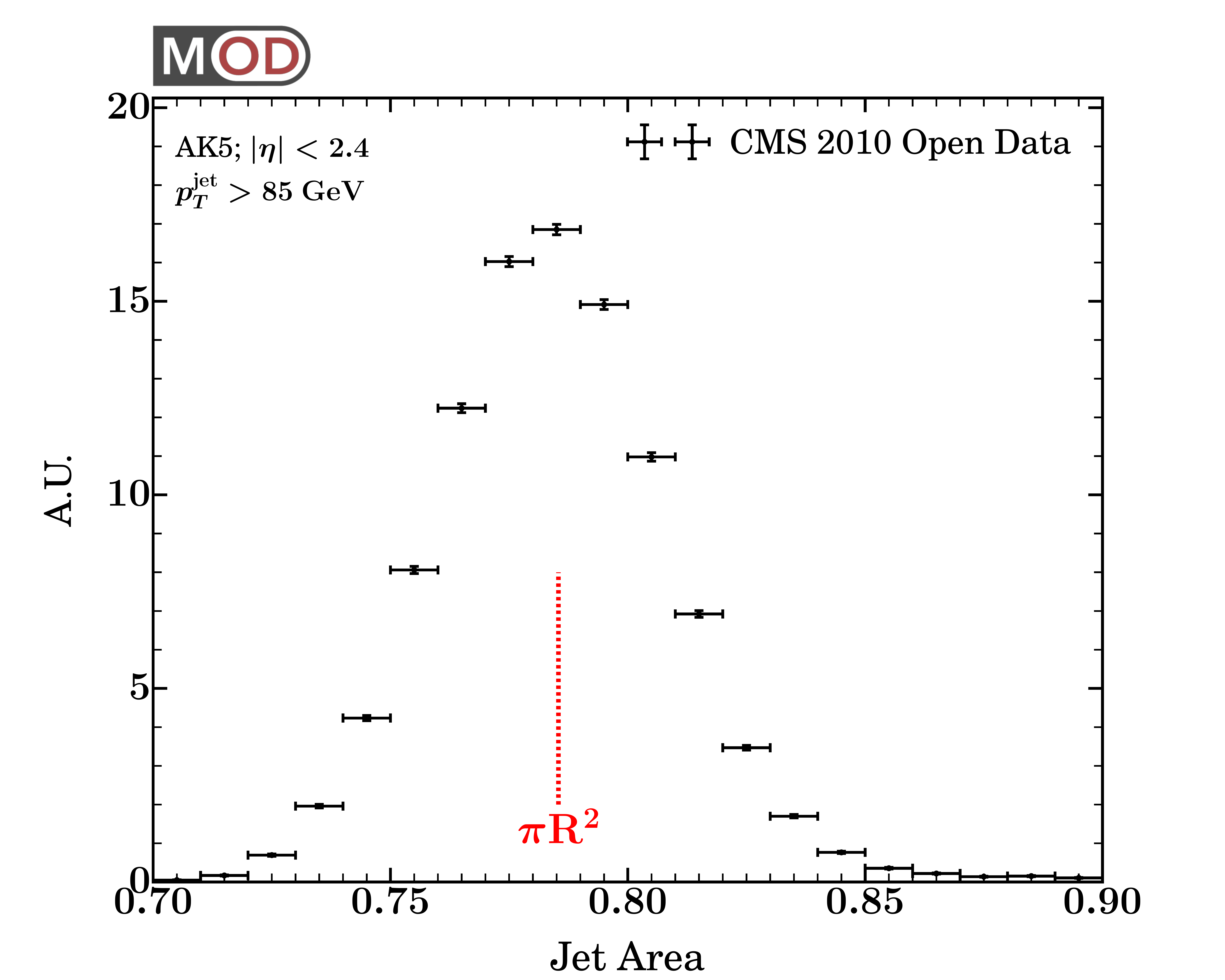}
}
\caption{Range of (a) JEC factors and (b) active jet areas \cite{Cacciari:2008gn} encountered for the hardest jet.}
\end{figure*}

\begin{table}[t]
\begin{tabular}{r@{$\quad$}c@{$\quad$}c@{$\quad$}c}
\hline
\hline
 & Loose & Medium & Tight \\
\hline
Neutral Hadron Fraction & $<0.99$ & $<0.95$ & $<0.90$ \\
Neutral EM Fraction & $<0.99$ & $<0.95$ & $<0.90$ \\
Number of Constituents & $>1$ & $>1$ & $>1$ \\
\hline
Charged Hadron Fraction & $>0.00$ & $>0.00$ & $>0.00$ \\
Charged EM Fraction & $<0.99$ & $<0.99$ & $<0.99$ \\
Charged Multiplicity & $>0$ & $>0$ & $>0$ \\
\hline
\hline
\end{tabular}
\caption{Recommended jet quality criteria provided by CMS for $|\eta| < 2.4$.  For $|\eta| > 2.4$, where no tracking is available, the last three requirements are not applied, and all jet constituents are treated as neutral.  For our analysis, we always impose the ``loose'' criteria.}
\label{tab:jet_quality}
\end{table}

\begin{table}[t]
\begin{tabular}{r@{$\quad$}r@{$\quad$}r@{$\quad$}r@{$\quad$}r}
\hline
\hline
& \multicolumn{2}{c@{$\quad$}}{Jet Primary Dataset} &  \multicolumn{2}{c}{Hardest Jet Selection} \\
$N_{\rm PV}$ & Events & Fraction & Events & Fraction\\
\hline
1 & 4,716,494 & 0.289 & 190,277 & 0.248\\
2 & 4,814,495 & 0.295 & 246,387 & 0.321\\
3 & 3,630,413 & 0.222 & 180,021 & 0.234\\
4 & 1,933,832 & 0.118 & 93,587 & 0.122\\
5 & 819,835 & 0.050 & 38,598 & 0.050\\
6 & 294,612 & 0.018 & 13,805 & 0.018\\
7 & 93,714 & 0.006 & 4,318 & 0.006\\
8 & 27,550 & 0.002 & 1,242 & 0.002\\
9 & 7,481 & 0.000 & 330 & 0.000\\
10 & 2,041 & 0.000 & 91 & 0.000\\
11 & 540 & 0.000 & 21 & 0.000\\
12 & 125 & 0.000 & 6 & 0.000\\
13 & 41 & 0.000 & 3 & 0.000\\
14 & 9 & 0.000 & 1 & 0.000\\
$\ge 15$ & 5 & 0.000 & 0 & 0.000\\
\hline
\hline
\end{tabular}
\caption{Number of primary interactions per bunch crossing.  Since Run 2010B was a relatively low luminosity run, a large fraction of the event sample has $N_{\rm  PV} = 1$, corresponding to no pileup contamination.}
\label{tab:NPV}
\end{table}

To properly select the hardest jet, we have to impose jet quality criteria and apply JEC factors.  The CMS-recommended jet quality criteria are shown in \Fig{tab:jet_quality}; we always use the ``loose'' selection in our analysis.  In \Fig{fig:JEC}, we show the distribution of JEC factors encouraged for the hardest jet.  These are multiplicative scaling factors that tend give a 5-10\% correction to the jet $p_T$.  In addition to accounting for detector effects, the JEC factor accounts for pileup through area subtraction \cite{Cacciari:2008gn}.  The distribution of jet areas for the hardest jet are shown in \Fig{fig:jet_area}, which peak at $\pi R^2$ for $R = 0.5$ as expected.  Note that the impact of pileup was minimal in Run 2010B, since as shown in \Tab{tab:NPV}, the number of primary interactions per bunch crossing was less than 5 (i.e.~effectively no pileup) for over 90\% of the events and never more than 15 for the selection used for our analysis (modest pileup).

\begin{figure*}
\subfloat[]{
\label{fig:pfc_pt_neutral_extended}
\includegraphics[width=\columnwidth, page=1]{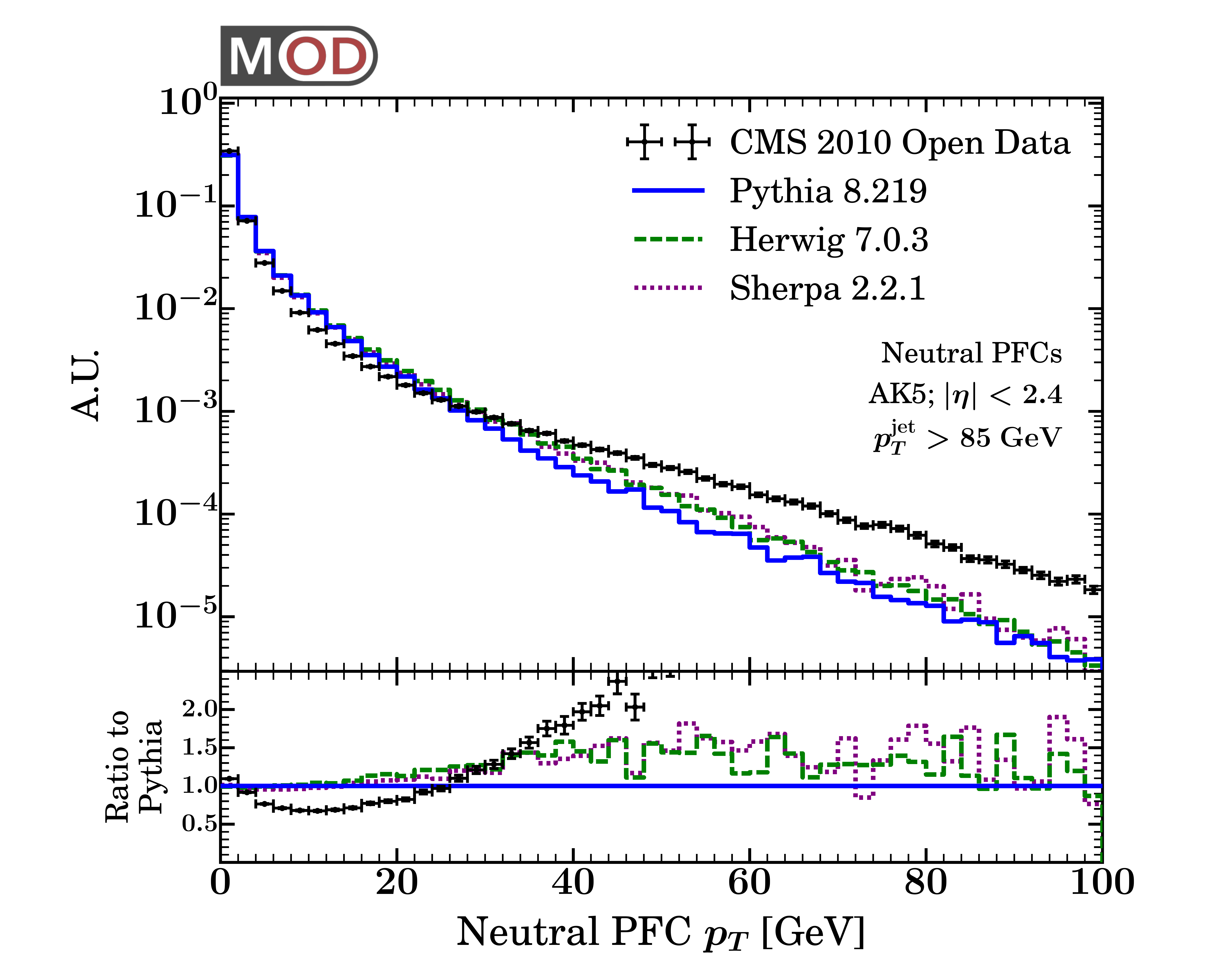}
}
\subfloat[]{
\label{fig:pfc_pt_track_extended}
\includegraphics[width=\columnwidth, page=1]{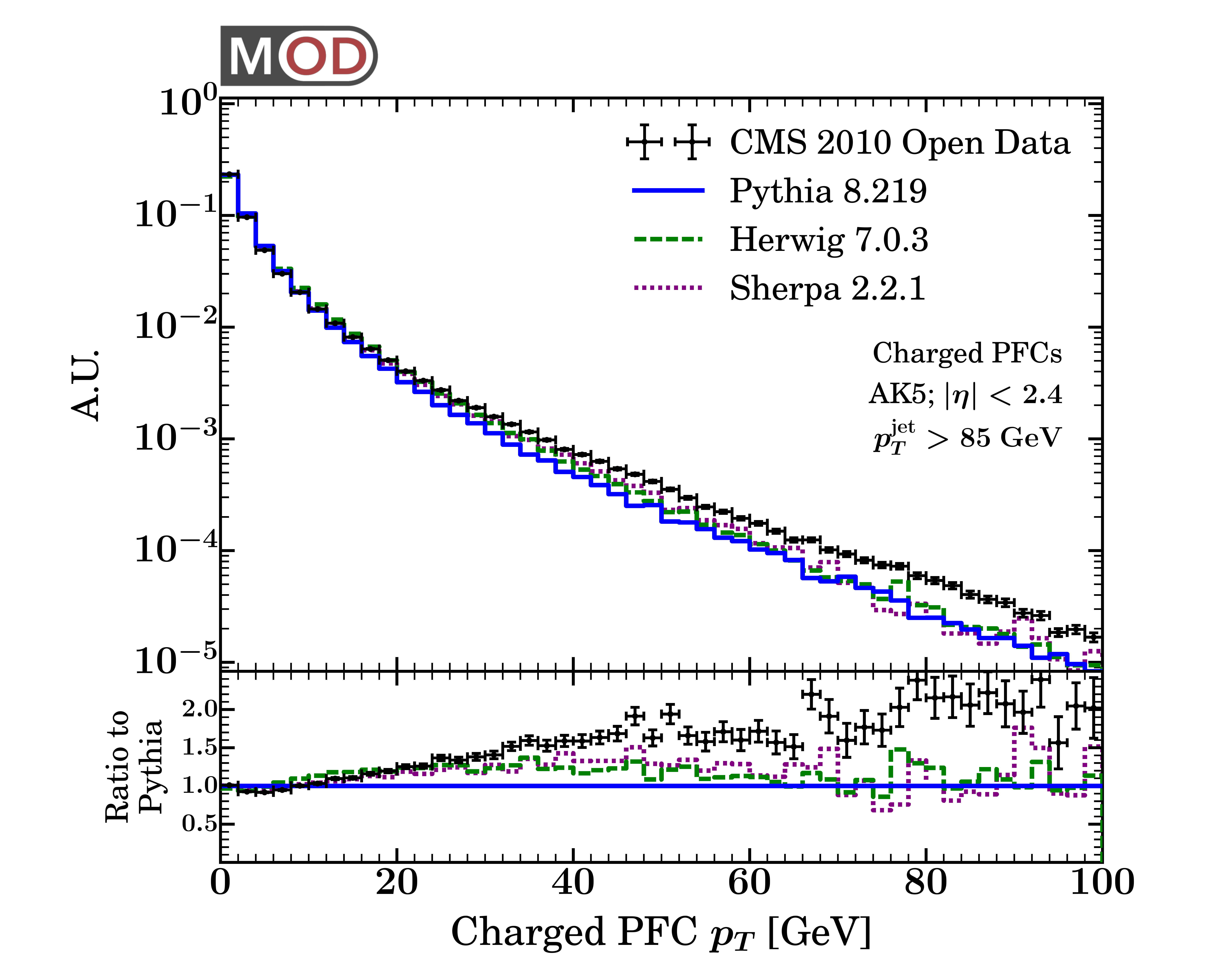}
}
\caption{Same as \Fig{fig:PFC}, but showing a wider range of PFC $p_T$ values.}
\label{fig:pfc_extended}
\end{figure*}

To partially account for detector effects in our substructure analysis, we impose a PFC cut of $p_T^{\rm min} = 1.0~\GeV$, motivated by \Fig{fig:PFC}.  In \Fig{fig:pfc_extended}, we plot the PFC $p_T$ spectrum over an extended range, again restricting to PFCs within the hardest jet.  For neutral particles, there is a growing difference between the CMS Open Data and the parton shower generators for constituents that carry a large fraction of the jet momentum, though this difference is reduced when considering only charged particles.

\section{Additional Soft-Dropped Distributions}
\label{app:yetmore}

\begin{figure}
\includegraphics[width=\columnwidth, page=1]{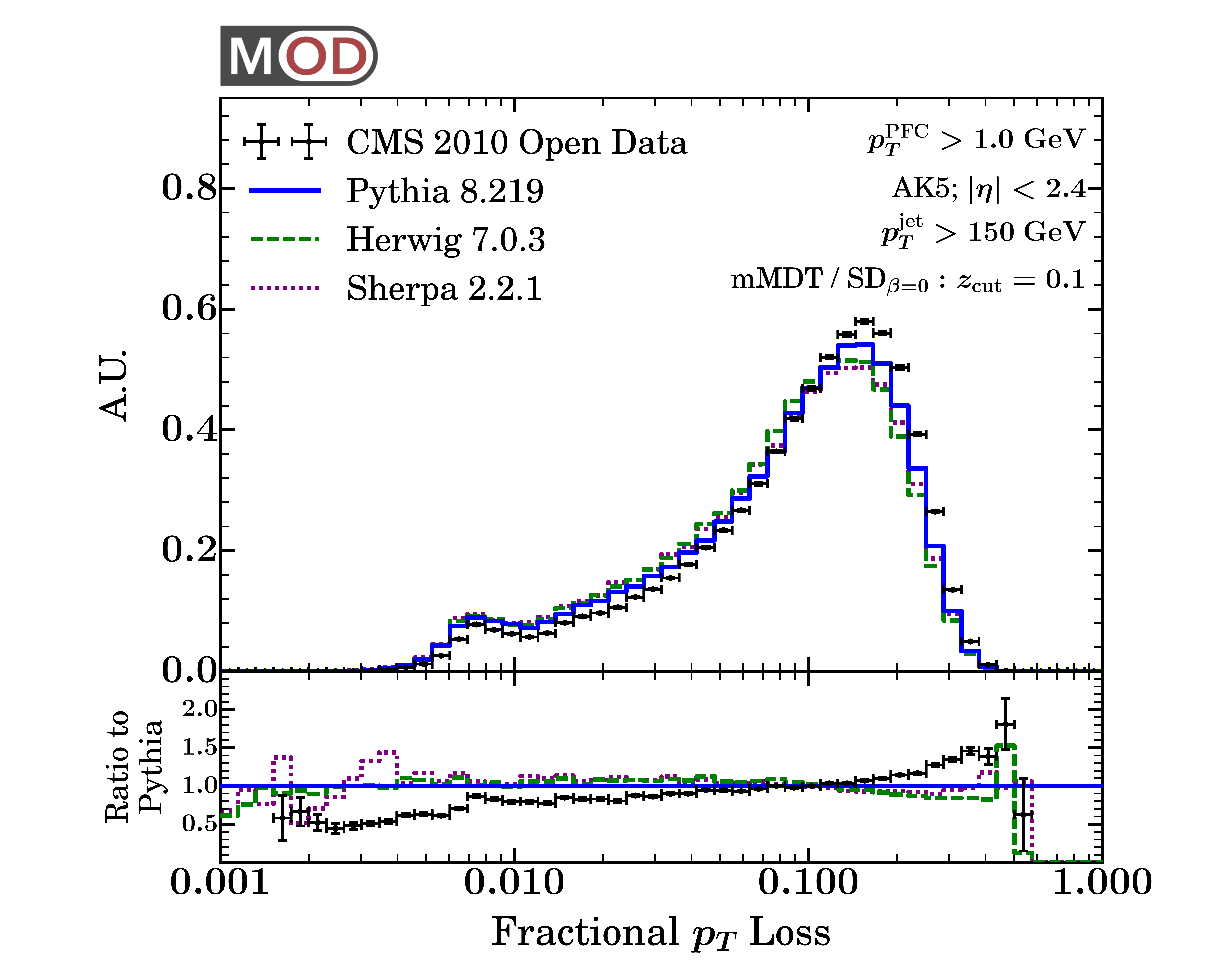}
\caption{Fraction of the orginal jet $p_T$ lost after performing soft drop declustering.  Because this is a fraction, no JEC factors are applied.}
\label{fig:softdrop_pT_loss}
\end{figure}

In this appendix, we show additional distributions obtained from soft drop declustering.  In \Fig{fig:softdrop_pT_loss}, we show the fraction of the original jet $p_T$ discarded after soft drop, plotted logarithmically.  This distribution was advocated in \Ref{Larkoski:2014wba} as an interesting example of a Sudakov safe \cite{Larkoski:2013paa,Larkoski:2015lea} observable, and we see good agreement between the CMS open data and parton showers. 

\begin{figure*}
\subfloat[]{
\label{fig:constituent_multiplicity_SD}
\includegraphics[width=0.9\columnwidth, page=1]{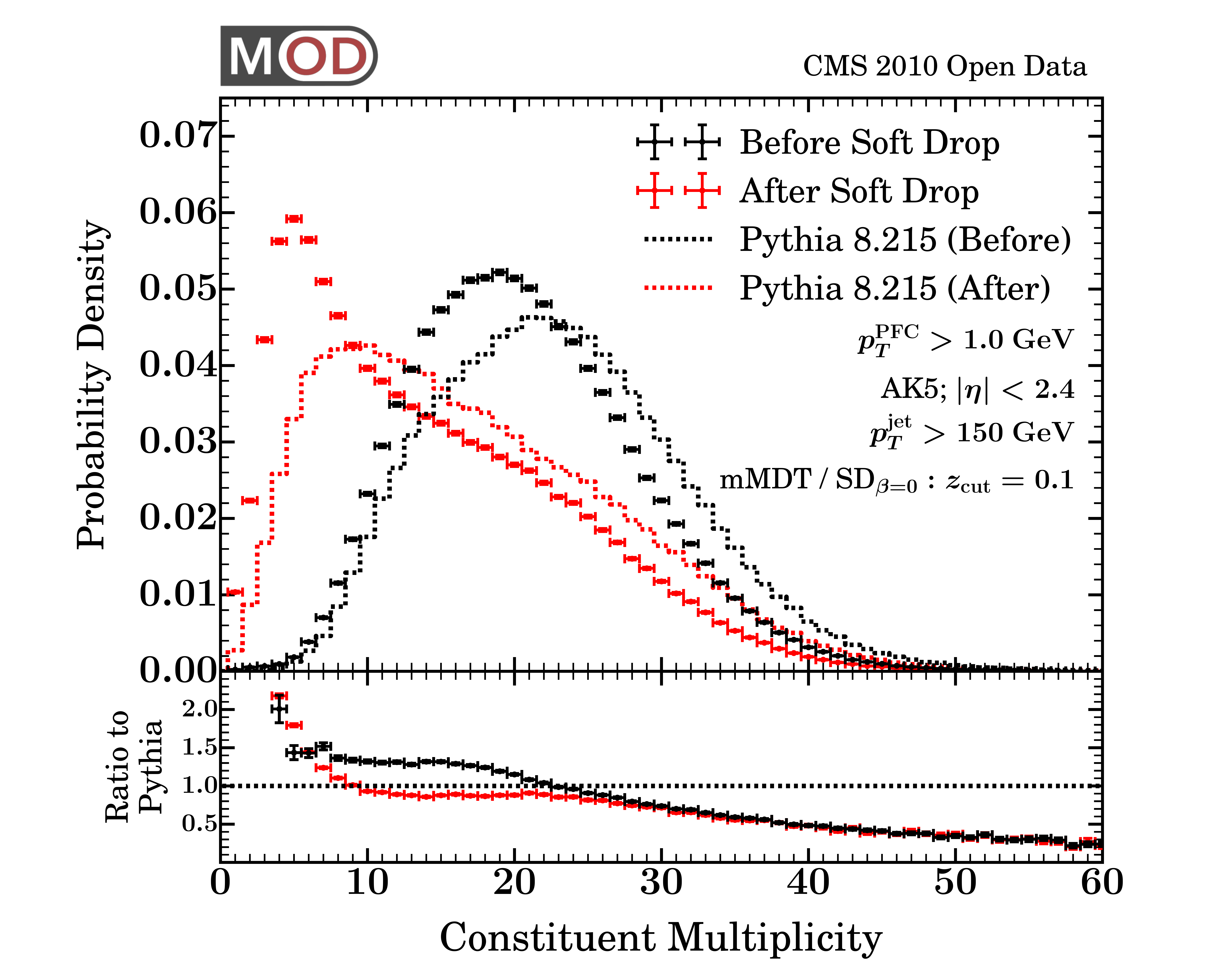}
}
\subfloat[]{
\label{fig:constituent_multiplicity_track_SD}
\includegraphics[width=0.9\columnwidth, page=1]{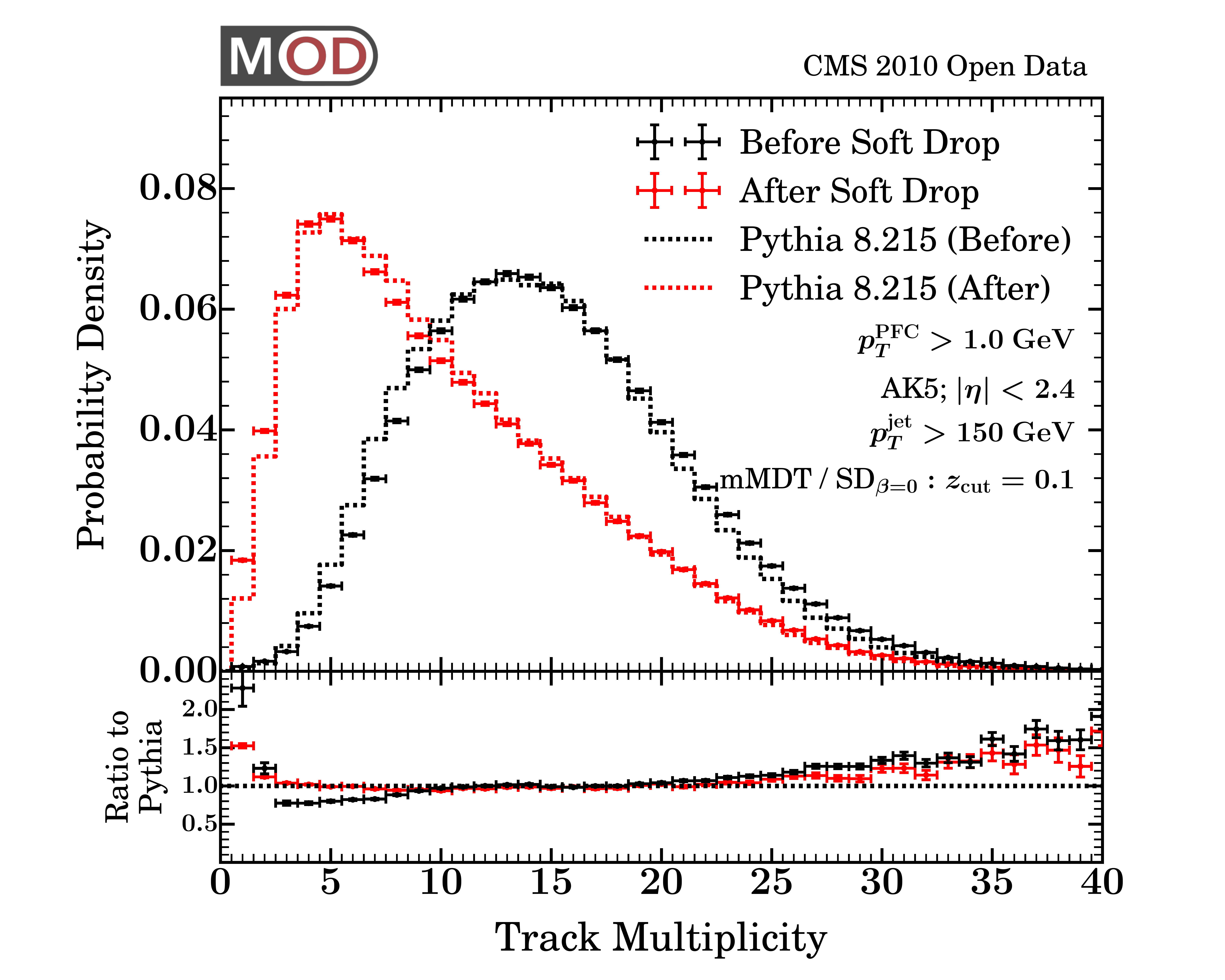}
}

\subfloat[]{
\label{fig:pTD_SD}
\includegraphics[width=0.9\columnwidth, page=1]{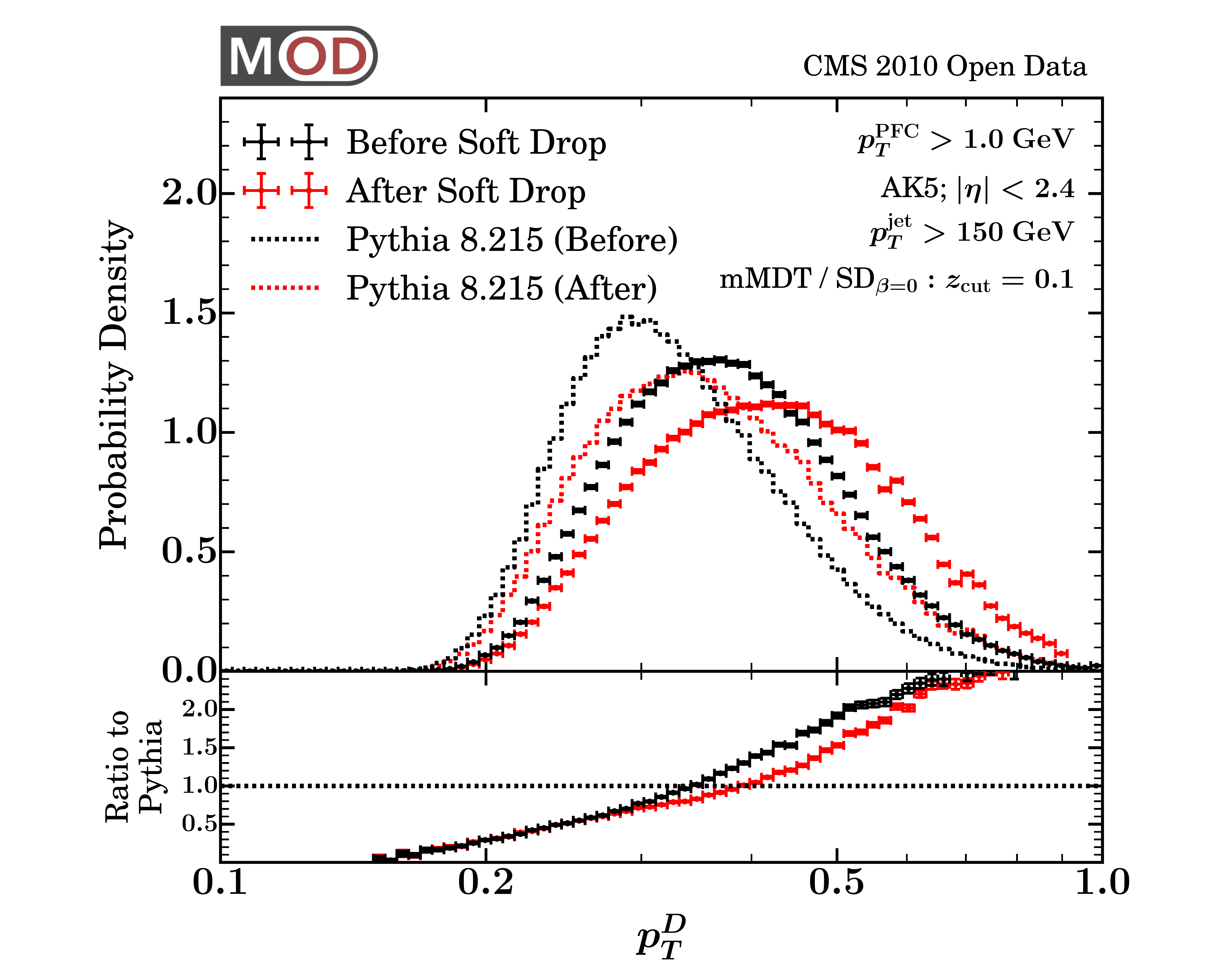}
}
\subfloat[]{
\label{fig:pTD_track_SD}
\includegraphics[width=0.9\columnwidth, page=1]{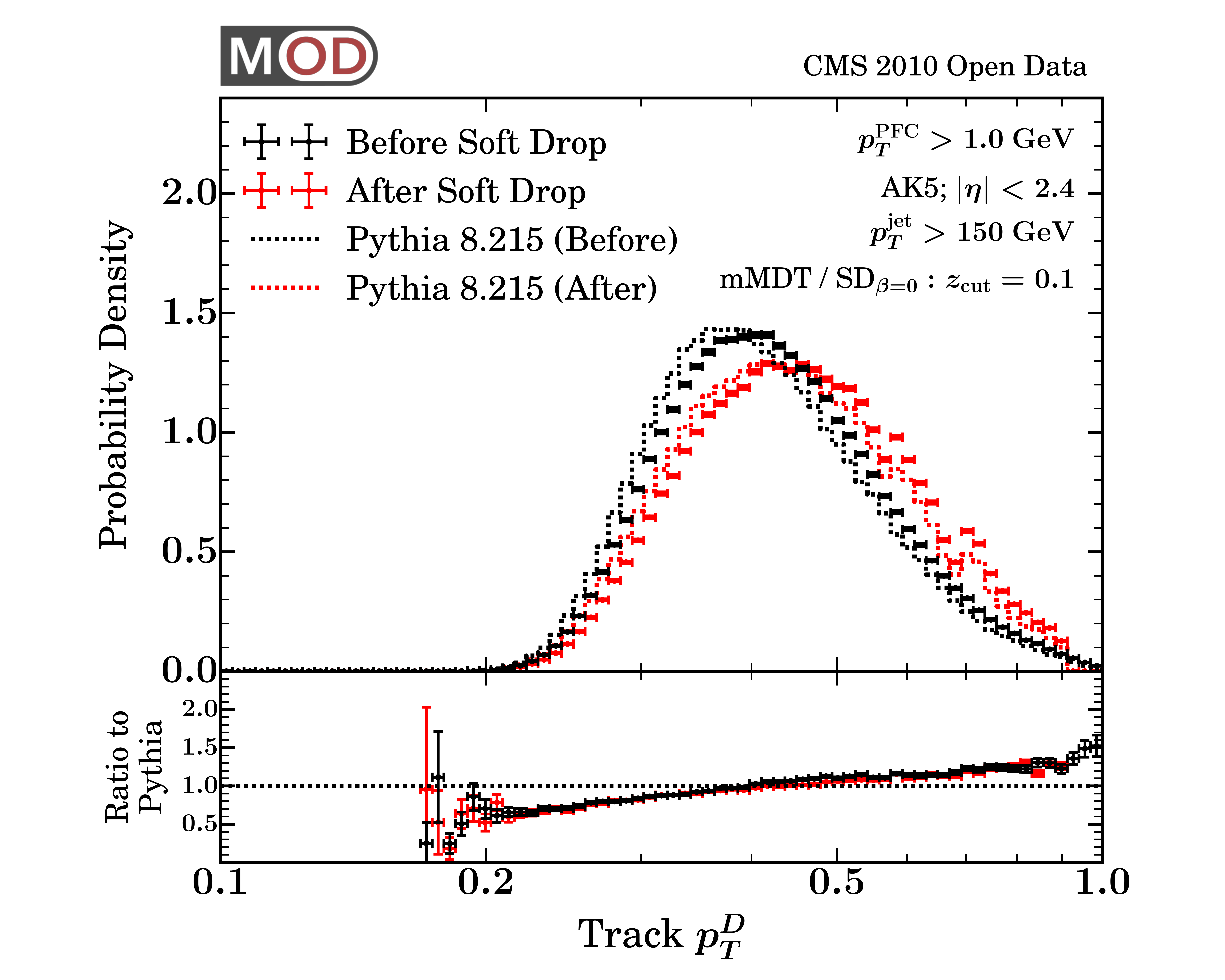}
}

\subfloat[]{
\label{fig:jet_mass_SD}
\includegraphics[width=0.9\columnwidth, page=1]{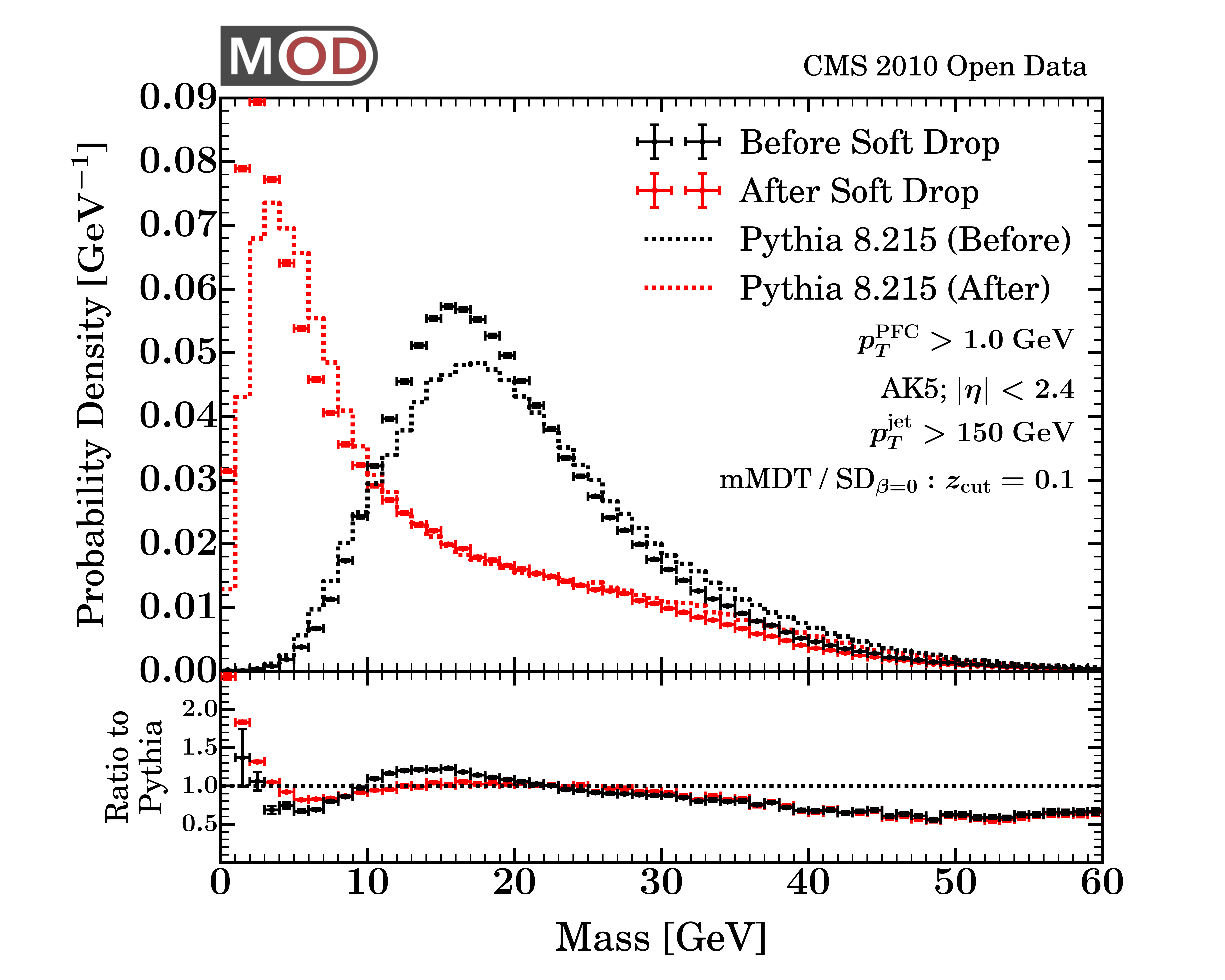}
}
\subfloat[]{
\label{fig:jet_mass_track_SD}
\includegraphics[width=0.9\columnwidth, page=1]{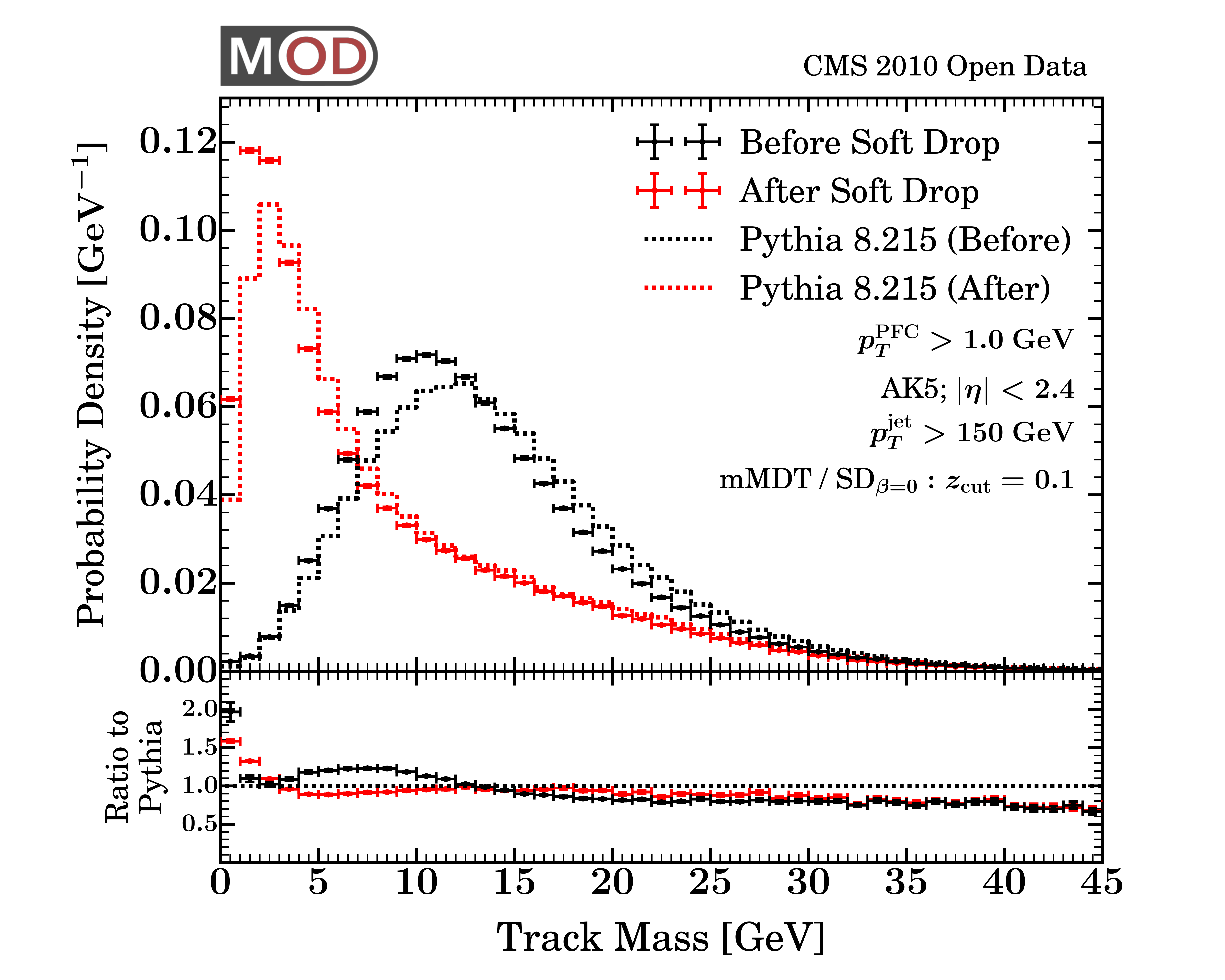}
}
\caption{Same observables as in \Fig{fig:basic_substructure}, but now showing the original distributions (black) compared to those obtained after soft drop declustering (red).}
\label{fig:basic_substructure_SD}
\end{figure*}

\begin{figure*}
\subfloat[]{
\label{fig:jet_LHA_SD}                            
\includegraphics[width=0.9\columnwidth, page=1]{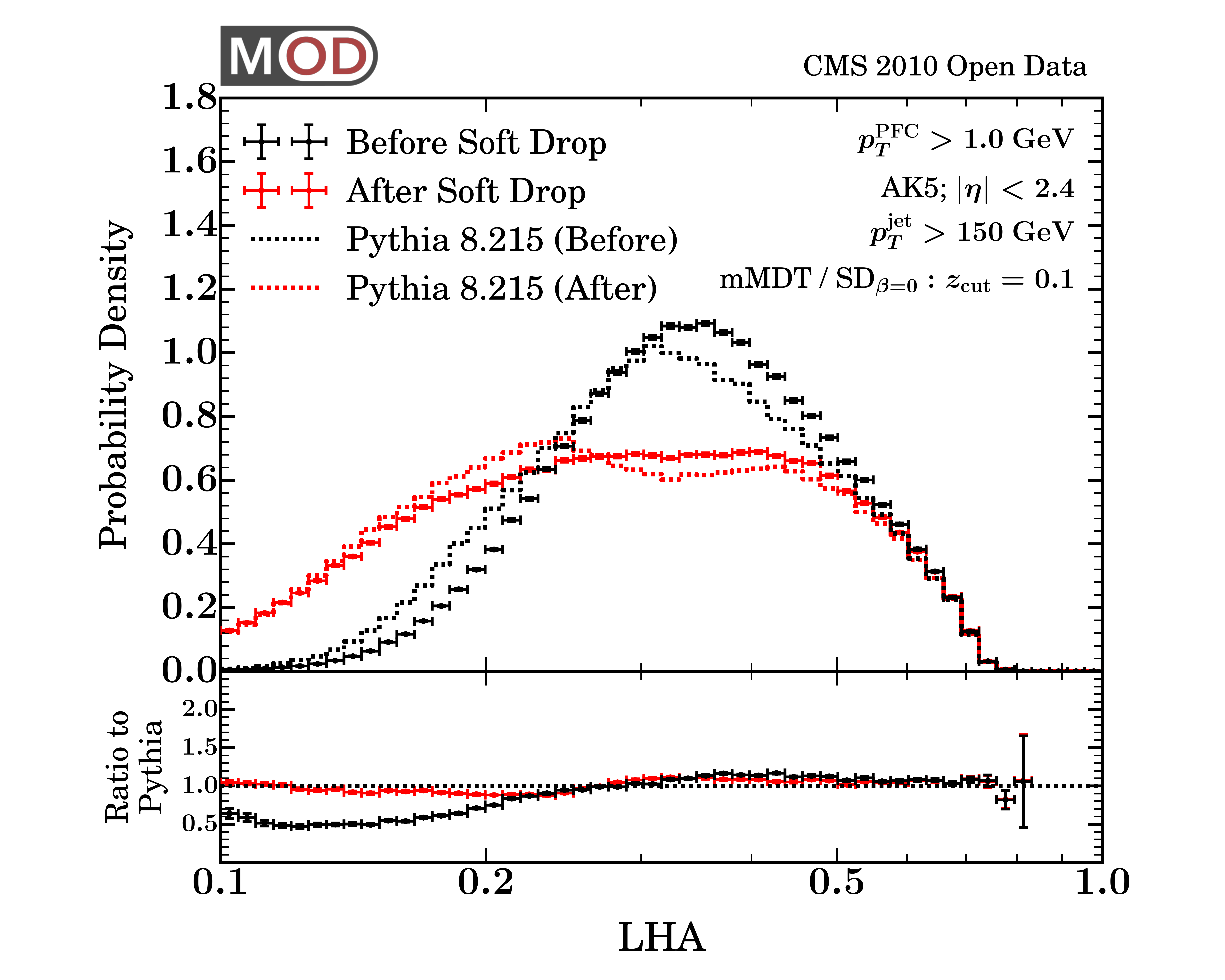}
}
\subfloat[]{
\label{fig:jet_LHA_track_SD}                            
\includegraphics[width=0.9\columnwidth, page=1]{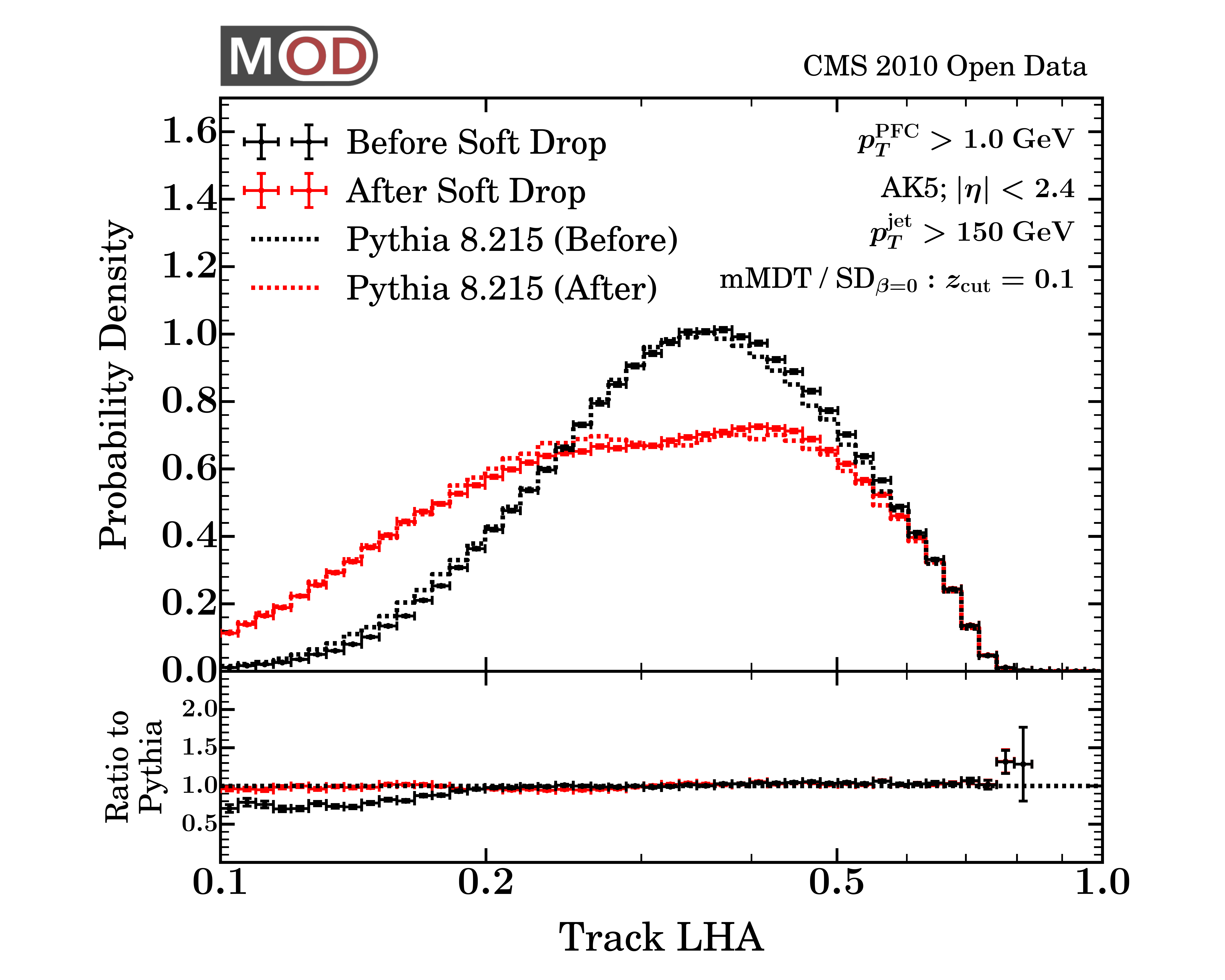}
}

\subfloat[]{
\label{fig:jet_width_SD}
\includegraphics[width=0.9\columnwidth, page=1]{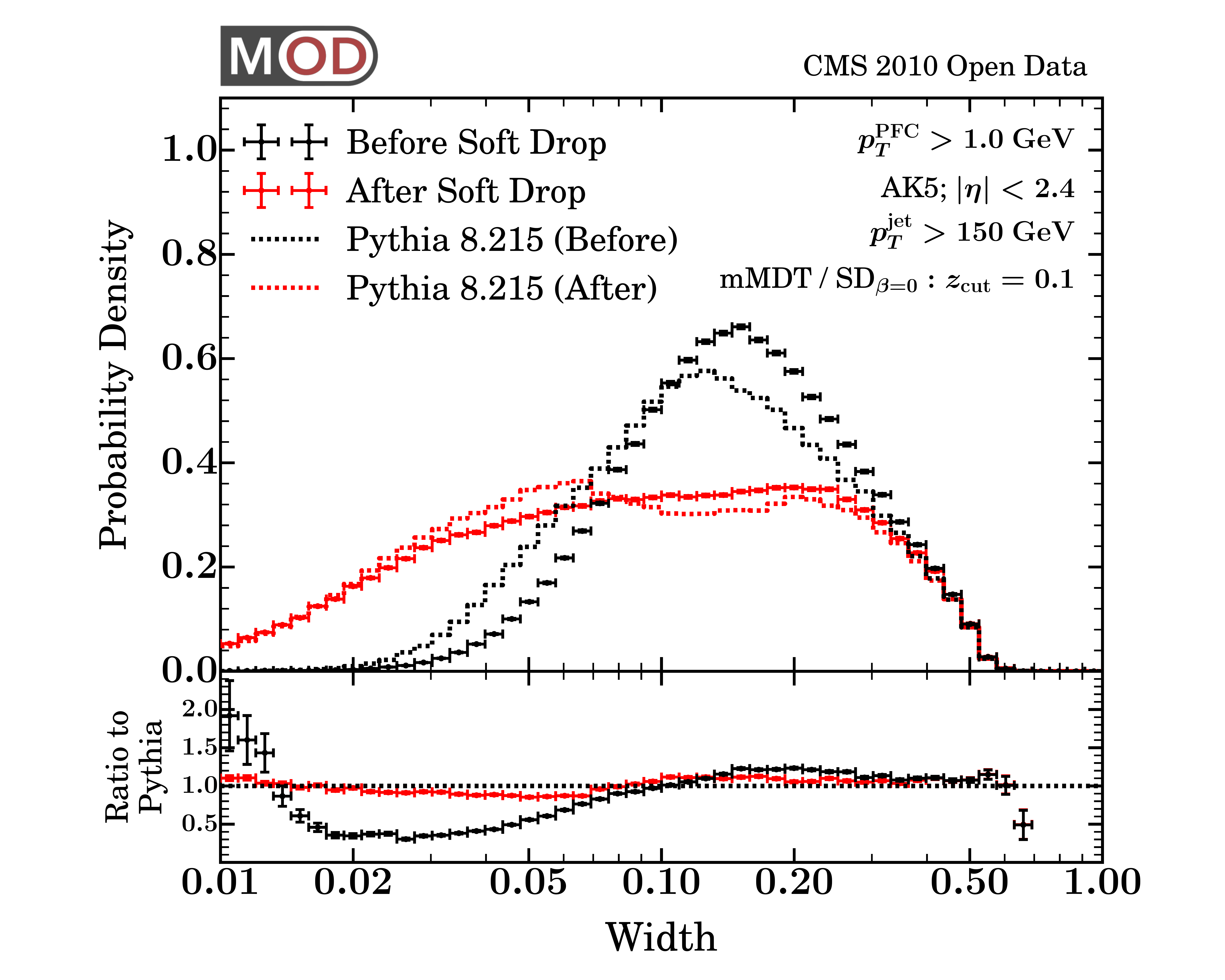}
}
\subfloat[]{
\label{fig:jet_width_track_SD}
\includegraphics[width=0.9\columnwidth, page=1]{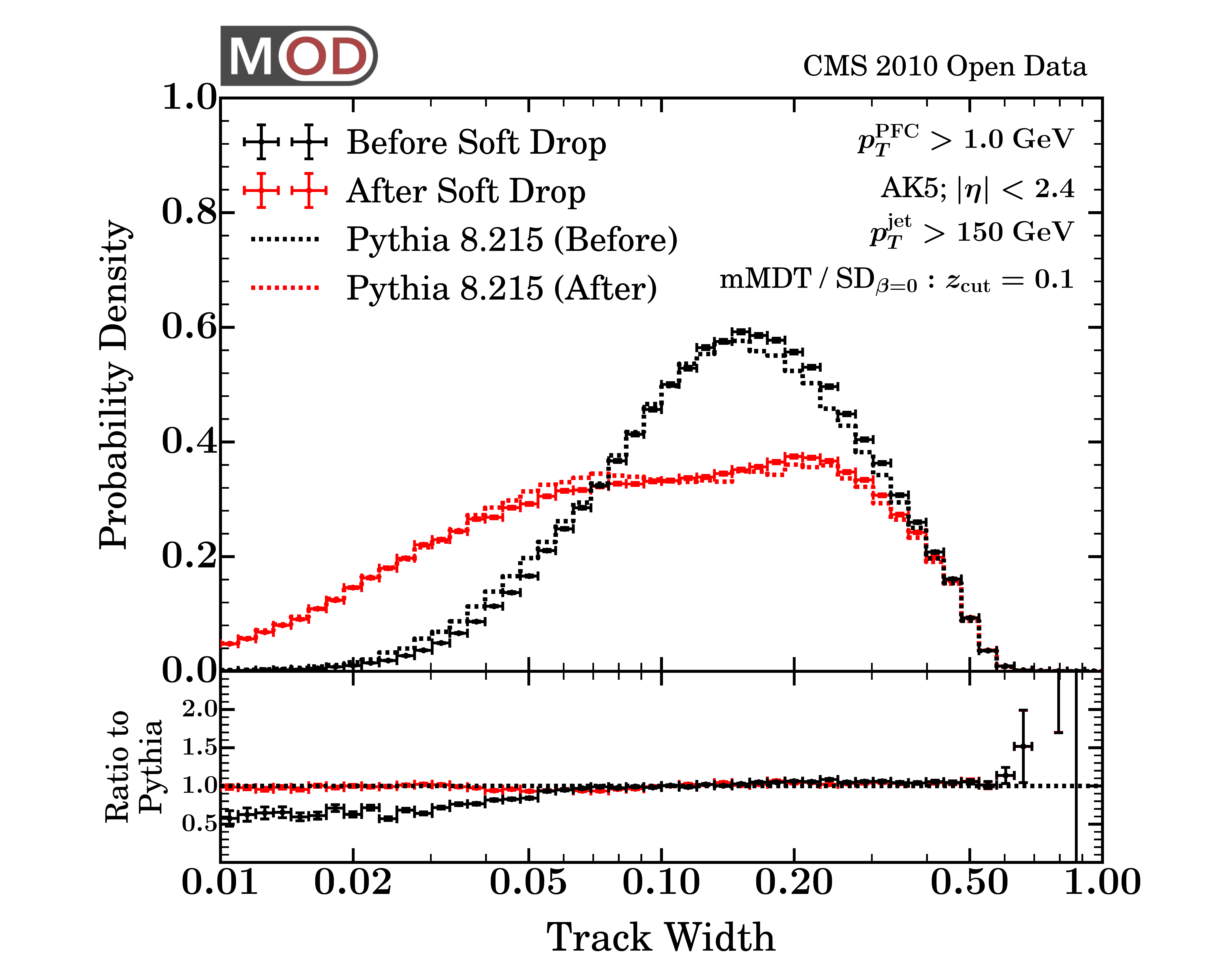}
}

\subfloat[]{
\label{fig:jet_thrust_SD}
\includegraphics[width=0.9\columnwidth, page=1]{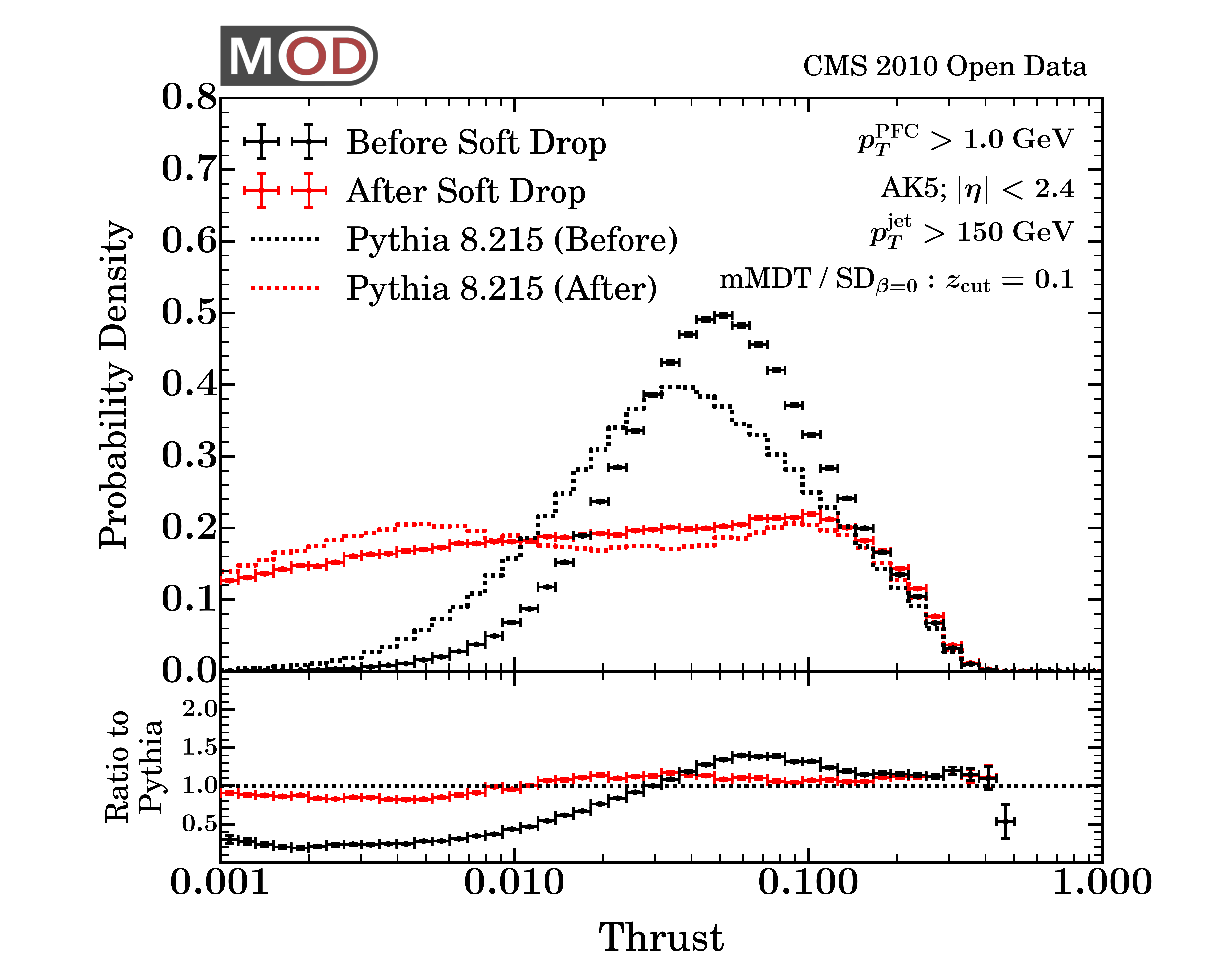}
}
\subfloat[]{
\label{fig:jet_thrust_track_SD}
\includegraphics[width=0.9\columnwidth, page=1]{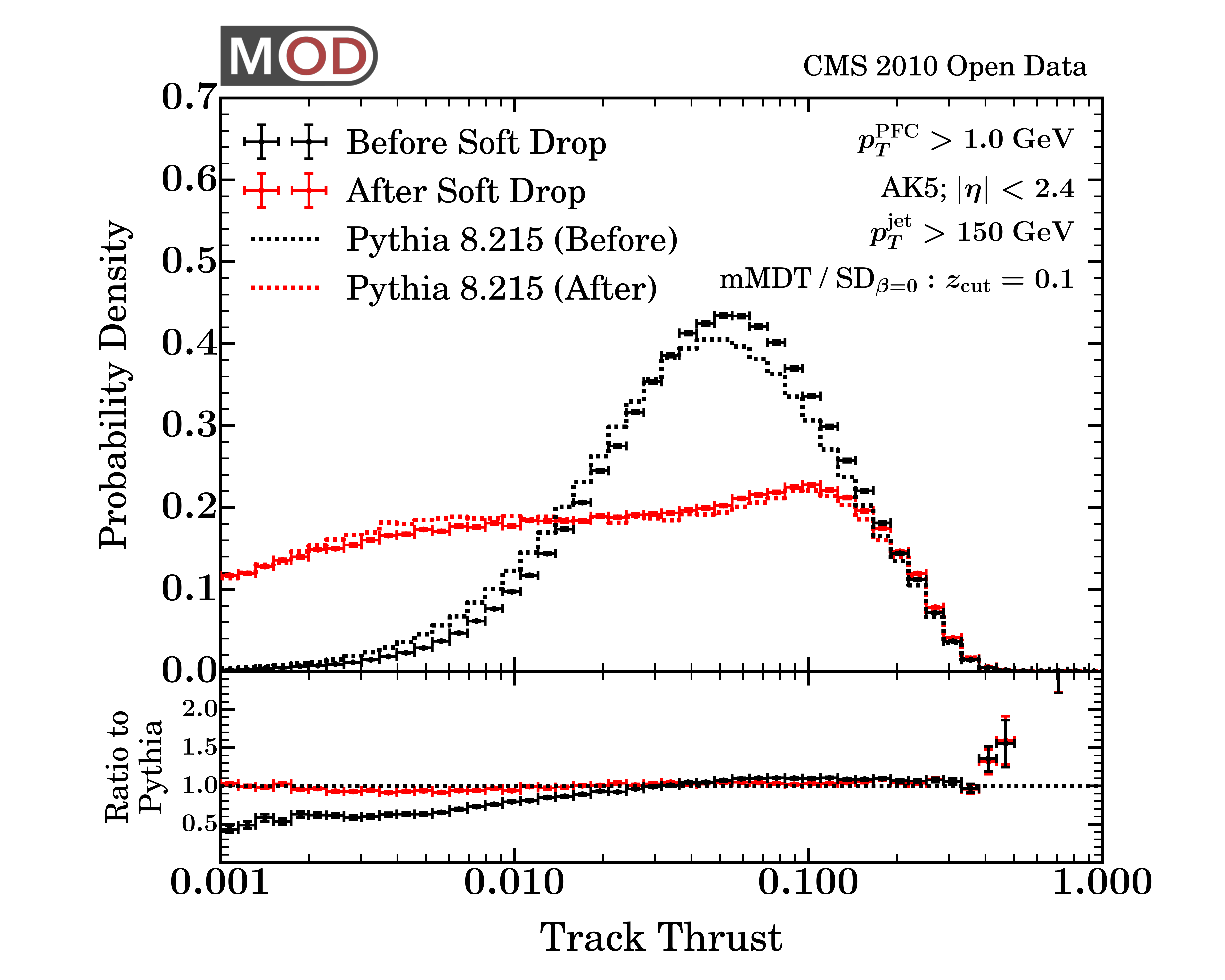}
}
\caption{Same observables as in \Fig{fig:jet_angularities}, but now showing the original distributions (black) compared to those obtained after soft drop declustering (red).}
\label{fig:jet_angularities_SD}
\end{figure*}

The distributions in \Sec{sec:hardest_jet} were obtained prior to applying any jet grooming.  In \Fig{fig:basic_substructure_SD}, we show the same basic substructure observables from \Fig{fig:basic_substructure}, but now showing the impact of soft drop.  Soft drop does not necessarily improve the agreement between the CMS Open Data and the \textsc{Pythia} parton shower, though it also does not make it any worse, and the track-based agreement is very good.  We perform a similar study in \Fig{fig:jet_angularities_SD} for the jet angularities from \Fig{fig:jet_angularities}.  There is good qualitative agreement between the open data and \textsc{Pythia}, but the track-only version has much better quantitative agreement as expected.

\bibliography{MOD_zg}

\end{document}